\documentclass [12pt,notitlepage]{article}
\usepackage{amsmath,amssymb,cite}
\usepackage{appendix}
\usepackage{feynmp}
\usepackage{float}
\usepackage[vcentermath,enableskew]{youngtab}
\usepackage{tabularx}
\usepackage[english]{babel}
\usepackage{graphicx}
\usepackage{caption}
\usepackage{bmpsize}
\usepackage{indentfirst}
\usepackage{epsfig}
\usepackage{epstopdf}
\usepackage[]{hyperref}
\usepackage[section]{placeins}
\usepackage[stable]{footmisc}
\usepackage{appendix}
\usepackage{tabularx}
\usepackage[english]{babel}
\usepackage{graphicx}
\usepackage{indentfirst}
\usepackage{epsfig}
\usepackage{slashed}
\usepackage{fancyhdr}
\numberwithin{equation}{section}

\fancypagestyle{plain}{\fancyhead{}
}

\begin{document}

\title{Quantum Black Holes
}

\author{Badis Ydri  \footnote{ydri@stp.dias.ie.}\\
Department of Physics, Faculty of Sciences, Annaba University,\\
 Annaba, Algeria.
}

\maketitle

\setcounter{page}{2} 

\begin{abstract}
This article is divided into three parts. First, a systematic derivation of the Hawking radiation is given in three different ways. The information loss problem is then discussed in great detail. The last part contains a concise discussion of black hole thermodynamics. This article was published as chapter $6$ of the IOP book "Lectures on General Relativity, Cosmology and Quantum Black Holes" (July $2017$).
\end{abstract}

\tableofcontents



\section{Introduction and Summary}
String theory provides one of the most deepest insights into quantum gravity. Its single most central and profound result is the AdS/CFT correspondence or gauge/gravity duality \cite{Maldacena:1997re}. See \cite{Natsuume:2014sfa,Nastase:2007kj} for a pedagogical introduction. As it turns out, this duality allows us to study in novel ways: i) the physics of strongly coupled gauge theory (QCD in particular and the existence of Yang-Mills theories in $4$ dimensions), as well as ii) the physics of black holes (the information loss paradox and the problem of the reconciliation of general relativity and quantum mechanics). String theory reduces therefore for us to the study of the AdS/CFT correspondence.

Indeed, the fundamental observation which drives the lectures in this chapter is that: ``BFSS matrix model \cite{Banks:1996vh} and the AdS/CFT duality \cite{Maldacena:1997re,Gubser:1998bc,Witten:1998qj} relates string theory in certain backgrounds to quantum mechanical systems and quantum field theories'' which is a quotation taken from Polchinski \cite{Polchinski:2016hrw}. The basic problem which is of paramount interest to quantum gravity is Hawking radiation of a black hole and the consequent evaporation of the hole and corresponding information loss \cite{Hawking:1974sw,Hawking:1976ra}. The BFSS and the AdS/CFT imply that there is no information loss paradox in the Hawking radiation of a black hole. This is the central question we would like to understand in great detail.

Towards this end, we need to understand first quantum black holes, before we can even touch the AdS/CFT correspondence, which requires in any case a great deal of conformal field theory and string theory as crucial ingredients. Thus, in this last chapter of this book we will only worry about black hole radiation, black hole thermodynamics and the information problem following \cite{Polchinski:2016hrw,Susskind:2005js,Page:1993up,Harlow:2014yka}.

The main reference, guideline and motivation behind these lectures is the lucid and elegant book by Susskind and Lindesay \cite{Susskind:2005js}. The lectures by Jacobson \cite{Jacobson:2003vx} and Harlow \cite{Harlow:2014yka} played also a major role in many crucial issues throughout. We have also benefited greatly from the books by Mukhanov \cite{Mukhanov:2007zz} and Carroll \cite{Carroll:2004st}. The reference list at the end of these lectures is very limited and only include articles that were actually consulted by the author in the preparation of this chapter. A far more extensive and exhaustive list of references can be found in Harlow \cite{Harlow:2014yka} and Jacobson \cite{Jacobson:2003vx}.   


We summarize the content of this article as follows

A systematic derivation of the Hawking radiation is given in three different ways. By employing the fact that the near-horizon geometry of Schwarzschild black hole is Rindler spacetime and then applying the Unruh effect in Rindler spacetime. Secondly, by considering the eternal black hole geometry and studying the properties of the Kruskal vacuum state with respect to the Schwarzschild observer. Thirdly, by considering a Schwarzschild black hole formed by gravitational collapse and deriving the actual incoming state known as the Unruh vacuum state. Although, the actual quantum state of the black hole is pure, the asymptotic Schwarzschild observer registers a thermal mixed state with temperature $T_H=1/(8\pi GM)$. Indeed, a correlated entangled pure state near the horizon gives rise to a thermal mixed state outside the horizon.

The information loss problem is then discussed in great detail. The black hole starts in a pure state and after its complete evaporation the Hawking radiation is also in a pure state. This is the assumption of unitarity. Thus, the entanglement entropy starts at zero value then it reaches a maximum value at the so-called Page time then drops to zero again. The Page time is the time at which the black hole evaporates around one half of its mass and the information starts to get out with the radiation. Before the Page time only energy gets out with the radiation with little or no information. The behavior of the entanglement entropy with time is called the Page curve and a nice rough derivation of this curve using the so-called Page theorem is outlined. 

The last part contains a discussion of the black hole thermodynamics. The thermal entropy is the maximum amount of information contained in the black hole. The entropy is mostly localized near the horizon, but quantum field theory (QFT) gives a divergent value, instead of the Bekenstein-Hawking value $S = A/4G$. QFT must be replaced by quantum gravity (QG) near the horizon and this separation of the QFT and QG degrees of freedom can be implemented by the stretched horizon which is a time like membrane, at a distance of one Planck length $l_P=\sqrt{G\hbar}$ from the actual horizon, and where the temperature gets very large and most of the black hole entropy accumulates.

\section{Rindler Spacetime and General Relativity}
\subsection{Rindler Spacetime}
We start with Minkowski spacetime with metric and interval 
\begin{eqnarray}
\eta_{\mu\nu}=(-1,+1,+1,+1)~,~ds^2=\eta_{\mu\nu}dx^{\mu}dx^{\nu}.
\end{eqnarray}
We recall the Planck length 
\begin{eqnarray}
l_P=\sqrt{\frac{\hbar G}{c^3}}.
\end{eqnarray}
Usually we will use the natural units $\hbar=c=1$.

We will first construct the so-called Rindler spacetime, i.e. a uniformly accelerating (non-inertial) reference frame with respect to (say) Minkowski spacetime. This is characterized by an artificial gravitational field which can be removed (the only known case of its kind) by a coordinates transformation. We will follow the presentation by 't Hooft \cite{tHooft}.
  
Let us consider an elevator in the vicinity of the Earth in free fall. The elevator is assumed to be sufficiently small so that the gravitational field inside can be taken to be uniform. By the equivalence principle all objects inside the elevator will accelerate in the same way. Thus, during the free fall of the elevator the observer inside will not experience any gravitational field at all since he is effectively weightless.

We consider the opposite situation in which an elevator in empty space, where there is no gravitational field, is uniformly accelerated upward. The observer inside will feel pressure from the floor as if he is near the Earth or any other planet. In other words, this observer will be experiencing an artificial uniform gravitational field  given precisely by the constant acceleration.   The question now is how does this observer inside the elevator sees spacetime?

Let $\xi^{\mu}$ be the coordinates system inside the elevator which is uniformly accelerated outward in the $x$ direction in outer space with an acceleration $a$. The motion of the elevator is given by the functions $x^{\mu}=x^{\mu}(\xi)$ where $x^{\mu}$ are the coordinates of Minkowski spacetime.  At time $\tau=0$, as measured by the observer inside the elevator, the two systems coincide. We take the origin to be at the middle floor of the elevator. 

During an infinitesimal time $d\tau$ the elevator can be assumed to have a constant velocity $v=ad\tau$. In other words, the motion of the elevator within this time is approximately inertial given by the Lorentz transformation 
\begin{eqnarray}
&&\xi^0=\gamma(x^0-\frac{v}{c}x^1)\Rightarrow d\tau \simeq x^0-a d\tau x^1\nonumber\\ 
&&\xi^1=\gamma(x^1-\frac{v}{c}x^0)\Rightarrow \xi^1\simeq x^1-ad\tau x^0\nonumber\\
&&\xi^2=x^2\nonumber\\
&&\xi^3=x^3.
\end{eqnarray}
We write this as (by suppressing the transverse directions)
\begin{eqnarray}
\left( \begin{array}{c}
x^0  \\
x^1  \end{array} \right)-\left( \begin{array}{c}
d\tau  \\
0  \end{array} \right)=\left( \begin{array}{cc}
1 & a d\tau \\
ad\tau  & 1 \end{array} \right)\left( \begin{array}{c}
0  \\
\xi^1  \end{array} \right).\label{LT}
\end{eqnarray}
This relates the coordinates $(\vec{\xi},d\tau)$ as measured by the observer in the elevator to the coordinates $(\vec{x},t)$ as measured by the Minkowski observer. The above transformation looks like a Poincar\'e transformation, i.e. a combination of a Lorentz transformation and a translation which is here in time. In many cases Poincar\'e transformations can be rewritten as Lorentz transformations with respect to a properly chosen reference point as the origin. The reference point here is given by 
\begin{eqnarray}
A^{\mu}=(0,1/a,0,0).
\end{eqnarray}
Indeed,
\begin{eqnarray}
\left( \begin{array}{c}
d\tau  \\
0  \end{array} \right)=\left( \begin{array}{cc}
0 & a d\tau \\
ad\tau  & 0 \end{array} \right)\left( \begin{array}{c}
0  \\
1/a  \end{array} \right).
\end{eqnarray}
Thus
\begin{eqnarray}
\left( \begin{array}{c}
x^0  \\
x^1 +1/a \end{array} \right)=\left( \begin{array}{cc}
1 & a d\tau \\
ad\tau  & 1 \end{array} \right)\left( \begin{array}{c}
0  \\
\xi^1+1/a  \end{array} \right).
\end{eqnarray}
We rewrite then the Lorentz transformation (\ref{LT}) as 
\begin{eqnarray}
\left( \begin{array}{c}
x^0  \\
\vec{x}+\vec{A} \end{array} \right)=(1+\delta L)\left( \begin{array}{c}
0  \\
\vec{\xi}+\vec{A}  \end{array} \right)~,~\delta L=\left( \begin{array}{cc}
0 & a d\tau \\
ad\tau  & 0 \end{array} \right).
\end{eqnarray}
We repeat this $N$ times. In other words, at time $\tau=N d\tau$ the Minkowski coordinates $x^{\mu}=(t,\vec{x})$ are related to the elevator coordinates $\xi^{\mu}=(\tau,\vec{\xi})$ by
 \begin{eqnarray}
\left( \begin{array}{c}
x^0  \\
\vec{x}+\vec{a}/a^2 \end{array} \right)=L(\tau)\left( \begin{array}{c}
0  \\
\vec{\xi}+\vec{a}/a^2  \end{array} \right)~,~L(\tau)=(1+\delta L)^N.
\end{eqnarray}
Then we have 
\begin{eqnarray}
L(\tau+d\tau)=(1+\delta L)L(\tau).
\end{eqnarray}
The solution can be put in the form (suppressing again transverse directions)
\begin{eqnarray}
L(\tau)=\left( \begin{array}{cc}
A(\tau) & B(\tau) \\
B(\tau)  & A(\tau) \end{array} \right).
\end{eqnarray}
The initial condition is 
\begin{eqnarray}
L(0)={\bf 1}\leftrightarrow A(0)=1~,~B(0)=0.
\end{eqnarray}
We have then the differential equation 
\begin{eqnarray}
\delta L. L(\tau)=L(\tau+d\tau)-L(\tau)=d\tau \left( \begin{array}{cc}
\frac{dA}{d\tau} & \frac{dB}{d\tau} \\
\frac{dB}{d\tau}  & \frac{dA}{d\tau} \end{array} \right).
\end{eqnarray}
Equivalently 
\begin{eqnarray}
\frac{dA}{d\tau}=aB~,~\frac{dB}{d\tau}=aA.
\end{eqnarray}
The solution is then 
\begin{eqnarray}
A=\cosh a\tau~,~B=\sinh a\tau.
\end{eqnarray}
Finally we get the coordinates 
\begin{eqnarray}
&&x^0=\sinh a\tau.(\xi^1+\frac{1}{a})\nonumber\\
&&x^1=\cosh a\tau.(\xi^1+\frac{1}{a})-\frac{1}{a}\nonumber\\
&&x^2=\xi^2\nonumber\\
&&x^3=\xi^3.\label{rindler}
\end{eqnarray}
We compute immediately 
\begin{eqnarray}
-(dx^0)^2+(dx^1)^2=-a^2(\xi^1+\frac{1}{a})^2 d\tau^2+(d\xi^1)^2.
\end{eqnarray}
Thus, the metric in Rindler spacetime is given by (with $\xi^0=\tau$)
\begin{eqnarray}
ds^2=g_{\mu\nu}d\xi^{\mu}d\xi^{\nu}=-a^2(\xi^1+\frac{1}{a})^2 d\tau^2+d\vec{\xi}^2.\label{Rindler}
\end{eqnarray}
This is one of the simplest Riemann spacetimes. More on this spacetime in the following discussion.

\subsection{Review of General Relativity}
We consider a Riemannian (curved) manifold ${\cal M}$ with a metric $g_{\mu\nu}$. A coordinates transformation is given by 
\begin{eqnarray}
x^{\mu}\longrightarrow x^{'\mu}=x^{'\mu}(x).
\end{eqnarray}
The vectors and one-forms on the manifold are quantities which are defined to transform under the above coordinates transformation respectively as follows 
\begin{eqnarray}
V^{'\mu}=\frac{\partial x^{'\mu}}{\partial x^{\nu}}V^{\nu}.\label{contr}
\end{eqnarray}
\begin{eqnarray}
V^{'}_{\mu}=\frac{\partial x^{\nu}}{\partial x^{'\mu}}V_{\nu}.\label{cov}
\end{eqnarray}
The spaces of vectors and one-forms are the tangent and co-tangent bundles. 

A tensor is a quantity with multiple indices (covariant and contravariant) transforming in a similar way, i.e. any contravariant index is transforming as (\ref{contr}) and any covariant index is transforming as (\ref{cov}). For example, the metric $g_{\mu\nu}$ is a second rank symmetric tensor which transforms as
 \begin{eqnarray}
g^{'}_{\mu\nu}(x^{'})=\frac{\partial x^{\alpha}}{\partial x^{'\mu}}\frac{\partial x^{\beta}}{\partial x^{'\nu}}g_{\alpha\beta}(x).
\end{eqnarray}
The interval $ds^2=g_{\mu\nu}dx^{\mu}dx^{\nu}$ is therefore invariant. In fact, all scalar quantities are invariant under coordinate transformations. For example, the volume element $d^4x\sqrt{-{\rm det}g}$ is a scalar under coordinate transformation. 

The derivative of a tensor does not transform as a tensor. However, the so-called covariant derivative of a tensor will transform as a tensor. The covariant derivatives of vectors and one-forms are given by  
\begin{eqnarray}
\nabla_{\mu}V^{\nu}=\partial_{\mu}V^{\nu}+\Gamma_{\alpha\mu}^{\nu}V^{\alpha}.
\end{eqnarray}
\begin{eqnarray}
\nabla_{\mu}V_{\nu}=\partial_{\mu}V_{\nu}-\Gamma_{\mu\nu}^{\alpha}V_{\alpha}.
\end{eqnarray}
These transforms indeed as tensors as one can easily check. Generalization to tensor is obvious. The Christoffel symbols $\Gamma_{\mu\nu}^{\alpha}$ are given in terms of the metric $g_{\mu\nu}$ by
\begin{eqnarray}
\Gamma_{\mu\nu}^{\alpha}=\frac{1}{2}g^{\alpha\beta}\big(\partial_{\mu}g_{\nu\beta}+\partial_{\nu}g_{\mu\beta}-\partial_{\beta}g_{\mu\nu}\big).
\end{eqnarray}
There exists a unique covariant derivative, and thus a unique choice of Christoffel symbols,  for which the metric is covariantly constant, viz
\begin{eqnarray}
\nabla_{\mu}g_{\alpha\beta}=0.
\end{eqnarray}
The straightest possible lines on the curved manifolds are given by the geodesics. A geodesic is a curve whose tangent vector is parallel transported along itself.  It is given explicitly by the Newton's second law on the curved manifold
 \begin{eqnarray}
\frac{d^2x^{\mu}}{d\lambda}+\Gamma^{\mu}_{\alpha\beta}\frac{dx^{\alpha}}{d\lambda}\frac{dx^{\beta}}{d\lambda}=0.
\end{eqnarray}
The $\lambda$ is an affine parameter along the curve. The time like geodesics define the trajectories of freely falling particles in the gravitational field encoded in the curvature of the Riemannian manifold. The Riemann curvature tensor $R_{\mu\nu\beta}^{\alpha}$ is defined in terms of the covariant derivative by 
\begin{eqnarray}
(\nabla_{\mu}\nabla_{\nu}-\nabla_{\nu}\nabla_{\mu})t^{\alpha}=-R_{\mu\nu\beta}^{\alpha}t^{\beta}.
\end{eqnarray}
The metric is determined by the Hilbert-Einstein action given by 
 \begin{eqnarray}
S=\frac{1}{16\pi G}\int d^4x\sqrt{-{\rm det}g}R,
\end{eqnarray}
where the Ricci scalar $R$ is defined from the Ricci tensor $R_{\mu\nu}$ by 
\begin{eqnarray}
R=g^{\mu\nu}R_{\mu\nu}.
\end{eqnarray}
\begin{eqnarray}
R_{\mu\nu}=R_{\mu\alpha\nu}^{\alpha}.
\end{eqnarray}
The Riemann tensor is given explicitly by
\begin{eqnarray}
R_{\mu\nu\rho}^{\alpha}=\partial_{\nu}\Gamma_{\mu\rho}^{\alpha}-\partial_{\rho}\Gamma_{\mu\nu}^{\alpha}+\Gamma_{\sigma\nu}^{\alpha}\Gamma_{\mu\rho}^{\sigma}-\Gamma_{\rho\sigma}^{\alpha}\Gamma_{\mu\nu}^{\sigma}.
\end{eqnarray}
Indeed, the Euler-Lagrange equations which follows from the above action are precisely the Einstein equations in vacuum, viz
 \begin{eqnarray}
\delta S=\frac{1}{16\pi G}\int d^4x\sqrt{-{\rm det}g}(R_{\mu\nu}-\frac{1}{2}g_{\mu\nu}R)\delta g^{\mu\nu}=0\Rightarrow R_{\mu\nu}-\frac{1}{2}g_{\mu\nu}R=0.
\end{eqnarray}
If we add matter action $S_{\rm matter}$ we obtain the full Einstein equations of motion, viz 
  \begin{eqnarray}
R_{\mu\nu}-\frac{1}{2}g_{\mu\nu}R=8\pi GT_{\mu\nu}.
\end{eqnarray}
The energy-momentum tensor is defined by the equation 
 \begin{eqnarray}
T_{\mu\nu}=-\frac{2}{\sqrt{-{\rm det} g}}\frac{\delta S_{\rm matter}}{\delta g^{\mu\nu}}.
\end{eqnarray}
The cosmological constant is one of the simplest matter action that one can add to the Hilbert-Einstein action. It is given by 
\begin{eqnarray}
S_{\rm cc}=-\frac{1}{8\pi G}\int d^4x\sqrt{-{\rm det}g}\Lambda.
\end{eqnarray}
In this case the energy-momentum tensor and the Einstein equations read
\begin{eqnarray}
T_{\mu\nu}=-\frac{\Lambda}{8\pi G}g_{\mu\nu}.
\end{eqnarray}
 \begin{eqnarray}
R_{\mu\nu}-\frac{1}{2}g_{\mu\nu}R+\Lambda g_{\mu\nu}=0.
\end{eqnarray}

\section{Schwarzschild Black Holes}
\subsection{Schwarzschild Black Holes}
Without further ado we present our first (eternal) black hole. The Schwarzschild black hole is given by the metric 
 \begin{eqnarray}
ds^2=-(1-\frac{2GM}{r})dt^2+(1-\frac{2GM}{r})^{-1}dr^2+r^2d\Omega^2.
\end{eqnarray}
The powerful Birkhoff's theorem states that the Schwarzschild metric is the unique vacuum solution (static or otherwise) to Einstein's equations which is spherically symmetric.

The Schwarzschild radius is given by 
 \begin{eqnarray}
r_s=2GM.
\end{eqnarray}
This is the event horizon. We remark that the Schwarzschild metric is apparently singular at $r=0$ and at $r=r_s$.  However, only the singularity at $r=0$ is a true singularity of the geometry. For example we can check that the scalar quantity $R^{\mu\nu\alpha\beta}R_{\mu\nu\alpha\beta}$ is divergent at $r=0$ whereas it is perfectly finite at $r=r_s$ since \cite{Carroll:1997ar}
\begin{eqnarray}
R^{\mu\nu\alpha\beta}R_{\mu\nu\alpha\beta}=\frac{48G^2M^2}{r^6}.
\end{eqnarray}
Indeed, the divergence of the Ricci scalar\footnote{Actually $R=0$ for the  Schwarzschild metric.} or any other higher order scalar such as $R^{\mu\nu}R_{\mu\nu}$, $R^{\mu\nu\alpha\beta}R_{\mu\nu\alpha\beta}$, etc at a point is a sufficient condition for that point to be singular.  We say that $r=0$ is an essential singularity. The  Schwarzschild radius  $r=r_s$ is not a true singularity of the metric and its appearance as such only reflects the fact that the chosen coordinates are behaving badly at  $r=r_s$. We say that $r=r_s$ is a coordinate singularity. Indeed, it should appear like any other point if we choose a more appropriate coordinates system. 
 
The Riemann tensor encodes the effect of tidal forces on freely falling objects. Thus, the singularity at $r=0$ corresponds to infinite tidal forces.

The motion of test particles in (Schwarzschild or otherwise) spacetime is given by the geodesic equation 
\begin{eqnarray}
\frac{d^2x^{\rho}}{d\lambda^2}+\Gamma^{\rho}~_{\mu\nu}\frac{dx^{\mu}}{d\lambda}\frac{dx^{\nu}}{d\lambda} =0.
\end{eqnarray}
The Schwarzschild metric is obviously invariant under time translations and space rotations. There will therefore be $4$ corresponding Killing vectors $K_{\mu}$ and $4$ conserved quantities (energy and angular momentum) given by
\begin{eqnarray}
Q=K_{\mu}\frac{dx^{\mu}}{d\lambda}.
\end{eqnarray}
The metric is independent of $x^0$ and $\phi$ and hence the corresponding Killing vectors are 
\begin{eqnarray}
K^{\mu}=(\partial_{x^0})^{\mu}=\delta^{\mu}_0=(1,0,0,0)~,~K_{\mu}=g_{\mu 0}=(-(1-\frac{R_s}{r}),0,0,0).
\end{eqnarray}
\begin{eqnarray}
R^{\mu}=(\partial_{\phi})^{\mu}=\delta^{\mu}_{\phi}=(0,0,0,1)~,~R_{\mu}=g_{\mu \phi}=(0,0,0,r^2\sin^2\theta).
\end{eqnarray}
The corresponding conserved quantities are the energy and the magnitude of the angular momentum given by
\begin{eqnarray}
E=-K_{\mu}\frac{dx^{\mu}}{d\lambda}=(1-\frac{r_s}{r})\frac{dx^0}{d\lambda}.\label{consE}
\end{eqnarray}
\begin{eqnarray}
L=R_{\mu}\frac{dx^{\mu}}{d\lambda}=r^2\sin^2\theta\frac{d\phi}{d\lambda}.
\end{eqnarray}
There is an extra conserved quantity along the geodesic given by (use the geodesic equation and the fact that the metric is covariantly constant)
\begin{eqnarray}
\epsilon=-g_{\mu\nu}\frac{dx^{\mu}}{d\lambda}\frac{dx^{\nu}}{d\lambda}.
\end{eqnarray}
Clearly,
\begin{eqnarray}
&& \epsilon=1~,~{\rm massive}~{\rm particle}.
\end{eqnarray}
\begin{eqnarray}
&&\epsilon=0~,~{\rm massless}~{\rm particle}.
\end{eqnarray}
This extra conserved quantity leads to the radial equation of motion

\begin{eqnarray}
\frac{1}{2}\big(\frac{dr}{d\lambda}\big)^2+V(r)={\cal E}~,~
{\cal E}=\frac{1}{2}(E^2-\epsilon).
\end{eqnarray}
The potential is given by
\begin{eqnarray}
V(r)
&=&-\frac{\epsilon GM}{r}+\frac{L^2}{2r^2}-\frac{GML^2}{r^3}.
\end{eqnarray}
This is the equation of a particle with unit mass and energy ${\cal E}$ in a potential $V(r)$. In this potential only the last term is new compared to Newtonian gravity. Clearly when $r\longrightarrow 0$ this potential will go to $-\infty$ whereas if the last term is absent (the case of Newtonian gravity) the potential will go to $+\infty$ when $r\longrightarrow 0$. 

For a radially (vertically) freely object we have $d\phi/d\lambda =0$ and thus the angular momentum is $0$, viz $L=0$. The radial equation of motion becomes
 \begin{eqnarray}
\big(\frac{dr}{d\lambda}\big)^2-\frac{2GM}{r}=E^2-1.\label{free1}
\end{eqnarray}
This is essentially the Newtonian equation of motion. The conserved energy is given by
 \begin{eqnarray}
E=(1-\frac{2GM}{r})\frac{dt}{d\lambda}.\label{free2}
\end{eqnarray}
We also consider the situation in which the particle was initially at rest at $r=r_i$, viz
\begin{eqnarray}
\frac{dr}{d\lambda}|_{r=r_i}=0.
\end{eqnarray}
This means in particular that
\begin{eqnarray}
E^2-1=-\frac{2GM}{r_i}.
\end{eqnarray}
The equation of motion becomes
 \begin{eqnarray}
\big(\frac{dr}{d\lambda}\big)^2=\frac{2GM}{r}-\frac{2GM}{r_i}.\label{lk}
\end{eqnarray}
We can identify the affine parameter $\lambda$ with the proper time for a massive particle. The proper time required to reach the point $r=r_f$ is 
 \begin{eqnarray}
\tau=\int_0^{\tau} d\lambda =-(2GM)^{-\frac{1}{2}}\int_{r_i}^{r_f} dr\sqrt{\frac{rr_i}{r_i-r}}.
\end{eqnarray}
The minus sign is due to the fact that in a free fall $dr/d\lambda <0$. By performing the change of variables $r=r_i(1+\cos\alpha)/2$ we find the closed result
  \begin{eqnarray}
\tau=\sqrt{\frac{r_i^3}{8GM}}(\alpha_f+\sin\alpha_f).
\end{eqnarray}
This is finite when $r_f\longrightarrow r_s=2GM$. Thus, a freely falling object will cross the  Schwarzschild radius in a finite proper time. 

We consider now a distant stationary observer hovering at a fixed radial distance  $r_{\infty}$. His proper time is 
\begin{eqnarray}
\tau_{\infty}=\sqrt{1-\frac{2GM}{r_{\infty}}} t.
\end{eqnarray}
By using equations (\ref{free1}) and (\ref{free2}) we can find $dr/dt$. We get
\begin{eqnarray}
\frac{dr}{dt}&=&-E^{\frac{1}{2}}\frac{d\lambda}{dt}(E-\frac{d\lambda}{dt})^{\frac{1}{2}}\nonumber\\
&=&-\frac{1}{{E}}(1-\frac{2GM}{r})\bigg(E^2-1+\frac{2GM}{r} \bigg)^{\frac{1}{2}}.
\end{eqnarray}
Near $r=2GM$ we have
\begin{eqnarray}
\frac{dr}{dt}&=&-\frac{1}{2GM}(r-r_s).\label{lk1}
\end{eqnarray}
The solution is
\begin{eqnarray}
r-r_s=\exp(-\frac{t}{2GM}).
\end{eqnarray}
Thus when $r\longrightarrow r_s=2GM$ we have $t\longrightarrow \infty$.

We see that with respect to a stationary distant observer at a fixed radial distance $r_{\infty}$ the elapsed time $\tau_{\infty}$ goes to infinity as  $r\longrightarrow 2GM$. The correct interpretation of this result is to say that the stationary distant observer can never see the particle actually crossing the  Schwarzschild radius $r_s=2GM$ although the particle does cross the  Schwarzschild radius in a finite proper time as seen by an observer falling with the particle.

\subsection{Near Horizon Coordinates}
A proper distance from the horizon can be defined by the formula
 \begin{eqnarray}
\rho=\int_{r_s}^r\sqrt{g_{rr}(r^{'})}dr^{'}&=&\int_{r_s}^r\frac{dr^{'}}{\sqrt{1-r_s/r}}\nonumber\\&=&\sqrt{r(r-r_s)}+r_s\sinh\sqrt{\frac{r}{r_s}-1}.
\end{eqnarray}
In terms of $\rho$ the metric becomes 
 \begin{eqnarray}
ds^2=-(1-\frac{r_s}{r(\rho)})dt^2+d\rho^2+r^2(\rho)d\Omega^2.
\end{eqnarray}
Very near the horizon we write $r=r_s+\delta$ and thus $\rho=2\sqrt{r_s\delta}$. We get then the metric 
 \begin{eqnarray}
ds^2=-\rho^2\frac{dt^2}{4r_s^2}+d\rho^2+r_s^2d\Omega^2.
\end{eqnarray}
The first two terms correspond to two-dimensional Minkowski flat space. Indeed, $\rho$ and $\omega=t/2r_s$ are radial and hyperbolic angle variables for Minkowski spacetime. The Minkowski coordinates $X$ and $T$ are defined by 
\begin{eqnarray}
X=\rho\cosh \frac{t}{2r_s}~,~T=\rho\sinh \frac{t}{2r_s}.
\end{eqnarray}
The metric becomes 
 \begin{eqnarray}
ds^2=-dT^2+dX^2+r_s^2d\Omega^2.
\end{eqnarray}
If we are only interested in small angular region of the horizon around $\theta=0$ we can replace the angular variables by Cartesian coordinates as follows
\begin{eqnarray}
Y=r_s\theta\cos\phi~,~Z=r_s\theta\sin\phi.
\end{eqnarray}
We have then the metric 
 \begin{eqnarray}
ds^2&=&-\rho^2d\omega^2+d\rho^2+dY^2+dZ^2\nonumber\\
&=&-dT^2+dX^2+dY^2+dZ^2.
\end{eqnarray}
By comparing with (\ref{Rindler}), we recognize the first line to be the Rindler metric with the identification $a\tau\leftrightarrow \omega$ and $\xi^1+1/a\leftrightarrow \rho$. The time $\omega$ is called Rindler time and the time translation $\omega\longrightarrow \omega+c$ corresponds to a Lorentz boost in Minkowski spacetime. This approximation  of the black hole near-horizon geometry (valid for $r\simeq r_s$ and small angular region) by a Minkowski spacetime is called the Rindler approximation. It shows explicitly that the event horizon is locally non-singular and in fact it is indistinguishable from flat Minkowski spacetime.

\begin{figure}[H]
\begin{center}
\includegraphics[width=10.0cm,angle=0]{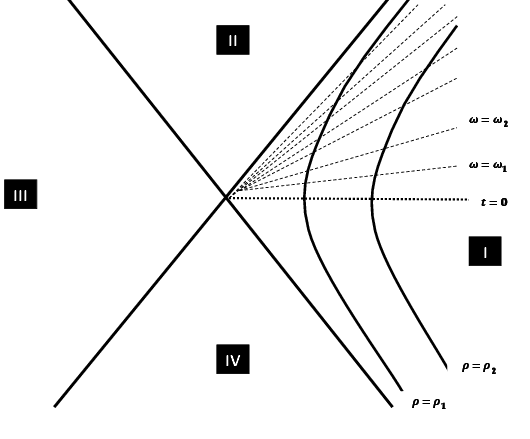}
\end{center}
\caption{Rindler spacetime.}\label{susskind}
\end{figure}

The relation between the Minkowski coordinates $X=\rho\cosh \omega$ and $T=\rho\sinh \omega$ and the Rindler coordinates $\rho$ and $\omega$ can also be rewritten as
 \begin{eqnarray}
\rho^2=X^2-T^2~,~\frac{T}{X}=\tanh \omega.
\end{eqnarray}
Obviously we must have $X>|T|$. This is called quadrant I or Rindler spacetime. This is the region outside the black hole. The lines of constant $\rho$ are hyperbolae while the lines of constant $\omega$ are straight lines through the origin. The horizon lies at the point $\rho=0$ or $T=X=0$. The horizon is actually a two-dimensional surface located at $r=r_s$ since $g_{00}=0$ there and as a consequence this surface has no time extension. See figure (\ref{susskind}).

\section{Kruskal-Szekres Diagram}
\subsection{Kruskal-Szekres Extension and Einstein-Rosen Bridge}

In this lecture we will follow \cite{Carroll:2004st}. The above  Schwarzschild geometry can be maximally extended as follows. For a radial null curve, which corresponds to a photon moving radially in  Schwarzschild spacetime,  the angles $\theta$ and $\phi$ are constants and $ds^2=0$, and thus
\begin{eqnarray}
0=-(1-\frac{2GM}{r})dt^2+(1-\frac{2GM}{r})^{-1}dr^2.
\end{eqnarray}
In other words,
\begin{eqnarray}
\frac{dt}{dr}=\pm \frac{1}{1-\frac{2GM}{r}}.
\end{eqnarray}
We integrate the above equation as follows
\begin{eqnarray}
t&=&\pm \int \frac{dr}{1-\frac{2GM}{r}}\nonumber\\
&=&\pm \bigg(r+ 2GM\log(\frac{r}{2GM}-1)\bigg)+{\rm constant}\nonumber\\
&=&\pm r_*+{\rm constant}.
\end{eqnarray}
We call $r_*$ the tortoise coordinate which makes sense only for $r>2GM$. The event horizon $r=2GM$ corresponds to $r_*\longrightarrow \infty$. We compute $dr_*=rdr/(r-2GM)$ and as a consequence the Schwarzschild metric becomes
\begin{eqnarray}
ds^2=(1-\frac{2GM}{r})(-dt^2+dr_*^2)+r^2d\Omega^2.
\end{eqnarray}
Next we define $v=t+r_*$ and $u=t-r_*$. Then
\begin{eqnarray}
ds^2=-(1-\frac{2GM}{r})dv du+r^2d\Omega^2.
\end{eqnarray}
For infalling radial null geodesics we have $t=-r_*$ or equivalently $v={\rm constant}$ whereas for outgoing  radial null geodesics we have $t=+r_*$ or equivalently $u={\rm constant}$. For every point in spacetime we have two solutions:
\begin{itemize}
\item {} For points outside the event horizon there are two solutions one infalling and one outgoing.
\item {} For points inside the event horizon there are two solutions which are both infalling.
\item {} For points on the event horizon there are two solutions one infalling and one trapped.
\end{itemize}
Next, we will give a maximal extension of the Schwarzschild solution by constructing a coordinate system valid everywhere in Schwarzschild spacetime. We start by noting that the radial coordinate $r$ should be given in terms of $u$ and $v$ by solving the equations 
\begin{eqnarray}
\frac{1}{2}(v-u)=r+2GM\log(\frac{r}{2GM}-1).
\end{eqnarray}
The event horizon $r=2GM$ is now either at $v=-\infty$ or $u=+\infty$. The coordinates of the event horizon can be pulled to finite values by defining new coordinates $u^{'}$ and $v^{'}$ as
\begin{eqnarray}
v^{'}&=&\exp(\frac{v}{4GM})\nonumber\\
&=&\sqrt{\frac{r}{2GM}-1}\exp(\frac{r+t}{4GM}).
\end{eqnarray}
\begin{eqnarray}
u^{'}&=&-\exp(-\frac{u}{4GM})\nonumber\\
&=&-\sqrt{\frac{r}{2GM}-1}\exp(\frac{r-t}{4GM}).
\end{eqnarray}
The Schwarzschild metric becomes 
\begin{eqnarray}
ds^2=-\frac{32G^3M^3}{r}\exp(-\frac{r}{2GM})dv^{'} du^{'}+r^2d\Omega^2.
\end{eqnarray}
It is clear that the coordinates $u$ and $v$ are null coordinates and thus 
$u^{'}$ and $v^{'}$ are also null coordinates. However, we prefer to work with a single time like  coordinate while we prefer the other coordinate to be space  like. We introduce therefore new coordinates $T$ and $R$ defined for $r>2GM$ by
\begin{eqnarray}
T=\frac{1}{2}(v^{'}+u^{'})=\sqrt{\frac{r}{2GM}-1}\exp(\frac{r}{4GM})\sinh\frac{t}{4GM}.
\end{eqnarray}
\begin{eqnarray}
R=\frac{1}{2}(v^{'}-u^{'})=\sqrt{\frac{r}{2GM}-1}\exp(\frac{r}{4GM})\cosh\frac{t}{4GM}.
\end{eqnarray}
Clearly, $T$ is time like while $R$ is space like. This can be confirmed by computing the metric. This is given by
\begin{eqnarray}
ds^2=\frac{32G^3M^3}{r}\exp(-\frac{r}{2GM})(-dT^2+dR^2)+r^2d\Omega^2.
\end{eqnarray}
We see that $T$ is always time like while $R$ is always space like since the sign of the components of the metric never get reversed.

\begin{figure}[H]
\begin{center}
\includegraphics[width=13.0cm,angle=0]{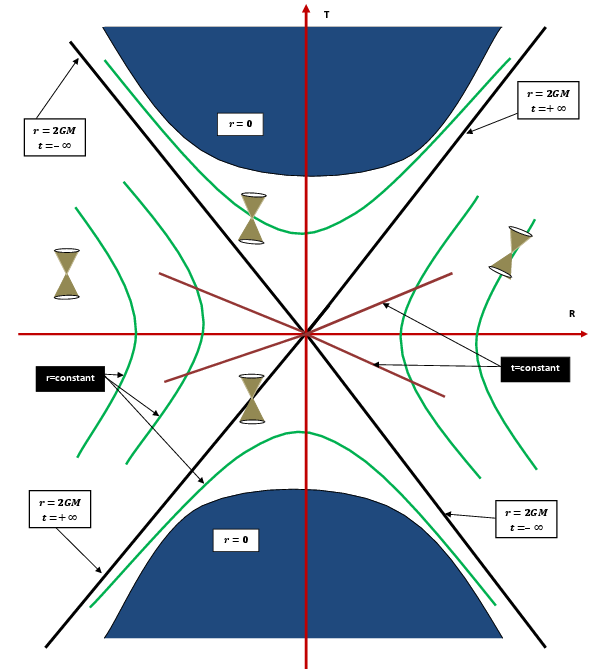}
\end{center}
\caption{Kruskal-Szekres diagram.}\label{carroll}
\end{figure}

We remark that
\begin{eqnarray}
T^2-R^2&=&v^{'}u^{'}\nonumber\\
&=&-\exp\frac{v-u}{4GM}\nonumber\\
&=&-\exp\frac{r+2GM\log(\frac{r}{2GM}-1)}{2GM}\nonumber\\
&=&(1-\frac{r}{2GM})\exp\frac{r}{2GM}.\label{impl}
\end{eqnarray}
The radial coordinate $r$ is determined implicitly in terms of $T$ and $R$ from this equation, i.e. equation (\ref{impl}). The coordinates $(T,R,\theta,\phi)$ are called Kruskal-Szekres coordinates. Remarks are now in order:
\begin{itemize}
\item{}The radial null curves in this system of coordinates are given by 
\begin{eqnarray}
T=\pm R+{\rm constant}.\label{horizon}
\end{eqnarray}
All light cones are at $\pm 45$ degrees. This $45$-degree property means in particular that the radial light cone in the Kruskal-Szekeres diagram has the same form as in special relativity.

\item{}The horizon defined by $r\longrightarrow 2GM$ is seen to appear at $T^2-R^2\longrightarrow 0$, i.e. at (\ref{horizon}) in the new coordinate system. This shows in an elegant way that the event horizon is a null surface.

\item{}The surfaces of constant $r$ are given from (\ref{impl}) by $T^2-R^2={\rm constant}$ which are hyperbolae in the $R-T$ plane. 
\item{}For $r>2GM$ the surfaces of constant $t$ are given by $T/R=\tanh \frac{t}{4GM}={\rm constant} $ which are straight lines through the origin. In the limit $t\longrightarrow \pm \infty$ we have $T/R\longrightarrow \pm 1$ which is precisely the horizon $r=2GM$.
\end{itemize}
The above solution defines region I of the so-called the Kruskal-Szekres diagram. This solution can be extended to the interior region of the black hole $r<2GM$ (region II of the Kruskal-Szekres diagram) as follows:
\begin{itemize}
\item{}For $r<2GM$ we have
\begin{eqnarray}
T=\frac{1}{2}(v^{'}+u^{'})=\sqrt{1-\frac{r}{2GM}}\exp(\frac{r}{4GM})\cosh\frac{t}{4GM}.
\end{eqnarray}
\begin{eqnarray}
R=\frac{1}{2}(v^{'}-u^{'})=\sqrt{1-\frac{r}{2GM}}\exp(\frac{r}{4GM})\sinh\frac{t}{4GM}.
\end{eqnarray}
The metric and the condition determining $r$ implicitly in terms of $T$ and $R$ do not change form in the $(T,R,\theta,\phi)$ system of coordinates and thus the radial null curves, the horizon as well as the surfaces of constant $r$ are given by the same equation as before. 
\item{}For $r<2GM$ the surfaces of constant $t$ are given by $R/T=\tanh \frac{t}{4GM}={\rm constant} $ which are straight lines through the origin. 
\item{}It is clear that the allowed range for $R$ and $T$ is (analytic continuation from the region $T^2-R^2<0$ ($r>2GM$) to the first singularity which occurs in the region $T^2-R^2<1$ ($r<2GM$))
\begin{eqnarray}
-\infty\leq R\leq +\infty~,~T^2-R^2< 1.
\end{eqnarray}
\end{itemize}
The  Kruskal-Szekres diagram gives the maximal extension of the Schwarzschild solution. A Kruskal-Szekres diagram is shown on figure  (\ref{carroll}). Every point in this diagram is actually a $2-$dimensional sphere since we are suppressing $\theta$ and $\phi$ and drawing only $R$ and $T$.  The Kruskal-Szekres diagram  represents the entire Schwarzschild spacetime. It can be divided into $4$ regions:

\begin{itemize}
\item{}Region I: Exterior of black hole with $r>2GM$ ($R>0$ and $T^2-R^2<0$). Clearly future directed time like (null) worldlines will lead to region II whereas past directed time like (null) worldlines can reach it from region IV. Regions I and III are connected by space like geodesics.
\item{}Region II: Inside of black hole with $r<2GM$ ($T>0$, $0<T^2-R^2<1$). Any future directed path in this region will hit the singularity. In this region $r$ becomes time like (while $t$ becomes space like) and thus we can not stop moving in the direction of decreasing $r$ in the same way that we can not stop time progression in region I. 
\item{}Region III: Parallel exterior region with $r>2GM$ ($R<0$, $T^2-R^2<0$). This is another asymptotically flat region of spacetime which we can not access along future or past directed paths. The Kruskal-Szekres coordinates inside this region are
\begin{eqnarray}
T=-\sqrt{\frac{r}{2GM}-1}\exp(\frac{r}{4GM})\sinh\frac{t}{4GM}.
\end{eqnarray}
\begin{eqnarray}
R=-\sqrt{\frac{r}{2GM}-1}\exp(\frac{r}{4GM})\cosh\frac{t}{4GM}.
\end{eqnarray}

\item{}Region IV: Inside of white hole with $r<2GM$ ($T<0$, $0<T^2-R^2<1$). The white hole is the time reverse of the black hole. This corresponds to a singularity in the past at which the universe originated. This is a part of spacetime from which observers can escape to reach us while we can not go there.

\end{itemize}

\begin{figure}[H]
\begin{center}
\includegraphics[width=12.0cm,angle=0]{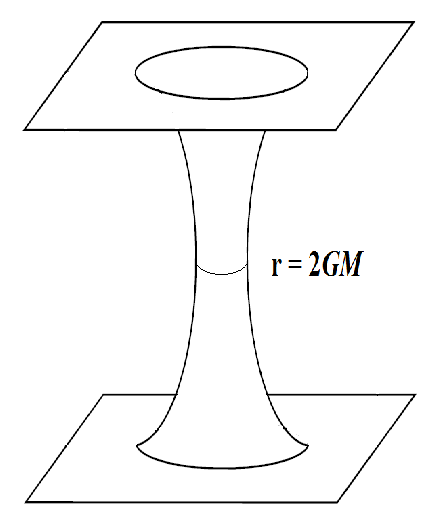}
\end{center}
\caption{The Einstein-Rosen bridge.}\label{carrolle}
\end{figure}

The full metric describes therefore  two asymptotically flat universes, regions I and III, which are connected by a non-traversable Einstein-Rosen bridge (a whormhole). This is easiest seen at $t=T=0$ in figure  (\ref{carrolle}).  However, for constant $T\neq 0$, it is seen that the two asymptotically flat universes disconnect and the wormhole closes up, and thus any time like observer can not cross from one region to the other. The singularity $r=0$ is equivalently given by the hyperboloid $T^2-R^2=1$ which consists of two connected components in regions II (black hole)  and IV (white hole) which are called future and past interiors respectively. The regions I and III are precisely the exterior regions.

\subsection{Euclidean Black Hole and Thermal Field Theory}
By analytic continuation to Euclidean time $t_E=it$ we obtain 
 \begin{eqnarray}
ds^2=\rho^2\frac{dt_E^2}{4r_s^2}+d\rho^2+r_s^2d\Omega^2.
\end{eqnarray}
The first two terms correspond to two-dimensional flat space, viz
\begin{eqnarray}
X=\rho\cos \frac{t_E}{2r_s}~,~Y=\rho\sin \frac{t_E}{2r_s}.
\end{eqnarray}
The metric becomes 
 \begin{eqnarray}
ds^2=dX^2+dY^2+r_s^2d\Omega^2.
\end{eqnarray}
In order for the Euclidean metric to be smooth the Euclidean time $t_E$ must be periodic with period $\beta=4\pi r_s$ otherwise the metric has a conical singularity at $\rho=0$.

In quantum mechanics, the transition amplitude between point $q$ at time $t$ and point $q^{'}$ at time $t^{'}$ is given by 
\begin{eqnarray}
<q^{'},t^{'}|q,t>=<q^{'}|\exp(-iH(t^{'}-t))|q>=\sum_n\psi_n(q^{'})\psi_n^*(q)\exp(-iE_n(t^{'}-t)).
\end{eqnarray}
This can also be given by the path integral 
\begin{eqnarray}
<q^{'},t^{'}|q,t>=\int {\cal D}q(t)\exp(iS[q(t)]).
\end{eqnarray}
The action $S$ is given in terms of Lagrangian $L$ by the formula 
\begin{eqnarray}
S=\int dt L(q,\dot{q}).
\end{eqnarray}
We perform Wick rotation to Euclidean time $t_E=it$ with $\beta=t_E^{'}-t_E=i(t^{'}-t)$ and we consider closed paths $q^{'}=q(t_E+\beta)=q=q(t_E)$. We get immediately the thermodynamical partition function 
\begin{eqnarray}
Z=\exp(-\beta F)&=&Tr \exp(-\beta H)\nonumber\\
&=&\int dq <q|\exp(-\beta H)|q>\nonumber\\
&=&\int dq <q,t^{'}|q,t>.
\end{eqnarray}
The corresponding path integral is (with  $iS=-S_E$)
\begin{eqnarray}
Z=\int_{q(t_E+\beta)=q(t_E)} {\cal D}q(t)\exp(-S_E[q(t)]).
\end{eqnarray}
The Euclidean action is given in terms of the Lagrangian $L_E=-L$ by the formula 
\begin{eqnarray}
S=\int_0^{\beta} dt_E L_E(q,\dot{q}).
\end{eqnarray}
Thus, a path integral with periodic Euclidean time generates the thermodynamic partition function $Tr\exp(-\beta H)$. This very general and very remarkable result can also be stated by saying that thermal equilibrium is equivalent to summing over all periodic configurations $q(t_E+\beta)=q(t_E)$ in Euclidean time. 

The path integral for quantum fields in Euclidean Schwarzschild black hole geometry corresponds to a periodic Euclidean time $t_E\longrightarrow t_E+\beta$ with $\beta=4\pi r_s$ and thus it describes a gas in equilibrium with the black hole at temperature 
\begin{eqnarray}
T_H=\frac{1}{4\pi r_s}.
\end{eqnarray}
The Schwarzschild black hole is thus at equilibrium at the temperature $T_H$ and hence it must emits as much particles as it absorbs.

\section{Density Matrix and Entanglement}
This section is taken mostly from \cite{Susskind:2005js} and \cite{zurek} but we also found the lecture of \cite{moore} on the density matrix very useful. 

\subsection{Density Matrix: Pure and Mixed States}
We consider a system consisting of two subsystems $A$ and $B$. The wave function of this system is written 
\begin{eqnarray}
\Psi=\Psi(\alpha,\beta).\label{o1}
\end{eqnarray}
The $\alpha$ and $\beta$ are two sets of commuting variables relevant for the subsystems $A$ and $B$ separately. If we are only interested in the subsystem $A$ then its complete description is encoded in the density matrix or density operator 
\begin{eqnarray}
\rho_A(\alpha,\alpha^{\prime})=\sum_{\beta}\Psi^*(\alpha,\beta)\Psi(\alpha^{\prime},\beta).\label{o2}
\end{eqnarray}
The expectation value of an $A-$operator $a$ is given by the rule 
\begin{eqnarray}
\langle a\rangle =Tr a\rho_A.\label{o3}
\end{eqnarray}
A density matrix $\rho$ satisfies: I) $Tr\rho =1$ (sum of probabilities is $1$), II) $\rho=\rho^+$, III) $\rho_i\geq 0$. The eigenvalue $\rho_i$ is the probability that the system $A$ is in the eigenstate $|i\rangle$. The density matrix $\rho_A$  describes therefore a mixed state of the subsystem $A$, i.e. a statistical ensemble of several quantum states, which arises from the entanglement of the two subsystems $A$ and $B$, and thus our lack of knowledge of the exact state in which the subsystem $A$ will be found. 

This should be contrasted with pure states which are represented by single vectors in Hilbert space. The density matrix associated with a pure state $|i\rangle$ is simply given by $|i\rangle \langle i|$. The complete system formed by $A$ and $B$ is in a pure system although the subsystems $A$ and $B$ are both in mixed states due to entanglement. Another example of a pure state is the case when the density matrix $\rho_A$ has only one non-zero eigenvalue $(\rho_{A})_j$ which can only arise from a state of the form 
\begin{eqnarray}
\Psi(\alpha,\beta)=\Psi_A(\alpha)\Psi_B(\beta).
\end{eqnarray}
In general, we can write the density matrix corresponding to a mixed state as a convex sum, i.e. a weighted sum with $\sum_ip_i=1$, of pure state density matrices as follows  
  \begin{eqnarray}
\rho_A^{\rm mixed}=\sum_ip_i\rho_i^{\rm pure}=\sum_ip_i|\psi_i\rangle \langle\psi_i|.
\end{eqnarray}
The states $|\psi_i\rangle $ do not need to be orthogonal. This density matrix satisfies the Liouville-Von Neumann equation
\begin{eqnarray}
\frac{\partial\rho }{\partial t} =-\frac{i}{\hbar}[H,\rho].\label{LVN}
\end{eqnarray}
It is better to take an example here. Let us consider a spin $1/2$ system. A general pure state of this system is given by 
\begin{eqnarray}
|\psi\rangle=\cos\frac{\theta}{2}|+\rangle+\exp(i\phi)\sin\frac{\theta}{2}|-\rangle.
\end{eqnarray}
This state is given by the point on the surface of the unit $2-$sphere defined by the vector 
\begin{eqnarray}
\vec{a}=(\sin\theta\cos\phi,\sin\theta\sin\phi,\cos\theta).
\end{eqnarray}
The corresponding density matrix is
\begin{eqnarray}
\rho_{\rm pure}=|\psi\rangle \langle\psi|=\frac{1}{2}\left( \begin{array}{cc}
1+\cos\theta & \exp(-i\phi)\sin\theta \\
\exp(i\phi)\sin\theta & 1-\cos\theta \end{array} \right)=\frac{1}{2}({\bf 1}_2+\vec{a}.\vec{\sigma}).
\end{eqnarray}
This is a projector operator, viz 
\begin{eqnarray}
\rho_{\rm pure}^2=\rho_{\rm pure}.
\end{eqnarray}
The vector $\vec{a}$ is called the Bloch vector and the corresponding sphere is called the Bloch sphere. This vector is precisely the expectation value of the spin, viz
 \begin{eqnarray}
\vec{a}=\langle \rho_{\rm pure}\vec{\sigma}\rangle.
\end{eqnarray}
Mixed states are given by points inside the Bloch sphere. The corresponding density matrices are given by 
 \begin{eqnarray}
\rho_{\rm mixed}=\frac{1}{2}({\bf 1}_2+\vec{a}.\vec{\sigma})\neq \rho_{\rm mixed}^2~,~\vec{a}^2<1.
\end{eqnarray}
We have then the criterion 
 \begin{eqnarray}
Tr \rho^2=1~,~{\rm pure~state}.
\end{eqnarray}
\begin{eqnarray}
0<Tr \rho^2=\frac{1+\vec{a}^2}{2}<1~,~{\rm mixed~state}.
\end{eqnarray}
The quantity $Tr\rho^2$ is called the purity of the state. 

For example, a totally mixed state can have a $50$ per cent probability that the electron is in the state $|+\rangle$ and $50$ per cent probability that the electron is in the state $|-\rangle$. This corresponds to a completely unpolarized beam, viz $\vec{a}=0$. The corresponding density matrix is 
\begin{eqnarray}
\rho_{\rm mixed}&=&\frac{1}{2}|+\rangle \langle+|+\frac{1}{2}|-\rangle\langle-|\nonumber\\
&=&\frac{1}{2}{\bf 1}_2.
\end{eqnarray}
This decomposition is not unique. For example,  another totally mixed state can have a $50$ per cent probability that the electron is in the state $|+\rangle_x$ and $50$ per cent probability that the electron is in the state $|-\rangle_x$, viz
\begin{eqnarray}
\rho_{\rm mixed}&=&\frac{1}{2}|+\rangle_x\langle+|_x+\frac{1}{2}|-\rangle_x\langle-|_x\nonumber\\
&=&\frac{1}{2}\frac{|+\rangle+|-\rangle}{\sqrt{2}}\frac{\langle +|+\langle-|}{\sqrt{2}}+\frac{1}{2}\frac{|+\rangle-|-\rangle}{\sqrt{2}}\frac{\langle +|-\langle-|}{\sqrt{2}}\nonumber\\
&=&\frac{1}{2}{\bf 1}_2.
\end{eqnarray}
Thus a single density matrix can represent many, infinitely many in fact, different state mixtures. 

A partially mixed state for example can have a $50$ per cent probability that the electron is in the state $|+\rangle$ and $50$ per cent probability that the electron is in the state $(|+\rangle+|-\rangle)/\sqrt{2}$, viz
\begin{eqnarray}
\rho_{\rm mixed}=\frac{1}{2}|+\rangle \langle+|+\frac{1}{2}(\frac{|+\rangle +|-\rangle}{\sqrt{2}})(\frac{\langle +|+\langle-|}{\sqrt{2}}.
\end{eqnarray}
A pure state $|\Phi_c\rangle=(|+\rangle-|-\rangle)/\sqrt{2}$ for example is given by the density matrix 
\begin{eqnarray}
\rho_{\rm pure}=\frac{|+\rangle-|-\rangle}{\sqrt{2}}\frac{\langle +|-\langle-|}{\sqrt{2}}.
\end{eqnarray}
Again this decomposition is not unique. This can be rewritten also as
\begin{eqnarray}
\rho_{\rm pure}=\frac{|+\rangle_x-|-\rangle_x}{\sqrt{2}}\frac{\langle +|_x-\langle-|_x}{\sqrt{2}},
\end{eqnarray}
since $|\Phi_c>=-(|+\rangle_x-|-\rangle_x)/\sqrt{2}$. Thus the density matrix allows many, possibly infinitely many, different states of the subsystems on the diagonal. This freedom is expected since, by recalling the experiments of Aspect et al which showed that this nonseparable quantum correlation given by the state $|\Phi_c\rangle$ violates Bell's inequalities, we can conclude that: The pure states of the system described by $|\Phi_c\rangle$ are not just unknown but in fact can not exist before measurement \cite{zurek}.

It is clear from these examples that the relative phases between the basis states in a mixed state are random as opposed to coherent superpositions (pure states). This point is explained in more detail in the following.

A coherent superposition of two states $|\psi_1\rangle$ and $|\psi_2\rangle$ is given by the density matrix
\begin{eqnarray}
\rho_{c}=|\alpha|^2|\psi_1\rangle\langle \psi_1|+|\beta|^2|\psi_2\rangle\langle \psi_2|+\alpha\beta^*|\psi_1\rangle \langle\psi_2|+\alpha^*\beta|\psi_2\rangle \langle\psi_1|.
\end{eqnarray}
However, in the above preceding discussion the mixing is a statistical mixture as opposed to a coherent superposition. A statistical mixture of a state $|\psi_1\rangle$ with a probability $p_1=|\alpha|^2$ and state $|\psi_2\rangle$ with a probability $p_2=|\beta|^2$ is given by the density operator 
\begin{eqnarray}
\rho_{r}=p_1|\psi_1\rangle \langle\psi_1|+p_2|\psi_2\rangle \langle\psi_2|.
\end{eqnarray}
In other words, it is either $|\psi_1\rangle$ or $|\psi_2\rangle$ whereas in a coherent superposition it is both $|\psi_1\rangle$ and $|\psi_2\rangle$ at the same time. In the first case there is no interference effect (behave as classical probability distribution) while in the second case there is quantum interference. Mixed states are relevant when the exact initial quantum state is not known. 

Remark that in the statistical superposition we can change $\alpha\longrightarrow \exp(i\theta)\alpha$ and $\beta\longrightarrow \exp(i\theta^{\prime})\beta$ without changing the density matrix for $\theta$ and $\theta^{\prime}$ arbitrary. In the coherent superposition we must have $\theta=\theta^{\prime}$.

The probability of obtaining the eigenvalue $a_n$ in the measurement of the observable $A$ is then given by 
\begin{eqnarray}
  p(a_n)=p_1|\langle a_n|\psi_1\rangle |^2+p_2|\langle a_n|\psi_2\rangle|^2=Tr\rho|a_n\rangle\langle a_n|.
\end{eqnarray}
In fact, mixed states are incoherent superpositions. The diagonal elements of the density matrix give the probabilities to be in the corresponding states. The off diagonal elements measure the amount of coherence between the states. The off diagonal elements are called coherences. Coherence is maximized in a pure state when for every $m$ and $n$ we have 
\begin{eqnarray}
\rho_{mn}\rho_{nm}=\rho_{mm}\rho_{nn}.
\end{eqnarray}
A partially mixed state is such that for at least one pair of $m$ and $n$ we have 
\begin{eqnarray}
0<\rho_{mn}\rho_{nm}<\rho_{mm}\rho_{nn}.
\end{eqnarray}
A totally mixed state is such that for at least one pair of $m$ and $n$ we have 
\begin{eqnarray}
\rho_{mn}=\rho_{nm}=0~,~\rho_{mm}\rho_{nn}\ne 0.
\end{eqnarray}
Coherent superposition means interference whereas incoherent (mixed) superposition means absence of superposition.  Let us take an example. We consider a system described by a coherent superposition of two momentum states $k$ and $-k$ given by the pure state
\begin{eqnarray}
|\psi\rangle=\frac{1}{\sqrt{2}}(|k\rangle+|-k\rangle).
\end{eqnarray}
This quantum coherent superposition corresponds to sending particles through both slits at once. The density matrix is 
\begin{eqnarray}
\rho=|\psi\rangle\langle\psi|=\frac{1}{2}|k\rangle\langle k|+\frac{1}{2}|-k\rangle\langle -k|+\frac{1}{2}|k\rangle \langle -k|+\frac{1}{2}|-k\rangle \langle k|.
\end{eqnarray} 
The probability of finding the system at $x$ is 
\begin{eqnarray}
P(x)=Tr\rho|x\rangle \langle x|=1+\cos 2kx.
\end{eqnarray}
These are precisely the fringes (information). If the system is in a mixed (incoherent) state given for example by the density matrix  
\begin{eqnarray}
\rho=\frac{1}{2}|k\rangle \langle k|+\frac{1}{2}|-k\rangle \langle -k|.
\end{eqnarray}
This corresponds to the  classical case of sending particles at random, i.e. at $50$ per cent chance, through either one of the slits (totally mixed state). We get now the probability  
\begin{eqnarray}
P(x)=Tr\rho|x\rangle \langle x|=1.
\end{eqnarray}
So there are no fringes in this case, i.e. the incoherent mixed superposition is characterized by the absence of interference (no information). In a totally mixed state all interference effects are eliminated.

\subsection{Entanglement, Decoherence and Von Neumann Entropy}
We are now in a position to understand better our original definitions (\ref{o1}), (\ref{o2}), (\ref{o3}). The state $\Psi(\alpha,\beta)$ corresponds to a pure state $|\Psi\rangle$, viz $\langle \alpha,\beta|\Psi\rangle=\Psi(\alpha,\beta)$. The corresponding density matrix is $\rho=|\psi\rangle\langle \psi|$. We consider an $A-$observable $O_A\equiv O_A\otimes {\bf 1}_B$. The expectation value of $O_A$ is given by 
\begin{eqnarray}
\langle O_A\rangle&=&Tr\rho O_A\otimes {\bf 1}_B\nonumber\\
&=&\sum_{\alpha,\mu}\rho_A(\alpha,\mu)\langle \mu|O_A|\alpha\rangle\nonumber\\
&=&Tr_A\rho_A O_A.
\end{eqnarray}
The reduced density matrix $\rho_A$ is precisely given by 
\begin{eqnarray}
\rho_A(\alpha,\mu)=\sum_{\beta}\Psi^*(\alpha,\beta)\Psi(\mu,\beta).
\end{eqnarray}
To finish this important point we consider a system which is initially in a pure state and decoupled from the environment. The initial state of system+environment is then 
\begin{eqnarray}
|\psi\rangle^{(s,e)}=(\sum_sc_s|s\rangle^{(s)})\otimes|\phi\rangle^{(e)}.
\end{eqnarray}
The coupling between the system and the  environment is given by a unitary operator $U^{(s,e)}$, viz
\begin{eqnarray}
|\psi^{\prime}\rangle^{(s,e)}=U^{(s,e)}|\psi\rangle^{(s,e)}.
\end{eqnarray}
We will assume that the interaction is non-dissipative, i.e. the system does not decay to lower energy states, viz
\begin{eqnarray}
U^{(s,e)}|s\rangle^{(s)}\otimes|\phi\rangle^{(e)}=|s\rangle^{(s)}\otimes|\phi_s\rangle^{(e)}.
\end{eqnarray}  
Also we assume that the interaction is such that the different system states $|s\rangle$ drive the environment into orthogonal states $|\phi_s(t)\rangle^{(e)}$, viz
\begin{eqnarray}
\langle \phi_s|\phi_{s^{'}}\rangle^{(e)}=\delta_{s,s^{'}}.
\end{eqnarray}
The state of the system+environment becomes 
\begin{eqnarray}
|\psi^{'}\rangle^{(s,e)}=\sum_sc_s|s\rangle^{(s)}\otimes|\phi_s\rangle^{(e)}.
\end{eqnarray}
This is a pure state with a corresponding density matrix $\rho^{(s,e)}=|\psi^{'}\rangle \langle\psi^{'}|^{(s,e)}$. However, due to entanglement the state of the system is mixed given by tracing over the degrees of freedom of the environment which gives the reduced density matrix
\begin{eqnarray}
\rho^{(s)}&=&Tr_e\rho^{(s,e)}\nonumber\\
&=&\sum_s|c_s|^2|s\rangle \langle s|^s.
\end{eqnarray}
The probability of obtaining the system in the state $|s\rangle $ is $|c_s|^2$ which is the Born's rule. Hence, entanglement seems to give rise to collapse. The density matrix undergoes therefore the decrease of information $\rho^{(s,e)}\longrightarrow \rho^{(s)}$, called also decoherence, through interaction with the environment.

From the above result, entanglement seems also to give rise to decoherence which is actually what is at the origin of the collapse. Indeed, the above state is totally mixed and thus fully decohered since the off diagonal elements of the density matrix, which are responsible for quantum correlations, are zero. The environment kills therefore the coherence of the state as measured by the off diagonal elements of the density matrix.  The original pure state of the system has evolved into a mixed state because it is an open system, as opposed of being closed, and as such it does not obey the simple form (\ref{LVN}) of the Liouville-Von Neumann equation, but it satisfies instead the so-called master equation which has additional terms, viz
\begin{eqnarray}
\frac{\partial\rho }{\partial t} =-\frac{i}{\hbar}[H,\rho]+....
\end{eqnarray}
The extra terms can be given for example by those found in equation $(17)$ of \cite{zurek}.

We define the Von Neumann entropy or the entanglement entropy by the formula 
\begin{eqnarray}
S=-Tr\rho \ln\rho=-\sum_i\rho_i\ln\rho_i.
\end{eqnarray}
For a pure state, i.e. when all eigenvalues with the exception of one vanish, we get $S=0$. For mixed states we have $S>0$. For example, in the case of a totally incoherent mixed density matrix in which all the eigenvalues are equal to $1/N$ where $N$ is the dimension of the Hilbert space we get the maximum value of the Von Neumann entropy given by 
\begin{eqnarray}
S=S_{\rm max}=\ln N.
\end{eqnarray}
In the case that $\rho$ is proportional to a projection operator onto a subspace of dimension $n$ we find 
\begin{eqnarray}
S=\ln n.
\end{eqnarray}
In other words, the Von Neumann entropy measures the number of important states in the statistical ensemble, i.e. those states which have an appreciable probability. This entropy is also a measure of the degree of entanglement between subsystems $A$ and $B$ and hence its other name entanglement entropy.

The Von Neumann entropy is different from the thermodynamic Boltzmann entropy given by the formula 
\begin{eqnarray}
S_{\rm thermal}=-Tr\rho_{\rm MB} \ln\rho_{\rm MB},
\end{eqnarray}
where $\rho_{\rm MB}$ is the usual Maxwell-Boltzmann probability distribution given in terms of the Hamiltonian $H$ and the temperature $T=1/\beta$ by the formula 
\begin{eqnarray}
\rho_{\rm MB}=\frac{1}{Z}\exp(-\beta H)~,~Z=Tr\exp(-\beta H).
\end{eqnarray}

\section{Rindler Decomposition and Unruh Effect}
This lecture is based on \cite{Susskind:2005js,Harlow:2014yka}.
\subsection{Rindler Decomposition}
We consider quantum field theory in Minkowski spacetime. We introduce Rindler decomposition of this spacetime. The quadrant I is Rindler spacetime. Quadrants II and III have no causal relations with quadrant I. Quadrant IV provides initial data for Rindler spacetime. Indeed, signals from region IV must cross the surface $t=-\infty$ ($\omega=-\infty$) in order to reach region I.

We will work near the horizon with the metric (with $\omega=t/4MG$, $T=\rho\sinh\omega$, $Z=\rho\cosh\omega$) 
\begin{eqnarray}
ds^2&=&\rho^2d\omega^2-d\rho^2-dX^2-dY^2\nonumber\\
&=&dT^2-dZ^2-dX^2-dY^2.
\end{eqnarray}
The light cone is at $X=\pm T$ or $\rho=0$, $\omega=\pm \infty$. This also corresponds to the event horizon separating between $r<2GM$ and $r>2GM$. Remark that $\omega\longrightarrow \infty$ corresponds to $t\longrightarrow\infty$ since an observer falling into the black hole is never seen actually crossing it. Since the Rindler space is only valid near the horizon we have $r\simeq r_s$ or $\delta\simeq 0$ and thus $\rho\simeq 0$. The Horizon is actually at $\rho\simeq 0$.

The surface $T=0$ is divided into two halves. The first half in region I and the second half in region III. The fields in region I ($Z>0$) act in the Hilbert space ${\cal H}_L$ and those in region III ($Z<0$) act in the Hilbert space ${\cal H}_R$. We have then 
\begin{eqnarray}
\phi(X,Y,Z)=\phi_L(X,Y,Z)~,~Z>0.
\end{eqnarray}
\begin{eqnarray}
\phi(X,Y,Z)=\phi_R(X,Y,Z)~,~Z<0.
\end{eqnarray}
The general wave functional of interest is 
\begin{eqnarray}
\Psi=\Psi(\phi_L,\phi_R).
\end{eqnarray}
This is a pure state. But we want to compute the density matrix used by the fiducial observers called FIDOS (static observers at fixed $(X,Y,Z)$ which all measure the time $T$) in the Rindler quadrant (quadrant I) to describe the system. In other words, we need to compute the reduced density matrix $\rho_R$ which corresponds to the Minkowski vacuum to the FIDOS in Rindler quadrant I.

We have obviously translation invariance along the $X$ and $Y$ axes and thus the reduced density matrix is expected to commute with the momentum operators in the $X$ and $Y$ directions, viz
\begin{eqnarray}
[\rho_R,P_X]=[\rho_R,P_Y]=0.
\end{eqnarray}
Recall also that a translation in the Rindler time $\omega\longrightarrow \omega +c$ corresponds to a Lorentz boost along the $Z$ direction in Minkowski spacetime. The reduced density matrix $\rho_R$, since it represents the Minkowski vacuum in quadrant I, must be invariant under Lorentz boosts  in quadrant I. In other words, we must have
\begin{eqnarray}
[\rho_R,H_R]=0.
\end{eqnarray}
$H_R$ is the generator of the Lorentz boosts  $\omega\longrightarrow \omega +c$  in quadrant I. This is precisely the Hamiltonian in quadrant I given by 
 \begin{eqnarray}
H_R=\int_{\rho=0}^{\rho=\infty}\rho d\rho dX dY T^{00}(\rho,X,Y).
\end{eqnarray}
$T^{00}$ is the Hamiltonian density with respect to the Minkowski observer given for example for a scalar field by
\begin{eqnarray}
T^{00}(\rho,X,Y)=\frac{1}{2}\pi^2+\frac{1}{2}(\nabla\phi)^2+V(\phi).
\end{eqnarray}
Recall that in the $T-X$ plane the lines of constant $\omega$ are straight lines through the origin. The proper time separation between these lines is $\delta\tau=\rho\delta\omega$. This is the origin of the $\rho$ factor multiplying $T^{00}$. Since $\pi=\dot{\phi}$, the above Hamiltonian corresponds to the action 

\begin{eqnarray}
I=\int d^3xdT[\frac{1}{2}\dot{\phi}^2-\frac{1}{2}(\nabla\phi)^2-V(\phi)].
\end{eqnarray}
After Euclidean rotation $T\longrightarrow iX^0$ we get 
\begin{eqnarray}
I_E=\int d^4X[\frac{1}{2}(\partial_{X}\phi)^2+V(\phi)].
\end{eqnarray}
The original Lorentz invariance is now four dimensional rotation invariance. In particular, the $\omega-$translation, which is actually a boost in the $Z-$direction, becomes a rotation in the $(Z,X^0)$ plane.
 
\begin{figure}[H]
\begin{center}
\includegraphics[width=10.0cm,angle=0]{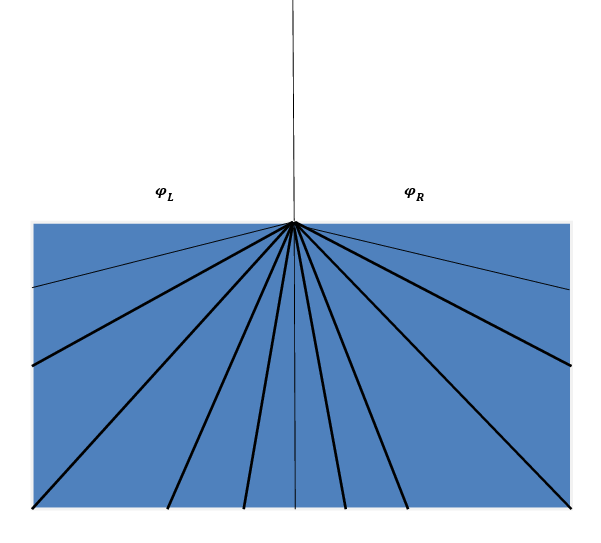}
\end{center}
\caption{Rindler decomposition.}\label{wedge}
\end{figure}

$\Psi(\phi_L,\phi_R)$ is the ground state of the Minkowski Hamiltonian which can be computed using Euclidean path integrals. We write this as $\Psi(\phi_L,\phi_R)=\langle \phi|\Omega\rangle$. The ground state $|\Omega\rangle$ can be obtained from any other state $|\chi\rangle$ by the action of the Hamiltonian as follows 
 \begin{eqnarray}
|\Omega\rangle =\frac{1}{\langle \Omega|\chi\rangle}{\rm Lim}_{T\longrightarrow \infty}\exp(-TH)|\chi\rangle.
\end{eqnarray}
Thus
 \begin{eqnarray}
\langle \phi|\Omega\rangle&=&\frac{1}{\langle \Omega|\chi\rangle}{\rm Lim}_{T\longrightarrow \infty}\langle \phi|\exp(-TH)|\chi\rangle\nonumber\\
&\propto &\int_{\hat{\phi}(t_E=-\infty)=0}^{\hat{\phi}(t_E=0)=\phi} {\cal D}\hat{\phi}\exp(-I_E).
\end{eqnarray}
The boundary condition at $t_E=0$, viz $\hat{\phi}(t_E=0)=\phi$, corresponds to the state $|\phi\rangle=|\phi_L\rangle|\phi_R\rangle$, because the states $\phi_L$ and $\phi_R$ correspond to $t_E=0$. The boundary condition at $t_E=-\infty$ is a choice. We could have chosen instead \cite{Susskind:2005js}
 \begin{eqnarray}
\langle \phi|\Omega\rangle
&\propto &\int_{\hat{\phi}(t_E=0)=\phi}^{\hat{\phi}(t_E=+\infty)=0} {\cal D}\hat{\phi}\exp(-I_E).
\end{eqnarray}
Let $\theta$ be the angle in the Euclidean plane $(Z,X^0)$ corresponding to the Rindler time $\omega$. We divide the region $T<0$ into infinitesimal wedges as in figure (\ref{wedge}). 

We integrate from the field $\phi_L$ at $\theta=0$ to the field $\phi_R$ at $\theta=\pi$. The  boost operator $K_x$ in the Euclidean plane generates rotations in the $(Z,X^0)$ plane. The restriction of this generator to the right Rindler wedge is given precisely by the Hamiltonian $H_R$. Thus we can write $\langle \phi|\Omega\rangle$ as a transition matrix element between initial state $|\phi_L\rangle$ and final state $|\phi_R\rangle$. In order to convert $|\phi_L\rangle$ back to a final state we act on it with the CPT operator $\Theta$ defined by 
\begin{eqnarray}
\Theta^+\phi(X^0,Z,X,Y)\Theta=\Phi^+(-X^0,-Z,X,Y).
\end{eqnarray}
This is an antiunitary operator which provides a map between the Hilbert spaces ${\cal H}_L$ and ${\cal H}_R$. 

The transfer matrix in an infinitesimal right wedge is $G=\exp(-\delta \theta H_R)$ but we have $n=\pi/\delta\theta$ wedges in total so the total transfer matrix is $G^n=\exp(-\pi H_R)$. In summary we have the result 
\begin{eqnarray}
\Psi(\phi_L,\phi_R)&=&\langle \phi_R|\langle\phi_L|\Omega\rangle\nonumber\\
&\propto &\langle \phi_R|\exp(-\pi H_R)\Theta|\phi_L\rangle.
\end{eqnarray}
This element is a transition matrix element in the right wedge. Let $|i_R\rangle$ be the eigenstates of $H_R$ with eigenvalues $E_i$. By inserting a complete set of such eigenstates we get
\begin{eqnarray}
\Psi(\phi_L,\phi_R)&=&\langle \phi_R|\langle\phi_L|\Omega\rangle\nonumber\\
&\propto &\sum_i e^{-\pi E_i}\langle \phi_R|i_R\rangle \langle i_R|\Theta|\phi_L\rangle.
\end{eqnarray}
$\Theta$ is an anti-unitary operator satisfying $\langle \Theta x|\Theta y\rangle=\langle y|x\rangle$ which should be contrasted with the unitarity property  $\langle \Theta x|A\Theta y\rangle =\langle x,y\rangle$. Thus we must have $\langle x|\Theta^+|y\rangle=\langle y|\Theta|x\rangle$. We define the state 
\begin{eqnarray}
\Theta^+|i_R\rangle=|i^*_L\rangle.
\end{eqnarray}
We get then the transition matrix element 
\begin{eqnarray}
\Psi(\phi_L,\phi_R)&=&\langle \phi_R|\langle\phi_L|\Omega\rangle\nonumber\\
&\propto &\sum_i e^{-\pi E_i}\langle \phi_R|i_R\rangle \langle\phi_L|i^*_L\rangle.
\end{eqnarray}
In other words, we get the ground state 
\begin{eqnarray}
|\Omega\rangle=\frac{1}{\sqrt{Z}}\sum_i e^{-\pi E_i}|i_R\rangle|i^*_L\rangle.
\end{eqnarray}
The entanglement between the left and right wedges is now fully manifest.

We can define immediately the reduced matrix $\rho_R$ by the relation 
\begin{eqnarray}
\rho_R(\phi_R,\phi_R^{'})&=&\int \Psi^*(\phi_L,\phi_R)\Psi(\phi_L,\phi_R^{'})d\phi_L\nonumber\\
&=&\frac{1}{Z}\sum_ie^{-2\pi E_i}\langle i_R|\phi_R\rangle\langle \phi_R^{'}|i_R\rangle\nonumber\\
&=&\frac{1}{Z}\langle \phi_R^{'}|e^{-2\pi H}|\phi_R\rangle.
\end{eqnarray}
In the second line we have used the identities
\begin{eqnarray}
\int |\phi_L\rangle \langle\phi_L|=1~,~\langle i_L^*|j_L^*\rangle=\delta_{ij}.
\end{eqnarray}
We get then the reduced density matrix 
\begin{eqnarray}
\rho_R
&=&\frac{1}{Z}e^{-2\pi H}.
\end{eqnarray}
Thus the fiducial observers FIDOS see the vacuum as a thermal ensemble with a Maxwell-Boltzmann distribution at a temperature 
\begin{eqnarray}
T_R=\frac{1}{2\pi}.
\end{eqnarray}
This is the Unruh effect.

Another derivation is as follows. The state $|\Omega\rangle$ is a pure state. The corresponding density matrix is $|\Omega\rangle \langle\Omega|$. By integrating over the degrees freedom of the left wedge we obtain a mixed state corresponding precisely to the reduced density matrix $\rho_R$, viz
 \begin{eqnarray}
\rho_R&=&\sum_i\langle i_L^*|\Omega\rangle\langle\Omega|i_L^*\rangle.
\end{eqnarray}
But
 \begin{eqnarray}
\langle i_L^*|\Omega\rangle=\frac{1}{\sqrt{Z}}e^{-\pi E_i}|i_R\rangle~,~\langle\Omega|i_L^*\rangle=\frac{1}{\sqrt{Z}}e^{-\pi E_i}\langle i_R|.
\end{eqnarray}
We get then 
 \begin{eqnarray}
\rho_R&=&\frac{1}{Z}\sum_ie^{-2\pi E_i}|i_R\rangle \langle i_R|.
\end{eqnarray}

\subsection{Unruh Temperature}
The temperature $T_R$ is dimensionless. We suppose a thermometer at rest with respect to the fiducial observer FIDOS at position $\rho$, i.e. it has the proper acceleration $a(\rho)=1/\rho$ (recall that $\rho\leftrightarrow \xi^3+1/a$ and $\omega\leftrightarrow a\tau$). The thermometer is also assumed to be in equilibrium with the quantum fields at temperature $T_R=1/2\pi$. If $\epsilon_i$ are the energy levels of the thermometer at rest then $\rho\epsilon_i$ are the Rindler energy levels of the thermometer. This is almost obvious from the form of the metric $ds^2=-\rho^2d\omega^2+d\rho^2+dX^2+dY^2$. We conclude therefore that the temperature measured by the thermometer is given by 
\begin{eqnarray}
T(\rho)=\frac{1}{2\pi\rho}=\frac{a(\rho)}{2\pi}.
\end{eqnarray}
Thus the FIDOS experiences a temperature which increases to infinity as we move towards the horizon at $\rho=0$. This temperature corresponds to virtual vacuum fluctuations given by particle pairs. Some of these virtual loops are conventional loops created in region I, some of them are of no importance to the FIDOS in region I since they are created in region III, but others are created around the horizon at $\rho=0$, and thus they are partly in region I partly in region III, and as a consequence cause non trivial entanglement between the degrees of freedom in regions I and III, which leads to a mixed density matrix in region I. Thus, the horizon behaves as a membrane which constantly emits and reabsorbs particles. This membrane is essentially the so-called stretched horizon.

\section{Quantum Field Theory in Curved Spacetime}
In this part we will follow briefly \cite{Mukhanov:2007zz,Wald:1984rg,Birrell:1982ix,Carroll:2004st}.

The action of a real scalar field coupled to the metric minimally is given by

 \begin{eqnarray}
S_M=\int d^4x\sqrt{-{\rm det}g}~\bigg(-\frac{1}{2}g^{\mu\nu}\nabla_{\mu}\phi\nabla_{\nu}\phi-V(\phi)\bigg).
\end{eqnarray}
If we are interested in an action which is at most quadratic in the scalar field then we must choose $V(\phi)=m^2\phi^2/2$. In curved spacetime there is another  term we can add which is quadratic in $\phi$  namely $R\phi^2$ where $R$ is the Ricci scalar. The full action should then read (in arbitrary dimension $n$)
 \begin{eqnarray}
S_M=\int d^nx\sqrt{-{\rm det}g}~\bigg(-\frac{1}{2}g^{\mu\nu}\nabla_{\mu}\phi\nabla_{\nu}\phi-\frac{1}{2}m^2\phi^2-\frac{1}{2}\zeta R\phi^2\bigg).
\end{eqnarray}
The choice $\zeta=(n-2)/(4(n-1))$ is called conformal coupling. At this value the action with $m^2=0$ is invariant under conformal transformations defined by
  \begin{eqnarray}
g_{\mu\nu}\longrightarrow \bar{g}_{\mu\nu}=\Omega^2(x)g_{\mu\nu}(x)~,~\phi\longrightarrow \bar{\phi}=\Omega^{\frac{2-n}{2}}(x)\phi(x).
\end{eqnarray}
The equation of motion derived from this  action are (we will keep in the following the metric arbitrary as long as possible)
 \begin{eqnarray}
\big(\nabla_{\mu}\nabla^{\mu} -m^2 -\zeta R\big)\phi=0.\label{Eof}
\end{eqnarray}
Let $\phi_1$ and $\phi_2$ be two solutions of this equation of motion. We define their inner product by
\begin{eqnarray}
(\phi_1,\phi_2)=-i\int_{\Sigma} \big(\phi_1\partial_{\mu}\phi_2^*-\partial_{\mu}\phi_1.\phi_2^*\big) d\Sigma n^{\mu}.\label{inner}
\end{eqnarray}
$d\Sigma$ is the volume element  in the space like hypersurface $\Sigma$ and $n^{\mu}$ is the time like unit vector which is normal to this hypersurface. This inner product is independent of the hypersurface $\Sigma$. 

Indeed let $\Sigma_1$ and $\Sigma_2$ be two non intersecting hypersurfaces and let $V$ be the four-volume bounded by $\Sigma_1$, $\Sigma_2$  and (if necessary) time like boundaries on which $\phi_1 = \phi_2 = 0$. We have from one hand
\begin{eqnarray}
i\int_{V} \nabla^{\mu}\big(\phi_1\partial_{\mu}\phi_2^*-\partial_{\mu}\phi_1.\phi_2^*\big) dV&=&i\oint_{\partial V} \big(\phi_1\partial_{\mu}\phi_2^*-\partial_{\mu}\phi_1.\phi_2^*\big) d\Sigma^{\mu}\nonumber\\
&=&(\phi_1,\phi_2)_{\Sigma_1}-(\phi_1,\phi_2)_{\Sigma_2}.
\end{eqnarray}
From the other hand 
\begin{eqnarray}
i\int_{V} \nabla^{\mu}\big(\phi_1\partial_{\mu}\phi_2^*-\partial_{\mu}\phi_1.\phi_2^*\big) dV&=&i\int_{V} \big(\phi_1\nabla^{\mu}\partial_{\mu}\phi_2^*-\nabla^{\mu}\partial_{\mu}\phi_1.\phi_2^*\big) dV\nonumber\\
&=&i\int_{V} \big(\phi_1(m^2+\xi R)\phi_2^*-(m^2+\xi R)\phi_1.\phi_2^*\big) dV\nonumber\\
&=&0.
\end{eqnarray}
Hence
\begin{eqnarray}
(\phi_1,\phi_2)_{\Sigma_1}-(\phi_1,\phi_2)_{\Sigma_2}=0.
\end{eqnarray}
There is always a complete set of solutions $u_i$ and $u_i^*$ of the equation of motion (\ref{Eof}) which are orthonormal in the inner product (\ref{inner}), i.e. satisfying 
\begin{eqnarray}
(u_i,u_j)=\delta_{ij}~,~(u_i^*,u_j^*)=-\delta_{ij}~,~(u_i,u_j^*)=0.
\end{eqnarray}
We can then expand the field as
\begin{eqnarray}
\phi=\sum_i(a_i u_i+a_i^* u_i^*).
\end{eqnarray}
We now canonically quantize this system. We choose a foliation of spacetime into space like hypersurfaces.
Let $\Sigma$ be a particular hypersurface with unit normal vector $n^{\mu}$ corresponding to a fixed value of the time coordinate $x^0=t$ and with induced metric $h_{ij}$. We write the action as $S_M=\int dx^0 L_M$ where $L_M=\int d^{n-1}x \sqrt{-{\rm det}g}~{\cal L}_M$. The canonical momentum $\pi$ is defined by 
\begin{eqnarray}
\pi=\frac{\delta {L}_M}{\delta( \partial_0{\phi})}&=&-\sqrt{-{\rm det}g}~g^{\mu 0}\partial_{\mu}\phi\nonumber\\
&=&-\sqrt{-{\rm det}h}~n^{\mu }\partial_{\mu}\phi.\label{cm0}
\end{eqnarray}
We promote $\phi$ and $\pi$ to hermitian operators $\hat{\phi}$ and $\hat{\pi}$ and then impose the equal time canonical commutation relations
\begin{eqnarray}
[\hat{\phi}(x^0,x^i),\hat{\pi}(x^0,y^i)]=i\delta^{n-1}(x^i-y^i).
\end{eqnarray}
The delta function satisfies the property 
\begin{eqnarray}
\int \delta^{n-1}(x^i-y^i)d^{n-1}y= 1 .
\end{eqnarray}
The coefficients $a_i$ and $a_i^*$ become annihilation and creation operators $\hat{a}_i$ and $\hat{a}_i^+$ satisfying the commutation relations 
\begin{eqnarray}
[\hat{a}_i,\hat{a}_j^+]=\delta_{ij}~,~[\hat{a}_i,\hat{a}_j]=[\hat{a}_i^+,\hat{a}_j^+]=0.
\end{eqnarray}
The vacuum state is given by a state $|0\rangle_u$ defined by
\begin{eqnarray}
\hat{a}_i|0_u\rangle=0.
\end{eqnarray}
The entire Fock basis of the Hilbert space can be constructed from the vacuum state by repeated application of the creation operators  $\hat{a}_i^+$.

The solutions $u_i$, $u_i^*$ are not unique and as a consequence the vacuum state  $|0\rangle_u$ is not unique. Let us consider another complete set of solutions $v_i$ and $v_i^*$ of the equation of motion (\ref{Eof}) which are orthonormal in the inner product (\ref{inner}).  We can then expand the field as
\begin{eqnarray}
\phi=\sum_i(b_i v_i+b_i^* v_i^*).
\end{eqnarray}
After canonical quantization the coefficients $b_i$ and $b_i^*$ become annihilation and creation operators $\hat{b}_i$ and $\hat{b}_i^+$ satisfying the standard commutation relations with a vacuum state given by $|0\rangle_v$ defined by
\begin{eqnarray}
\hat{b}_i|0_v\rangle=0.
\end{eqnarray}
We introduce the so-called Bogolubov transformation as the transformation from the set $\{u_i,u_i^*\}$ (which are the set of modes seen by some observer) to the set $\{v_i,v_i^*\}$ (which are the set of modes seen by another observer) as
\begin{eqnarray}
v_i=\sum_j(\alpha_{ij}u_j+\beta_{ij}u_j^*).\label{Bogolubov}
\end{eqnarray}
By using orthonormality conditions we find that
\begin{eqnarray}
\alpha_{ij}=(v_i,u_j)~,~\beta_{ij}=-(v_i,u_j^*).
\end{eqnarray}
We can also write
\begin{eqnarray}
u_i=\sum_j(\alpha_{ji}^*v_j-\beta_{ji}v_j^*).
\end{eqnarray}
The Bogolubov coefficients $\alpha$ and $\beta$ satisfy the normalization conditions 
\begin{eqnarray}
\sum_k(\alpha_{ik}\alpha_{jk}-\beta_{ik}\beta_{jk})=\delta_{ij}~,~\sum_k(\alpha_{ik}\beta_{jk}^*-\beta_{ik}\alpha_{jk}^*)=0.
\end{eqnarray}
The  Bogolubov coefficients $\alpha$ and $\beta$ transform also between the creation and annihilation operators $\hat{a}$, $\hat{a}^+$ and $\hat{b}$, $\hat{b}^+$. We find
\begin{eqnarray}
\hat{a}_k=\sum_i(\alpha_{ik}\hat{b}_i+\beta_{ik}^*\hat{b}_i^+)~,~\hat{b}_k=\sum_i(\alpha_{ki}^*\hat{a}_i-\beta_{ki}^*\hat{a}_i^+).
\end{eqnarray}
Let $N_u$ be the number operator with respect to the $u$-observer, viz $N_{u}=\sum _k \hat{a}_k^+\hat{a}_k$. Clearly
\begin{eqnarray}
\langle 0_u|N_u|0_u\rangle=0.
\end{eqnarray}
We compute
\begin{eqnarray}
\langle 0_v|\hat{a}_k^+\hat{a}_k|0_v\rangle=\sum_i \beta_{ik}\beta_{ik}^*.
\end{eqnarray}
Thus
\begin{eqnarray}
\langle 0_v|N_u|0_v\rangle=tr \beta\beta^+.
\end{eqnarray}
In other words with respect to the   $v$-observer the vacuum state $|0_u\rangle$ is not empty but filled with particles. This opens the door to the possibility of particle creation by a gravitational field.

\section{Hawking Radiation}
 \subsection{The Unruh Effect Revisited}
In this first part we will follow mostly \cite{Carroll:2004st}. We consider $2-$dimensional spacetime with metric 
\begin{eqnarray}
ds^2=-dt^2+dx^2.
\end{eqnarray}
We consider a uniformly accelerated, i.e. a Rindler, observer in this spacetime with acceleration $\alpha$. The trajectory of the Rindler observer is given by the equations (set $\xi^1=0$ in (\ref{rindler}) and discard the constant term $1/a$ in $x^1$, i.e. the Rindler does not coincide with Minkowski at $\tau=0$)  
\begin{eqnarray}
t=\frac{1}{\alpha}\sinh \alpha\tau~,~x=\frac{1}{\alpha}\cosh \alpha\tau.
\end{eqnarray}
The trajectory is then a hyperboloid given by 
\begin{eqnarray}
x^2=t^2+\frac{1}{\alpha^2}.
\end{eqnarray}
Thus the Rindler observer moves from the past null infinity $x=-t$ to the future null infinity $x=+t$ as opposed to the motion of geodesic observers which reaches timelike infinity.

\begin{figure}[H]
\begin{center}
\includegraphics[width=10.0cm,angle=0]{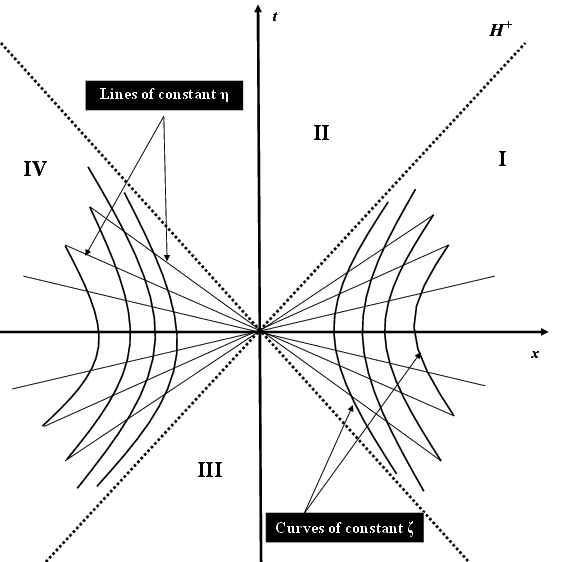}
\end{center}
\caption{Rindler space in two dimensions.}\label{wedge2}
\end{figure}

We introduce coordinates in Rindler space (quadrant or wedge I in figure (\ref{wedge2})) by 
\begin{eqnarray}
t=\frac{1}{a}\exp(a\xi)\sinh a\eta~,~x=\frac{1}{a}\exp(a\xi)\cosh a\eta~, ~x>|t|~,~-\infty<\eta,\xi<+\infty.
\end{eqnarray}
The trajectory of the Rindler observer in these coordinates read 
\begin{eqnarray}
\eta=\frac{\alpha}{a}\tau~,~\xi=\frac{1}{a}\ln \frac{a}{\alpha}.
\end{eqnarray}
In other words,
\begin{eqnarray}
a=\alpha\Rightarrow \eta=\tau~,~\xi=0.
\end{eqnarray}
The metric in Rindler space reads 
\begin{eqnarray}
ds^2=\exp(2a\xi)(-d\eta^2+d\xi^2).
\end{eqnarray}
The metric is independent of $\eta$ and thus $\partial_{\eta}$ is a Killing vector. This is given explicitly by 
\begin{eqnarray}
\partial_{\eta}=a(x\partial_t+t\partial_x).
\end{eqnarray}
This is then obviously the Killing field associated with a boost in the $x-$direction. This extends to regions II and III where it is spacelike while in region IV it is timelike past-directed \footnote{The lables III and IV are reversed here as compared with the previous discussion.}. The horizons $x=\pm t$ are actually Killing horizons. Every Killing horizon is associated with an acceleration called the surface gravity $\kappa$ which is here given exactly by 
\begin{eqnarray}
\kappa=a.
\end{eqnarray}
We will also need the coordinates $\eta$ and $\xi$ in the quadrant IV. They are given by
\begin{eqnarray}
t=-\frac{1}{a}\exp(a\xi)\sinh a\eta~,~x=-\frac{1}{a}\exp(a\xi)\cosh a\eta~, ~x<|t|.
\end{eqnarray}
The Klein-Gordon equation in Rindler space is (with $m^2=\zeta=0$)
\begin{eqnarray}
0&=&\nabla_{\mu}\nabla^{\mu}\phi\nonumber\\
&=&\frac{1}{\sqrt{-{\rm det}g}}\partial_{\mu}\bigg(\sqrt{-{\rm det}g}\partial^{\mu}\phi\bigg)\nonumber\\
&=&e^{-2a\xi}(-\partial^2_{\eta}+\partial_{\xi}^2)\phi.
\end{eqnarray}
A positive frequency normalized plane wave solution in region I is given by 
 \begin{eqnarray}
&&g_k^{(1)}=\frac{1}{\sqrt{4\pi \omega}}\exp(-i\omega \eta+ik\xi)~,~{\rm I}\nonumber\\
&&g_k^{(1)}=0~,~{\rm IV}.\label{pos}
\end{eqnarray}
Indeed, 
 \begin{eqnarray}
&&\partial_{\eta}g_k^{(1)}=-i\omega g_k^{(1)}~,~\omega=|k|.
\end{eqnarray}
$\partial_{\eta}$ is a future-director timelike Killing vector in region I. But it is a past-directed timelike Killing vector in region IV. Thus in region IV we should consider the Killing vector $\partial_{-\eta}=-\partial_{\eta}$ which is future-directed there. A positive frequency normalized plane wave solution in region II is thus given by 
 \begin{eqnarray}
&&g_k^{(2)}=0~,~{\rm I}\nonumber\\
&&g_k^{(2)}=\frac{1}{\sqrt{4\pi \omega}}\exp(i\omega \eta+ik\xi)~,~{\rm IV}.
\end{eqnarray}
Indeed, 
 \begin{eqnarray}
&&\partial_{-\eta}g_k^{(2)}=-i\omega g_k^{(2)}~,~\omega=|k|.
\end{eqnarray}
These two sets of positive frequency modes, together with their negative frequency conjugates, provide a complete set of basis elements for the expansion of any solution of the Klein-Gordon wave equation through spacetime.
 We denote the associated annihilation operators by $\hat{b}_k^{(1)}$ and $\hat{b}_k^{(2)}$. A general solution of the Klein-Gordon equation takes then the form 
 \begin{eqnarray}
\phi=\int_k \big(\hat{b}_k^{(1)}g_k^{(1)}+\hat{b}_k^{(2)}g_k^{(2)}+{\rm h.c}\big).
\end{eqnarray}
This should be contrasted with the expansion of the same solution in terms of the Minkowski modes $f_k\propto \exp(-i(\omega t-kx))$ with $\omega=|k|$ which we will write as
\begin{eqnarray}
\phi=\int_k \big(\hat{a}_k^{}f_k^{}+{\rm h.c}\big).
\end{eqnarray}
The above Rindler modes $g_k^{(1)}$ and $g_k^{(2)}$ are normalized according  to the inner product (\ref{inner}), viz
\begin{eqnarray}
(\phi_1,\phi_2)=-i\int_{\Sigma} \big(\phi_1\partial_{\mu}\phi_2^*-\partial_{\mu}\phi_1.\phi_2^*\big) d\Sigma n^{\mu}.
\end{eqnarray}
$d\Sigma$ is the volume element  in the spacelike hypersurface $\Sigma$ and $n^{\mu}$ is the timelike unit vector which is normal to this hypersurface. Thus $d\Sigma=\sqrt{{\rm det}\gamma}d^{n-1}x$. In our case, the timelike surface $\eta=0$ has a unit vector $n^{\mu}$ such as $g_{\mu\nu}n^{\mu}n^{\nu}=-1$ and thus $n^0=\exp(-a\xi)$. Also we have $\sqrt{{\rm det}\gamma}=\exp(a\xi)$ and $x\leftrightarrow\xi$. Hence the inner product becomes 
\begin{eqnarray}
(\phi_1,\phi_2)=-i\int \big(\phi_1\partial_{\eta}\phi_2^*-\partial_{\eta}\phi_1.\phi_2^*\big) d\xi.\label{inner1}
\end{eqnarray}
We compute for example 
\begin{eqnarray}
(g_{k_1}^{(1)},g_{k_2}^{(1)})&=&-\frac{i}{4\pi\sqrt{\omega_1\omega_2}}\int \bigg(i\omega_2 e^{-i\omega_1\eta+ik_1\xi}e^{i\omega_2\eta-ik_2\xi}
+i\omega_1e^{-i\omega_1\eta+ik_1\xi}e^{i\omega_2\eta-ik_2\xi}\bigg) d\xi\nonumber\\
&=&\frac{1}{4\pi}.4\pi\delta(k_1-k_2).
\end{eqnarray}
We also show 
\begin{eqnarray}
(g_{k_1}^{(2)},g_{k_2}^{(2)})
&=&\delta(k_1-k_2).
\end{eqnarray}
\begin{eqnarray}
(g_{k_1}^{(1)},g_{k_2}^{(2)})
&=&0.
\end{eqnarray}
The Minkowski vacuum $|0_M\rangle$ and the Rindler vacuum $|0_R\rangle$ are defined obviously by 
\begin{eqnarray}
\hat{a}_k|0_M\rangle=0.
\end{eqnarray}
\begin{eqnarray}
\hat{b}_k^{(1)}|0_R\rangle=\hat{b}_k^{(2)}|0_R\rangle=0.
\end{eqnarray}
However, the Hilbert space is the same. For the Rindler observer the Minkowski vacuum $|0_R\rangle$ is seen as a multi-particle state since she is traveling in Minkowski spacetime with a uniform acceleration, i.e. she is not an inertial observer. The expectation value of the Rindler number operator in the Minkowski vacuum can be calculated using the Bogolubov coefficients as we explained in the previous section. 

An alternative method due to Unruh consists in extending the positive frequency modes $g_k^{(1)}$ and $g_k^{(2)}$ to the entire spacetime and thus replacing the corresponding annihilation operators $\hat{b}_k^{(1)}$ and $\hat{b}_k^{(2)}$ by new annihilation operators $\hat{c}_k^{(1)}$ and $\hat{c}_k^{(2)}$ which annihilate the Minkowski vacuum $|0_M>$.

First, the coordinates $(t,x)$ and $(\eta,\xi)$ are related by
\begin{eqnarray}
-t+x=\frac{1}{a}e^{a(\xi-\eta)}\Rightarrow e^{-a(\eta-\xi)}=a(-t+x)~,~{\rm I}.
\end{eqnarray}
\begin{eqnarray}
t-x=\frac{1}{a}e^{a(\xi-\eta)}\Rightarrow e^{-a(\eta-\xi)}=a(t-x)~,~{\rm IV}.
\end{eqnarray}
Similarly, 
\begin{eqnarray}
e^{a(\eta+\xi)}=a(t+x)~,~{\rm I}.
\end{eqnarray}
\begin{eqnarray}
e^{a(\eta+\xi)}=a(-t-x)~,~{\rm IV}.
\end{eqnarray}
Thus if we choose $k>0$ we have in region I ($x>0$) 
\begin{eqnarray}
\sqrt{4\pi\omega}g_k^{(1)}&=&\exp(-i\omega (\eta-\xi))\nonumber\\
&=&e^{i\frac{\omega}{a}}(-t+x)^{i\frac{\omega}{a}}.
\end{eqnarray}
In region IV ($x<0$) we should instead consider 
\begin{eqnarray}
\sqrt{4\pi\omega}g_{-k}^{(2)*}&=&\exp(-i\omega (\eta-\xi))\nonumber\\
&=&e^{i\frac{\omega}{a}}(t-x)^{i\frac{\omega}{a}}\nonumber\\
&=&e^{i\frac{\omega}{a}}e^{\frac{\pi \omega}{a}}(-t+x)^{i\frac{\omega}{a}}.
\end{eqnarray}
Thus for all $x$, i.e. along the surface $t=0$, we should consider for $k>0$ the combination 
\begin{eqnarray}
\sqrt{4\pi\omega}\big(g_k^{(1)}+e^{-\frac{\pi\omega}{a}}g_{-k}^{(2)*}\big)
&=&e^{i\frac{\omega}{a}}(-t+x)^{i\frac{\omega}{a}}.
\end{eqnarray}
We get the same result for $k<0$. A normalized analytic extension to the entire spacetime of the positive frequency modes $g_k^{(1)}$ is given by the modes
\begin{eqnarray}
h_k^{(1)}&=&\frac{1}{\sqrt{2\sinh \frac{\pi\omega}{a}}}\big(e^{\frac{\pi\omega}{2a}} g_k^{(1)}+e^{-\frac{\pi\omega}{2a}}g_{-k}^{(2)*}\big).\label{h1}
\end{eqnarray}
Similarly, a normalized analytic extension to the entire spacetime of the positive frequency modes $g_k^{(2)}$ is given by the modes
\begin{eqnarray}
h_k^{(2)}&=&\frac{1}{\sqrt{2\sinh \frac{\pi\omega}{a}}}\big(e^{\frac{\pi\omega}{2a}} g_k^{(2)}+e^{-\frac{\pi\omega}{2a}}g_{-k}^{(1)*}\big).\label{h2}
\end{eqnarray}
The field operator can then be expanded in these modes as 
\begin{eqnarray}
\phi=\int_k \big(\hat{c}_k^{(1)}h_k^{(1)}+\hat{c}_k^{(2)}h_k^{(2)}+{\rm h.c}\big).
\end{eqnarray}
The relation between the annihilation operators $\hat{b}$ and the annihilation operators $\hat{c}$ is given by the same relation between the modes $h$ and the modes $g$, viz 
\begin{eqnarray}
\hat{b}_k^{(1)}&=&\frac{1}{\sqrt{2\sinh \frac{\pi\omega}{a}}}\big(e^{\frac{\pi\omega}{2a}} \hat{c}_k^{(1)}+e^{-\frac{\pi\omega}{2a}}\hat{c}_{-k}^{(2)+}\big).
\end{eqnarray}
\begin{eqnarray}
\hat{b}_k^{(2)}&=&\frac{1}{\sqrt{2\sinh \frac{\pi\omega}{a}}}\big(e^{\frac{\pi\omega}{2a}} \hat{c}_k^{(2)}+e^{-\frac{\pi\omega}{2a}}\hat{c}_{-k}^{(1)+}\big).
\end{eqnarray}
The modes $h_k^{(1)}$ and $h_k^{(2)}$ are positive frequency modes defined on the entire spacetime and thus they can be expressed entirely in terms of the positive frequency modes of Minkowski spacetime given by the plane waves $f_k\propto \exp(-i(\omega t-k x))$, $\omega=|k|$, where $k>0$ correspond to right moving modes and $k<0$ correspond to left moving modes. In other words, the modes $h_k^{(1)}$ and $h_k^{(2)}$ share with $f_k$ the same Minkowski vacuum $|0_M\rangle $, viz
\begin{eqnarray}
\hat{c}_k^{(1)}|0_M\rangle=\hat{c}_k^{(2)}|0_M\rangle =0.
\end{eqnarray}
The Rindler number operator in region I is defined by 
\begin{eqnarray}
\hat{N}_R^{(1)}(k)=\hat{b}_k^{(1)+}\hat{b}_k^{(1)}.
\end{eqnarray}
We can now immediately compute the expectation value of the Rindler number operator in region I in the Minkowski vacuum to find
\begin{eqnarray}
\langle 0_M|\hat{N}_R^{(1)}(k)|0_M\rangle&=&\langle 0_M|\hat{b}_k^{(1)+}\hat{b}_k^{(1)}|0_M\rangle\nonumber\\
&=&\frac{e^{-\frac{\pi\omega}{a}}}{2\sinh\frac{\pi\omega}{2}}\langle 0_M|\hat{c}_{-k}^{(2)}\hat{c}_{-k}^{(2)+}|0_M\rangle\nonumber\\
&=&\frac{1}{e^{\frac{2\pi\omega}{a}}-1}\delta(0).
\end{eqnarray}
This is a blackbody Planck spectrum corresponding to the temperature 
\begin{eqnarray}
T=\frac{a}{2\pi}.
\end{eqnarray}
Indeed, this spectrum corresponds to a thermal radiation, i.e. to a mixed state, without any correlations. This is the Unruh effect: A uniformly accelerated observer in the Minkowski vacuum  observes a thermal spectrum \cite{Unruh:1976db}.

\subsection{From Quantum Scalar Field Theory in Rindler Background}
We follow in this section the presentation of \cite{Susskind:2005js}. We consider Schwarzschild metric in tortoise coordinates, viz
\begin{eqnarray}
&&ds^²=F(r_*)(-dt^2+dr_*^2)+r^2d\Omega^2\nonumber\\
&&F(r_*)=1-\frac{2GM}{r}\nonumber\\
&&r_*=r+2GM\log(\frac{r}{2GM}-1).
\end{eqnarray}
We consider the action of a massless scalar field $\phi$ in this background given by (with $\psi=r\phi$)
\begin{eqnarray}
I&=&\int \sqrt{-{\rm det}g}d^4x\frac{1}{2}\partial_{\mu}\phi\partial^{\mu}\phi\nonumber\\
&=&\int Fr^2\sin\theta dtdr_*d\theta d\phi\frac{1}{2}\bigg(-\frac{1}{F}(\partial_t\phi)^2+\frac{1}{F}(\partial_{r_*}\phi)^2+\frac{1}{r^2}(\partial_{\theta}\phi)^2+\frac{1}{r^2\sin^2\theta}(\partial_{\phi}\phi)^2\bigg)\nonumber\\
&=&\int \sin\theta dtdr_*d\theta d\phi\frac{1}{2}\bigg(-(\partial_t\psi)^2+(\partial_{r_*}\psi-\partial_{r_*}\ln r.\psi)^2+\frac{F}{r^2}(\partial_{\theta}\psi)^2+\frac{F}{r^2\sin^2\theta}(\partial_{\phi}\psi)^2\bigg)\nonumber\\
&=&\int \sin\theta dtdr_*d\theta d\phi\frac{1}{2}\bigg(-(\partial_t\psi)^2+(\partial_{r_*}\psi-\partial_{r_*}\ln r.\psi)^2+\frac{F}{r^2}\psi{\cal L}^2\psi\bigg),\nonumber\\
\end{eqnarray}
where we have used 
\begin{eqnarray}
-{\cal L}^2=\frac{1}{\sin\theta}\frac{\partial}{\partial\theta}(\sin\theta\frac{\partial}{\partial\theta})+\frac{1}{\sin^2\theta}\frac{\partial^2}{\partial\phi^2}.
\end{eqnarray}
We expand now in spherical coordinates as
\begin{eqnarray}
\psi=\sum_{lm}\psi_{lm}Y_{lm}.
\end{eqnarray}
We get then 
\begin{eqnarray}
I
&=&\int dtdr_*\frac{1}{2}\sum_{lm}\psi_{lm}^*\bigg(\partial_t^2\psi_{lm}-\partial_{r_*}^2\psi_{lm}+\big(\partial_{r_*}^2\ln r+(\partial_{r_*}\ln r)^2\big)\psi_{lm}+\frac{F}{r^2}l(l+1)\psi_{lm}\bigg)\nonumber\\
&=&\int dtdr_*\frac{1}{2}\sum_{lm}\psi_{lm}^*\bigg(\partial_t^2\psi_{lm}-\partial_{r_*}^2\psi_{lm}+V(r_*)\psi_{lm}\bigg).
\end{eqnarray}
The potential is given by 
\begin{eqnarray}
V(r_*)&=&\partial_{r_*}^2\ln r+(\partial_{r_*}\ln r)^2+\frac{F}{r^2}l(l+1)\nonumber\\
&=&\frac{1}{r}\frac{\partial^2r}{\partial r_*^2}+\frac{F}{r^2}l(l+1)\nonumber\\
&=&\frac{r-2GM}{r}\bigg(\frac{2GM}{r^3}+\frac{l(l+1)}{r^2}\bigg).\label{VS}
\end{eqnarray}
The equation of motion reads 
\begin{eqnarray}
\partial_t^2\psi_{lm}=\partial_{r_*}^2\psi_{lm}-V(r_*)\psi_{lm}.
\end{eqnarray}
The stationary solutions are $\psi_{lm}=\exp(i\nu t)\tilde{\psi}_{lm}$ such that 
\begin{eqnarray}
-\tilde\partial_{r_*}^2\tilde\psi_{lm}+V(r_*)\tilde\psi_{lm}=\nu^2\tilde{\psi}_{lm}.
\end{eqnarray}
The potential vanishes at the horizon $r=2GM$ (where the solutions are given by free plane waves) and also vanishes at infinity. Thus it must pass through a maximum given by the condition 
\begin{eqnarray}
\frac{dV}{dr}=\frac{1}{r^5}\bigg(-2l(l+1).r^2-6GM(1-l(l+1)).r+16G^2M^2\bigg)=0.
\end{eqnarray}
We get the solutions 
\begin{eqnarray}
r_{\pm}=3GM\bigg(\frac{1}{2}-\frac{1}{2l(l+1)}\pm\frac{1}{2}\sqrt{1+\frac{7l^2+7l+4}{4l^2(l+1)^2}}\bigg).
\end{eqnarray}
Obviously, the physical solution is 
\begin{eqnarray}
r_{\rm max}=3GM\bigg(\frac{1}{2}-\frac{1}{2l(l+1)}+\frac{1}{2}\sqrt{1+\frac{7l^2+7l+4}{4l^2(l+1)^2}}\bigg).
\end{eqnarray}
Thus
\begin{eqnarray}
r_{\rm max}(l=\infty)=3GM.
\end{eqnarray}
For very large angular momentum $l$ the  maximum of the potential lies at $3GM$. For $r>>3GM$ the potential is repulsive, given by a generalization of the centrifugal potential $l(l+1)/r^2$,  whereas for $r<3GM$ (the region of thermal atmosphere) gravity dominates and the potential becomes attractive. Thus any particle in this region  with a zero initial velocity will spiral into the horizon eventually. 

The above equation is effectively Schrodinger equation with potential $V$ and energy $\nu^2$. Thus an $s$-wave ($l=0$) approaching the barrier $r=3GM$ from the inside (horizon) with energy satisfying $\omega>V_{\rm max}$ will be able to escape whereas if approaching from the outside it will be able to penetrate the barrier and reach the horizon. For energy  $\omega<V_{\rm max}$ the wave needs to tunnel through the barrier. 

For higher angular momentum the maximum of the potential is very large proportional to $l^2$ and thus it is more difficult to escape or penetrate the barrier.

Near horizon geometry is given by the metric (with $u=\ln \rho$ the tortoise coordinate in this case)
\begin{eqnarray}
ds^2&=&-\rho^2d\omega^2+d\rho^2+dY^2+dZ^2\nonumber\\
&=&e^{2u}(-d\omega^2+du^2)+dY^2+dZ^2.
\end{eqnarray}
The action of a scalar field is given immediately by 
\begin{eqnarray}
I=\int d\omega du dYdZ\frac{1}{2}\bigg(-(\partial_{\omega}\phi)^2+(\partial_u\phi)^2+e^{2u}(\partial_Y\phi)^2+e^{2u}(\partial_Z\phi)^2\bigg).
\end{eqnarray}
We expand the field into transverse plane waves as 
\begin{eqnarray}
\phi=\int \frac{dk_2}{2\pi}\frac{dk_3}{2\pi}e^{i(k_2Y+k_3Z)}\psi(k_2,k_3,\omega,u).
\end{eqnarray}
We get then the action 
\begin{eqnarray}
I=\int d\omega du \frac{1}{2}\psi^*\bigg(\partial_{\omega}^2\psi-\partial_u^2\psi+e^{2u}\vec{k}^2\psi\bigg).
\end{eqnarray}
The potential is then given by 
\begin{eqnarray}
V=e^{2u}\vec{k}^2.\label{VR}
\end{eqnarray}
This is proportional to $l^2$ since $l=|k|r=|k|.2MG$ and thus this approximation is not expected to work for small angular momentum. Thus in approximating  sums over $l$ and $m$ by integrals over $k$ we should for consistency employ the infrared cutoff $|k|\sim 1/MG$. 

The Rindler potential $V=\rho^2\vec{k}^2$, for $|k|\ne 0$,  is confining to the region near the horizon. This is also the situation in the  Schwarzschild black hole where the potential confines particles to the region near the horizon. However, in the  Schwarzschild black hole the potential becomes repulsive for $r>3MG$ which is equivalent to $\rho>MG$. Thus the potential barrier for  Schwarzschild black hole is cutoff for $\rho>MG$ as opposed to Rindler space which keeps increasing without bound as $\rho^2$.

Since $(1)$ a Schwarzschild black hole near the horizon will appear as Rindler, and $(2)$ the Rindler observer will see the Minkowski vacuum as a thermal canonical ensemble with a temperature given by $T=1/2\pi$, it is expected that an identical thermal effect should be observed near the horizon of the Schwarzschild black hole.

However, there is a crucial difference. In the case of Rindler the thermal atmosphere is fully confined by the potential (\ref{VR}) as opposed to the case of the Schwarzschild black hole where the thermal atmosphere is not fully confined by the potential (\ref{VS}). This means in particular that particles leak out of the thermal atmosphere in the case of the Schwarzschild black hole and as a consequence the black hole evaporates.

Let us find the temperature as seen by the Schwarzschild observer. The Rindler time $\omega$ is related to the Schwarzschild time $t$ by the relation 
\begin{eqnarray}
\omega=\frac{t}{4GM}.
\end{eqnarray}
This leads immediately to the fact that frequency as measured  by the Schwarzschild observer $\nu$ is red shifted compared to the frequency $\nu_R$ measured by the Rindler observer given by
\begin{eqnarray}
\nu_R=4GM.\nu\Rightarrow \nu=\frac{\nu_R}{4GM}.
\end{eqnarray}
Hence the temperature as measured by the Schwarzschild observer is also red shifted as 
\begin{eqnarray}
T_R=4GM.T\Rightarrow T=\frac{T_R}{4GM}=\frac{1}{8\pi GM}.
\end{eqnarray}
This is precisely Hawking temperature.

Next we show how the black hole can radiate particles. The potential  (\ref{VS}) is not fully confining and it contains only a barrier around $r\simeq 3GM$. The height of the barrier for modes with angular momentum $l=0$, which corresponds to $r_{\rm max}(l=0)=8GM/3$, is given by
\begin{eqnarray}
V_{\rm max}(l=0)=\frac{27}{1024G^2M^2}.
\end{eqnarray}
The energy in the potential (\ref{VS}) is $E=\nu^2$. Thus the modes $l=0$ will escape the potential barrier coming from the horizon if 
\begin{eqnarray}
E\geq V_{\rm max}(l=0)\Rightarrow \nu\geq \frac{3\pi \sqrt{3}}{4}T.
\end{eqnarray}
However, these modes since they are in a thermal state with temperature $T=1/8\pi MG$, their energies is of the order of $T$, and hence they can quite easily escape the potential barrier. The height of the barrier for modes with higher angular momentum $l$ goes as $l^2/G^2M^2$, i.e. it is very high compared to the thermal scale set by Hawking radiation, and hence these modes do not escape as easily as the zero modes. This is Hawking radiation.

\subsection{Summary}
In summary, since 
\begin{itemize}
\item
$(1)$ a Schwarzschild black hole near the horizon will appear as Rindler, and 
\item
$(2)$ the Rindler observer will see the Minkowski vacuum as a thermal canonical ensemble with a temperature given by $T=1/2\pi$ (Unruh effect), 
\end{itemize}
an identical thermal effect is observed near the horizon of the Schwarzschild black hole.

\begin{figure}[H]
\begin{center}
\includegraphics[width=13.0cm,angle=0]{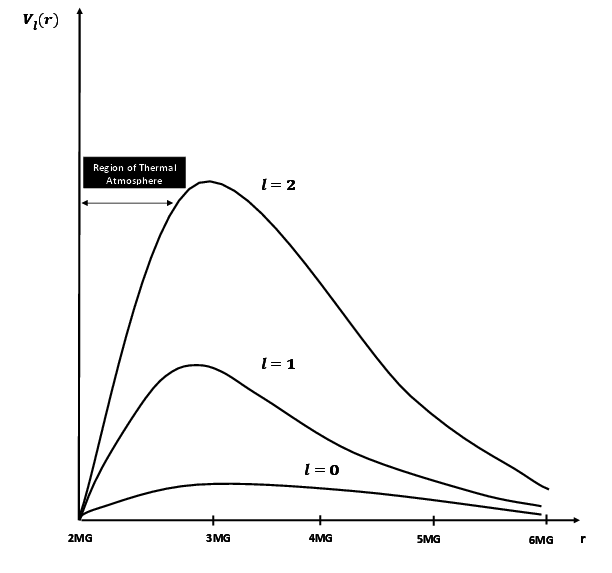}
\end{center}
\caption{Schwarzschild potential.}\label{QFTS}
\end{figure}

Indeed, to a distant observer the Schwarzschild black hole appears as a body with energy given by its mass $M$ and a temperature $T$ given by Hawking temperature 
\begin{eqnarray}
T=\frac{1}{8\pi GM}.\nonumber
\end{eqnarray}
However, there is a crucial difference between Rindler space and  Schwarzschild black hole. In the case of Rindler the thermal atmosphere (the particles near the horizon) is fully confined by the Rindler potential as opposed to the case of the Schwarzschild black hole where the thermal atmosphere is not fully confined by the  Schwarzschild potential. This means in particular that particles leaks out of the thermal atmosphere in the case of the Schwarzschild black hole and as a consequence the black hole evaporates. The particles which can escape the black hole have zero angular momentum for which the height of the potential barrier at around $r\simeq 3MG$ is of the same order as the thermal scale set by Hawking temperature while particles with larger angular momentum can not escape because for them the height of the potential barrier is much larger than the thermal scale. See figure (\ref{QFTS}).

The thermodynamical entropy $S$ is related to the energy and the temperature by the formula $dU=T dS$. Thus we obtain for the black hole the entropy 
\begin{eqnarray}
dS=\frac{dM}{T}=8\pi GM dM\Rightarrow S=4\pi GM^2.
\end{eqnarray}
However, the radius of the event horizon of the Schwarzschild black hole is $r_s=2MG$, and thus the area of the event horizon (which is a sphere) is 
\begin{eqnarray}
A=4\pi (2MG)^2.
\end{eqnarray}
By dividing the above two equations we get 
\begin{eqnarray}
S=\frac{A}{4G}.
\end{eqnarray}
The entropy of the black hole is proportional to its area. This is the famous Bekenstein-Hawking entropy formula.

\section{Hawking Radiation from QFT in Schwarzschild Background}

The original derivation of the Hawking radiation is found in \cite{Hawking:1974sw,Hawking:1976ra}. In here we will follow \cite{Mukhanov:2007zz,Polchinski:2016hrw} and to a lesser degree \cite{Giddings:1992ff,Jacobson:2003vx,Traschen:1999zr,Deeg}.
\subsection{Kruskal and Schwarzschild (Boulware) Observers and Field Expansions}
Let us start by recalling some formulas. The metric is  
\begin{eqnarray}
ds^2=-(1-\frac{2GM}{r})dt^2+\frac{dr^2}{1-\frac{2GM}{r}}+r^2d\Omega^2.
\end{eqnarray}
We define the Kruskal ingoing and outgoing null coordinates $U$ and $V$ (scaled versions of our previous $u^{'}$ and $v^{'}$) in region I as 
\begin{eqnarray}
U=r_su^{'}=-\sqrt{r_s(r-r_s)}e^{\frac{r-t}{2r_s}}~,~V=r_sv^{'}=\sqrt{r_s(r-r_s)}e^{\frac{r+t}{2r_s}}.
\end{eqnarray}
They satisfy 
\begin{eqnarray}
UV=r_s(r_s-r)e^{\frac{r}{r_s}}~,~\frac{U}{V}=-e^{-\frac{t}{r_s}}.
\end{eqnarray}
The metric becomes 
\begin{eqnarray}
ds^2=-\frac{4r_s}{r}e^{-\frac{r}{r_s}}dU dV+r^2d\Omega^2.
\end{eqnarray}
This form is valid throughout the spacetime and not only in region I. 

We consider now an inertial observer falling through the horizon $r_s=2GM$. This freely falling observer will cross the horizon in a finite proper time given by (with $r_s=r_i(1+\cos\alpha_s)/2$)
 \begin{eqnarray}
\tau=\sqrt{\frac{r_i^3}{4r_s}}(\alpha_s+\sin\alpha_s).
\end{eqnarray}
However, with respect to the Schwarzschild observer the radius $r$  of the freely falling object is related to its time $t$ by the formula (near the horizon)
\begin{eqnarray}
r-r_s=e^{-\frac{t}{r_s}}.
\end{eqnarray}
A distant inertial observer assumed to be hovering at a fixed radial distance $r_{\infty}$ will observe a proper time $\tau_{\infty}$ related to  Schwarzschild time $t$ by the equation 
\begin{eqnarray}
\tau_{\infty}=\sqrt{1-\frac{r_s}{r_{\infty}}}t.
\end{eqnarray}
Thus, this distant observer will then measure $\tau_{\infty}\longrightarrow\infty$ as $r\longrightarrow r_s$, i.e. she will never see the falling object actually crossing the horizon. Thus this observer may be interpreted as ending at the horizon. 

The discrepancy between the worldviews of the above two inertial observers (the freely falling and the asymptotic fixed observer) is what is at the source of Hawking radiation and all its related paradoxes \cite{Polchinski:2016hrw}.

We reduce the problem to $2$ dimensions, viz 
\begin{eqnarray}
ds^2=-(1-\frac{2GM}{r})dt^2+\frac{dr^2}{1-\frac{2GM}{r}}=-\frac{4r_s}{r}e^{-\frac{r}{r_s}}dU dV.
\end{eqnarray}
The tortoise coordinate (corresponding to a conformally flat metric) is defined by
  \begin{eqnarray}
dr=(1-\frac{r_s}{r})dr_*\rightarrow r_*=r-r_s+r_s\ln(\frac{r}{r_s}-1).
\end{eqnarray}
We will also work with the ingoing and outgoing null coordinates $u$ and $v$ defined only in quadrant I given by
\begin{eqnarray}
u&=&t-r_*\nonumber\\
&=&t-r-r_s\ln (\frac{r}{r_s}-1)+r_s\nonumber\\
&=&-2r_s\ln (\frac{-U}{r_s})+r_s.
\end{eqnarray}  
\begin{eqnarray}
v&=&t+r_*\nonumber\\
&=&t+r+r_s\ln (\frac{r}{r_s}-1)-r_s\nonumber\\
&=&2r_s\ln (\frac{V}{r_s})-r_s.
\end{eqnarray} 
The metric in this system is
\begin{eqnarray}
ds^2=-(1-\frac{2GM}{r})dudv=-\frac{r_s}{r}e^{\frac{v-u}{2r_s}}e^{-\frac{r-r_s}{r_s}}dudv.
\end{eqnarray}  
We will expand the field in modes as usual. The following important points should be taken into consideration.
\begin{itemize}
\item For the asymptotic inertial observer the modes will be denoted by the frequencies $\omega$ and they are clearly associated with the Schwarzschild time $t$ or equivalently $u=t-r_*$. This what corresponds to the exterior degrees of freedom. 
\item For the freely falling inertial observer the time is obviously given by the proper time $\tau$.  From equation (\ref{lk}) (with $\lambda=\tau$) and (\ref{lk1}) we obtain near the horizon 
\begin{eqnarray}
\frac{d\tau}{dt}\sim r-r_s\sim \exp(-t/r_s)\Rightarrow d\tau\sim \exp(-t/r_s)dt\Rightarrow \tau\sim -r_s\exp(-t/r_s)+\tau_0.
\end{eqnarray} 
We get then near the horizon 
\begin{eqnarray}
U\sim \frac{1}{\sqrt{r_s}}\exp(r/2r_s)(\tau-\tau_0)~,~V\sim \sqrt{r_s}\exp(r/2r_s).
\end{eqnarray} 
Thus $U\longrightarrow 0$ and  $V\longrightarrow ~{\rm constant}$. Also we conclude that the proper time $\tau$ is equivalent to the coordinate $U$ with frequencies denoted by $\nu$. Since $U$ is defined throughout spacetime the frequency $\nu$ is what corresponds to the interior degrees of freedom. 
\item
We know already that in the Schwarzschild geometry the solutions of the equation of motion are spherically symmetric which read
\begin{eqnarray}
\psi=\sum_{lm}Y_{lm}\psi_{lm}.
\end{eqnarray} 
The $\psi_{lm}$ solves schr\"odinger equation, viz
\begin{eqnarray}
(\partial_t^2-\partial_{r_*}^2+V(r_*))\psi_{lm}=0,
\end{eqnarray} 
with a potential function in the tortoise coordinates $r_*$ of the form 
\begin{eqnarray}
V(r_*)=\frac{r-r_s}{r}(\frac{r_s}{r}+\frac{l(l+1)}{r^2}).
\end{eqnarray} 
In the limit $r\longrightarrow \infty$ (the asymptotically flat spacetime limit) the tortoise coordinate behaves as $r_*\longrightarrow \infty$ and the potential goes to zero as $V\simeq l(l+1)/r^2$. The particle is therefore free in this limit. Similarly, in the near horizon limit $r\longrightarrow r_s$ the tortoise coordinate behave as $r_*\longrightarrow -\infty$ and the potential goes to zero again but now as $V\simeq (r-r_s)/r\sim \exp((r_*-r)/r_s)$. The particle is also free in this regime. 

Thus near infinity and near the horizon the solutions are plane waves of the form $\exp(ik(t\pm r_*)$ or equivalently $\exp(iku)$ and $\exp(ikv)$.

\item The scalar field action is 
\begin{eqnarray}
I&=&\frac{1}{2}\int d^2x\sqrt{-{\rm det}g}g^{\mu\nu}\partial_{\mu}\phi\partial_{\nu}\phi\nonumber\\
&=&-\int dU dV\partial_U\phi\partial_V\phi\nonumber\\
&=&-\int du dv\partial_u\phi\partial_v\phi=\frac{1}{2}\int dtdr_*\big(-(\partial_t\phi)^2+(\partial_{r_*}\phi)^2\big).
\end{eqnarray}
The equation of motion is 
\begin{eqnarray}
\partial_u\partial_v\phi=\partial_U\partial_V\phi=0.
\end{eqnarray}
The solution is 
\begin{eqnarray}
\phi&=&\phi_L(u)+\phi_R(v)\nonumber\\
&=&\phi_L(U)+\phi_R(V).
\end{eqnarray}
We will only consider the right moving part.
\item We consider a particular foliation of the near horizon geometry. For example, the coordinates $u$ and $v$ in region I 
are replaced by $\eta=t=(u+v)/2$ and $\xi=r_*=-(u-v)/2$ where $\eta$ is time. These coordinates near the horizon in region I define the metric of Rindler quadrant with acceleration given formally by $a=1/2r_s$, viz
\begin{eqnarray}
ds^2=\exp(2a \xi)(-d\eta^2+d\xi^2).
\end{eqnarray}
Thus, the Klein-Gordon inner product is precisely given by the formula (\ref{inner1}), viz
\begin{eqnarray}
(\phi_1,\phi_2)=-i\int \big(\phi_1\partial_{\eta}\phi_2^*-\partial_{\eta}\phi_1.\phi_2^*\big) d\xi.
\end{eqnarray}
We can check immediately that 
\begin{eqnarray}
(\phi_1,\phi_2)=-(\phi_2^*,\phi_1^*)~,~(\phi_1^*,\phi_2^*)=-(\phi_1,\phi_2)^*.
\end{eqnarray}
The positive frequency normalized modes in region I have been already computed. They are given by (\ref{pos})
\begin{eqnarray}
g_k^{(1)}&=&\frac{1}{\sqrt{4\pi \Omega}}\exp(-i\Omega\eta+ik\xi)~,~\Omega=|k|.
\end{eqnarray}
The right moving part of this positive frequency mode corresponds to $k>0$ and it is given explicitly by
\begin{eqnarray}
g_k^{(1)}&=&\frac{1}{\sqrt{4\pi \Omega}}\exp(-i\Omega u).
\end{eqnarray}
The right moving part with negative frequency corresponds therefore to $g_k^{(1)*}$. A right moving field will then be expanded as 
\begin{eqnarray}
\phi_R(u)=\int_0^{\infty}dk\bigg(\frac{b_k}{\sqrt{4\pi \Omega}}\exp(-i\Omega u)+\frac{b_k^+}{\sqrt{4\pi \Omega}}\exp(i\Omega u)\bigg).
\end{eqnarray}
After a change of variable $\omega=k $ and $b_{k}={b}_{\omega}/\sqrt{2\pi}$ we get 
\begin{eqnarray}
\phi_R(u)=\int_{0}^{\infty}\frac{d\omega}{2\pi}\bigg(\frac{b_{\omega}}{\sqrt{2\omega}}\exp(-i\omega u)+\frac{b_{\omega}^+}{\sqrt{2\omega}}\exp(i\omega u)\bigg).
\end{eqnarray}
Since $g_k^{(1)}$ are normalized such that $(g_k^{(1)},g_{k^{'}}^{(1)})=\delta (k-k^{'})$ the annihilation and creation operators $b_{k}$ and $b_{k}^+$ must satisfy $[b_k,b_{k^{'}}^+]=\delta (k-k^{'})$ and thus $[b_{\omega},b_{\omega^{'}}^+]=2 \pi \delta (\omega-\omega^{'})$.

\item From the above considerations, the field operator in the Schwarzschild tortoise coordinates $(t,r_*)$ is given by the formula 
 \begin{eqnarray}
\phi(t,r_*)=\int_{-\infty}^{+\infty}\frac{dk}{2\pi}\bigg(\frac{b_{k}}{\sqrt{2|k|}}\exp(-i|k|t+ikr_*)+\frac{b_{k}^+}{\sqrt{2|k|}}\exp(i|k|t-ikr_*)\bigg).
\end{eqnarray}
 \begin{eqnarray}
[b_k,b_{k^{\prime}}^+]=2\pi \delta(k-k^{\prime}).
\end{eqnarray}
The frequency is $\omega=|k|$ and $t$ is the proper time at infinity where  Schwarzschild becomes Minkowski. The momentum operator is 
\begin{eqnarray}
\pi(t,r_*)=\frac{\partial L}{\partial (\partial_t\phi)}&=&\partial_t\phi(t,r_*)\nonumber\\
&=&\int_{-\infty}^{+\infty}\frac{dk}{2\pi}\bigg(\frac{-i |k| b_{k}}{\sqrt{2|k|}}\exp(-i|k|t+ikr_*)+\frac{i|k| b_{k}^+}{\sqrt{2|k|}}\exp(i|k|t-ikr_*)\bigg).\nonumber\\
\end{eqnarray}
We compute immediately 
\begin{eqnarray}
[\phi(t,r_*),\pi(t,r_*^{\prime})]&=&i \int_{-\infty}^{+\infty}\frac{dk}{2\pi}e^{ik(r_*-r_*^{\prime})}=i\delta(r_*-r_*^{\prime}).
\end{eqnarray}
This confirms our normalization. 

The vacuum with respect to the inertial asymptotic tortoise Schwarzschild observer, also called the Boulware vacuum, is given by 
\begin{eqnarray}
b_k|0_T>=0~,~\forall k.
\end{eqnarray}
\item For an obvious reason, the mode expansion in the Kruskal coordinates $(U,V)$, with proper time given by $T=(U+V)/2$ and space like coordinate given by $X=-(U-V)/2$, is similar to the above expansion, viz
\begin{eqnarray}
\phi(T,X)
&=&\int_{-\infty}^{+\infty}\frac{dk}{2\pi}\bigg(\frac{a_{k}}{\sqrt{2|k|}}\exp(-i|k|T+ikX)+\frac{a_{k}^+}{\sqrt{2|k|}}\exp(i|k|T-ikX)\bigg).\nonumber\\
\end{eqnarray}
The frequency here is $\nu=|k|$ and $U$ is equivalent to the proper time $\tau$ of an infalling observer. The Kruskal vacuum is defined by 
\begin{eqnarray}
a_k|0_K>=0~,~\forall k.
\end{eqnarray}
\item The field decomposes into right moving field and left moving field or in the terminology of four dimensions into ingoing and outgoing fields. The right moving (outgoing) field corresponds to $k>0$ and the left moving (ingoing) field corresponds to $k<0$. We write the field as
\begin{eqnarray}
\phi(T,X)
&=&\int_{0}^{+\infty}\frac{d\nu }{2\pi}\bigg(\frac{a_{\nu}}{\sqrt{2\nu}}\exp(-i\nu U)+\frac{a_{-\nu}}{\sqrt{2\nu}}\exp(-i\nu V)+{\rm h.c}\bigg).
\end{eqnarray}
Similarly, 
\begin{eqnarray}
\phi(t,r_*)
&=&\int_{0}^{+\infty}\frac{d\omega }{2\pi}\bigg(\frac{b_{\omega}}{\sqrt{2\omega}}\exp(-i\omega u)+\frac{b_{-\omega}}{\sqrt{2\omega}}\exp(-i\omega v)+{\rm h.c}\bigg).
\end{eqnarray}

\end{itemize}

\subsection{Bogolubov Coefficients}
Let us summarize our main points. We have two observers: the asymptotic Schwarzschild tortoise observer and the freely falling Kruskal observer. The Schwarzschild observer defined for $r>r_s$ is the analogue of the accelerating Rindler observer with acceleration given by $a=1/2r_s$, whereas the Kruskal observer corresponds to the inertial Minkowski observer defined throughout the spacetime manifold.

The asymptotic observer at fixed $r$ ($r>r_s$) expands the right moving field in terms of the modes $v_{\omega}$ as
\begin{eqnarray}
\phi_R(u)=\int_0^{\infty}d\omega (v_{\omega}b_{\omega}+v_{\omega}^*b_{\omega}^{\dagger})~,~v_{\omega}=\frac{1}{\sqrt{4\pi\omega}}\exp(-i\omega u).
\end{eqnarray}
We have the normalization 
\begin{eqnarray}
(v_{\omega_1},v_{\omega_2})=\delta(\omega_1-\omega_2)~,~[b_{\omega},b_{\omega^{\prime}}^+]=\delta(\omega-\omega^{\prime}).
\end{eqnarray}
This observer sees the  Schwarzschild tortoise vacuum
\begin{eqnarray}
b_{\omega}|0_T>=0.
\end{eqnarray}
The freely falling observer expands the right moving field in terms of the modes $u_{\nu}$ as
\begin{eqnarray}
\phi_R(U)=\int_0^{\infty}d\nu (u_{\nu}a_{\nu}+u_{\nu}^*a_{\nu}^{\dagger})~,~u_{\nu}=\frac{1}{\sqrt{4\pi\nu}}\exp(-i\nu U).
\end{eqnarray}
We have the normalization 
\begin{eqnarray}
(u_{\nu_1},u_{\nu_2})=\delta(\nu_1-\nu_2)~,~[a_{\nu},a_{\nu^{\prime}}^+]=\delta(\nu-\nu^{\prime}).
\end{eqnarray}
This observer sees the Kruskal vacuum
\begin{eqnarray}
a_{\nu}|0_K>=0.
\end{eqnarray}
The asymptotic and freely falling objects are related through the Bogolubov transformations  

\begin{eqnarray}
v_{\omega}=\int_0^{\infty}d\nu (\alpha_{\omega\nu}u_{\nu}+\beta_{\omega\nu}u_{\nu}^*)~,~u_{\nu}=\int_0^{\infty}d\omega (\alpha_{\omega\nu}^*v_{\omega}-\beta_{\omega\nu}v_{\omega}^*).
\end{eqnarray}
\begin{eqnarray}
a_{\nu}=\int_0^{\infty}d\omega (\alpha_{\omega\nu}b_{\omega}+\beta_{\omega\nu}^*b_{\omega}^{\dagger})~,~b_{\omega}=\int_0^{\infty}d\nu (\alpha_{\omega\nu}^*a_{\nu}-\beta_{\omega\nu}^*a_{\nu}^{\dagger}).\label{next}
\end{eqnarray}
The first equation should be corrected by the introduction of the interior degrees of freedom (see next lecture). Nevertheless, the  Bogolubov coefficients are 
\begin{eqnarray}
\alpha_{\omega\nu}=(v_{\omega},u_{\nu})~,~\beta_{\omega\nu}=-(v_{\omega},u_{\nu}^*).
\end{eqnarray}
We calculate immediately 
\begin{eqnarray}
\alpha_{\omega\nu}=(v_{\omega},u_{\nu})&=&-i\int \frac{2}{4\pi\sqrt{\omega\nu}}(i\omega\partial_{\eta}u)e^{-i\omega u}e^{i\nu U}dr_*\nonumber\\
&=&-\int_{-\infty}^{+\infty} \frac{du}{2\pi}\sqrt{\frac{\omega}{\nu}}e^{-i\omega u}e^{i\nu U}\Rightarrow \alpha_{\omega\nu}^*=-\sqrt{\frac{\omega}{\nu}}F(\omega,\nu).
\end{eqnarray}
Similarly,
\begin{eqnarray}
\beta_{\omega\nu}&=&-(v_{\omega},u_{\nu}^*)\nonumber\\
&=&\int_{-\infty}^{+\infty} \frac{du}{2\pi}\sqrt{\frac{\omega}{\nu}}e^{-i\omega u}e^{-i\nu U}\Rightarrow -\beta_{\omega\nu}^*=-\sqrt{\frac{\omega}{\nu}}F(\omega,-\nu).
\end{eqnarray}
The function $F$ is given by (with $U=U_0e^{-a u}$ where $U_0=-\sqrt{e}/{2a}$)
\begin{eqnarray}
F(\omega,\nu)
&=&\int_{-\infty}^{+\infty} \frac{du}{2\pi}e^{i\omega u}e^{-i\nu U}=\int_{-\infty}^{+\infty}\frac{du}{2\pi}e^{i\omega u-i\nu U_0 e^{-a u}}.
\end{eqnarray}
This is the Euler gamma function. Indeed, if we make the change of variable $u\longrightarrow z=i\nu U_0 e^{-au}$ we immediately reach the formula 
\begin{eqnarray}
F(\omega,\nu)
&=&\frac{1}{2\pi a}\exp(\frac{i\omega}{a}\ln i\nu U_0)\int_{0}^{+\infty} e^{-\frac{i\omega}{a}-1} e^{-z} dz=\frac{1}{2\pi a}\exp(\frac{i\omega}{a}\ln i\nu U_0)\Gamma(-\frac{i\omega}{a}).\nonumber\\
\end{eqnarray}
The number of $b$ particles of frequency $\omega$ as seen by the Schwarzschild asymptotic fixed observer is given by the expectation value of the number operator $N_{\omega}=b_{\omega}^+b_{\omega}$. Obviously, the expectation value of this number operator in the tortoise vacuum is zero, viz $\langle 0_T|N_{\omega}|0_T\rangle=0$. However, the actual vacuum state of lowest energy of the quantum scalar field in the presence of a classical black hole is given by the freely falling Kruskal vacuum $|0_K\rangle$. This is because Schwarzschild is the analogue of Rindler whereas Kruskal is the analogue of Minkowski. See the nice discussion in \cite{Mukhanov:2007zz}. Also, in consideration of the gravitational collapse of a star onto a black hole it has been shown that before the collapse the vacuum state is that of Minkowski and after the collapse the vacuum state becomes that of Kruskal \cite{Hawking:1974sw,Hawking:1976ra}. Thus, the vacuum state is $|0_K\rangle$ and it does actually contain $b$ particles as seen by the asymptotic  Schwarzschild observer since 
\begin{eqnarray}
\langle 0_K|N_{\omega}|0_K\rangle &=&\langle 0_K|b_{\omega}^+b_{\omega}|0_K\rangle\nonumber\\
&=&\int_0^{\infty} d\nu|\beta_{\omega\nu}|^2.
\end{eqnarray}

\begin{figure}[H]
\begin{center}
\includegraphics[width=10.0cm,angle=0]{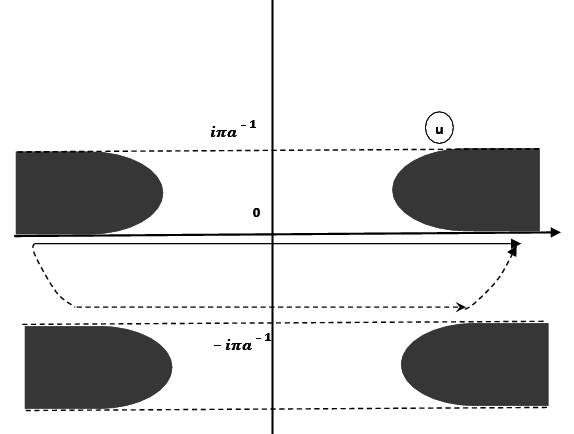}
\end{center}
\caption{The contour of integration.}\label{contour}
\end{figure}
\subsection{Hawking Radiation and Hawking Temperature}

The Bogolubov coefficient can be expressed in terms of Euler gamma function as shown above and then integrated over. However, the method outlined in \cite{Mukhanov:2007zz} is more illuminating.

We deform the $u$ integral from $-\infty$ to $+\infty$ to the $t$ integral from $-\infty -i\pi/a$ to $+\infty -i\pi/a$ where $u=t+i\pi/a$. See figure (\ref{contour}). The integral is not changed because $(i)$ the integrand has no poles which is obvious, $(i)$ the lateral segments are limited in length which is also obvious and $(i)$ the integrand vanishes for $t\longrightarrow \pm \infty -i\alpha$ where $0< \alpha< \pi/a$. The last point is shown as follows. Firstly, 
\begin{eqnarray}
{\rm lim}_{t\longrightarrow -\infty-i\alpha}{\rm Re}\bigg(i\nu U_0 e^{-a t}\bigg)&=&{\rm lim}_{u\longrightarrow -\infty}{\rm Re}\bigg(i\nu U_0 e^{ia\alpha} e^{-a u}\bigg)\nonumber\\
&=&-{\rm lim}_{t\longrightarrow -\infty}{\rm Re}\bigg(\nu U_0 \sin a\alpha e^{-a u}\bigg)\nonumber\\
&=&-\infty.\label{sinalpha}
\end{eqnarray}
For the limit  $t\longrightarrow + \infty -i\alpha$ the integral diverges and we need to regularize it for example as
\begin{eqnarray}
F(\omega,\nu)
=\int_{-\infty}^{+\infty}\frac{du}{2\pi}e^{i\omega u-i\nu U_0 e^{-a u}}e^{-bu^2}~,~b> 0.
\end{eqnarray}
This integral then for $b$ positive is zero in the limit  $t\longrightarrow + \infty -i\alpha$. Since there are no poles inside the closed contour formed by the original contour and the shifted one as in the figure below we conclude immediately that $F$ can be given by the integral 
\begin{eqnarray}
F(\omega,\nu)
&=&\int_{-\infty-\frac{i\pi}{a}}^{+\infty-\frac{i\pi}{a}}\frac{dt}{2\pi}e^{i\omega t-i\nu U_0 e^{-a t}}\nonumber\\
&=&\exp(\frac{\omega \pi}{a})F(\omega,-\nu).\label{lf}
\end{eqnarray}
This result should be understood in the sense of distribution. The exhibited contour is the unique possibility allowed to us since we can not deform the contour to $u=t-i(\pi +2\pi n)/a$ with $n\neq 0$ because the $\sin a\alpha$ in (\ref{sinalpha}) will change sign.

We use now the last formula (\ref{lf}) to compute the expectation value of the Schwarzschild asymptotic observer number operator $N_{\omega}$ in the Kruskal (black hole) vacuum $|0\rangle $ as follows. We start from the normalization condition 
\begin{eqnarray}
\delta(\omega-\omega^{\prime})&=&(v_{\omega},v_{{\omega}^{\prime}})\nonumber\\
&=&\int_0^{\infty}d\nu \big[\alpha_{\omega\nu}(u_{\nu},v_{{\omega}^{\prime}})+\beta_{\omega\nu}(u_{\nu}^*,v_{{\omega}^{\prime}})\big]\nonumber\\
&=&\int_0^{\infty}d\nu\bigg[\alpha_{\omega\nu}\alpha_{\omega^{\prime}\nu}^*-\beta_{\omega\nu}\beta^*_{\omega^{\prime}\nu}\bigg]\nonumber\\
&=&\int_0^{\infty}d\nu\frac{\sqrt{\omega\omega^{\prime}}}{\nu}\bigg[F^*(\omega,\nu)F(\omega^{\prime},\nu)-F^*(\omega,-\nu)F(\omega^{\prime},-\nu)\bigg]\nonumber\\
&=&\bigg(e^{\frac{\pi(\omega+\omega^{\prime})}{a}}-1\bigg)\int_0^{\infty}d\nu\frac{\sqrt{\omega\omega^{\prime}}}{\nu}F^*(\omega,-\nu)F(\omega^{\prime},-\nu).
\end{eqnarray}
We write this equation as 
\begin{eqnarray}
\int_0^{\infty}d\nu\frac{\sqrt{\omega\omega^{\prime}}}{\nu}F^*(\omega,-\nu)F(\omega^{\prime},-\nu)&=&\frac{\delta(\omega-\omega^{\prime})}{e^{\frac{\pi(\omega+\omega^{\prime})}{a}}-1}.
\end{eqnarray}
For $\omega=\omega^{'}$ we get precisely the desired result 
 \begin{eqnarray}
\int_0^{\infty}d\nu\frac{\omega}{\nu}|F(\omega,-\nu)|^2&=&\frac{\delta(0)}{e^{\frac{2\pi\omega}{a}}-1}.
\end{eqnarray}
In other words, 
\begin{eqnarray}
\langle 0_K|N_{\omega}|0_K\rangle &=&\langle 0_K|b_{\omega}^+b_{\omega}|0_K\rangle\nonumber\\
&=&\int_0^{\infty} d\nu|\beta_{\omega\nu}|^2\nonumber\\
&=&\frac{\delta(0)}{\exp(\frac{2\pi\omega}{a})-1}.
\end{eqnarray}
The density of $b$ particles in the black hole vacuum state $|0_K>$ is therefore given by \footnote{In $1+3$ dimensions using box normalization we have $(2\pi)^3\delta^3(0)=V$ where $V$ is the volume of spacetime. In the current $1+1$ dimensional case we have $(2\pi)\delta (0)=L$.}
 \begin{eqnarray}
n_{\omega}&=&\frac{1}{2\pi}\frac{1}{\exp(\frac{2\pi\omega}{a})-1}.
\end{eqnarray}
This a blackbody Planck spectrum with the temperature 
 \begin{eqnarray}
T_H=\frac{a}{2\pi}=\frac{1}{4\pi r_s}=\frac{1}{8\pi GM}.
\end{eqnarray}
By inserting SI units we obtain 
\begin{eqnarray}
T_H=\frac{\hbar c^3}{8\pi GM k_B}.
\end{eqnarray}
This is the famous Hawking temperature. The black hole as seen by a distant observer is radiating energy, thus its mass decreases, and as a consequence its temperature increases, i.e. the black hole becomes hotter, which indicates a negative specific heat.

\section{The Unruh vs Boulware Vacua: Pure to Mixed}
We will follow here the excellent pedagogical presentation of \cite{Jacobson:2003vx}.

The first type of information loss is by falling across the event horizon. The second type which is intimately related concerns Hawking radiation and is equivalent to the evolution of pure states to mixed states which is a process forbidden by quantum mechanics.

\subsection{The Adiabatic Principle and Trans-Planckian Reservoir}

We start with the Rindler space (which is the cleanest of the two cases) where we have obtained the Unruh effect by two methods. By computing the density matrix and also the flux formula with respect to the Rindler observer. By using quantum information, we have found that we can put the density matrix into the form  
 \begin{eqnarray}
\rho_R&=&\frac{1}{Z}\exp(-\frac{2\pi}{a}H_R)=\frac{1}{Z}\sum_i\exp(-\frac{2\pi}{a} E_i)|i_R\rangle \langle i_R|.
\end{eqnarray}
This is a mixed (thermal,random) state obtained by integrating out the left wedge degrees of freedom in the vacuum pure (entangled, correlated) state 
\begin{eqnarray}
|\Omega\rangle=\frac{1}{\sqrt{Z}}\sum_i \exp(-\pi E_i)|i_R\rangle|i^*_L\rangle.
\end{eqnarray}
On the other hand, by using QFT in curved backgrounds we calculated the number of particles with energy $\omega=|k|$ seen by the Rindler observer in the vacuum Minkowski state $|0_M\rangle \equiv |\Omega\rangle$ to be given by the blackbody spectrum
 \begin{eqnarray}
\langle 0_M|\hat{N}_R^{(1)}(k)|0_M\rangle=\frac{1}{\exp(\frac{2\pi\omega}{a})-1}\delta(0).
\end{eqnarray}
Since the near horizon geometry of the Schwarzschild black hole is Rindler a similar result is expected to hold in the Schwarzschild black hole geometry. Indeed, this is the result of \cite{Wald:1975kc} which we will try to derive here following  \cite{Jacobson:2003vx}.

In discussing the Hawking radiation so far we have omitted several points. First, we have only considered the exterior region. Second,  we did not talk about greybody factors and furthermore we have not mentioned at all the underlying adiabatic approximation or the trans-Planckian problem and its so-called nice-slice resolution. All these issues can be remedied somewhat by considering black holes as forming from collapsing shell of matter in some pure quantum state $|\psi\rangle$.

We consider therefore a black hole which had formed during gravitational collapse in a quantum state $|\psi\rangle$. The out state corresponds to an outgoing Killing null wave packet $P$ centered around some positive frequency $\omega$ with support only at large radii $r$ at late times $t\longrightarrow +\infty$. Recall that $\omega$, $r$ and $t$ relate to the Schwarzschild tortoise coordinates. Obviously, this wave packet is a solution of the Klein-Gordon equation which behaves at infinity as $\exp(-i\omega t)$ and thus near the horizon it can only depend on the outgoing (right moving) coordinates $u=t-r_*$, viz $P\propto \exp(-i\omega u)$. This wave packet $P$ corresponds to an annihilation operator $a(P)$ given in terms of the field operator $\phi$, which solves the Klein-Gordon equation, by the Klein-Gordon inner product 
\begin{eqnarray}
a(P)=(\phi,P).
\end{eqnarray}
We run this wave packet backwards in time towards the black hole. A reflected part $R$ will scatter off the black hole and return to large radii and a transmitted part $T$ with support only immediately outside the event horizon. We write 
\begin{eqnarray}
P=R+T.
\end{eqnarray}

\begin{figure}[H]
\begin{center}
\includegraphics[width=12.0cm,angle=0]{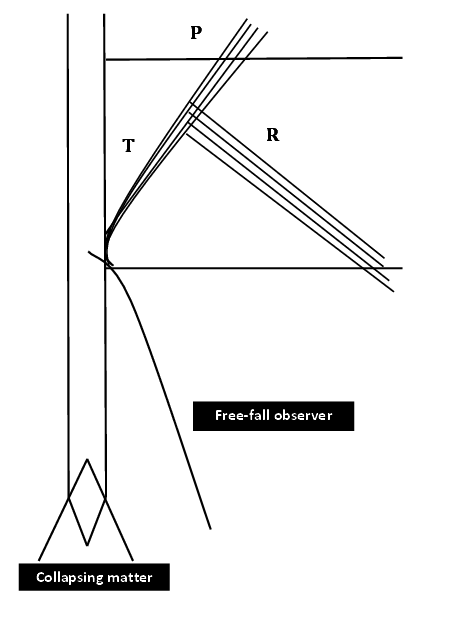}
\end{center}
\caption{The wave packets $P$, $R$ and $T$ near the horizon.}
\end{figure}

The wave packets $R$ and $T$ have the same positive Killing frequency with respect to the asymptotic Schwarzschild observer as the outgoing wave packet $P$ because the black hole metric is stationary. But with respect to a freely falling observer who intersects the trajectory of the transmitted wave packet $T$ at the event horizon both positive and negative frequency modes will be seen in $T$. The annihilation operator $a(P)$ decomposes in an almost obvious way as
\begin{eqnarray}
a(P)=a(R)+a(T).
\end{eqnarray}
Since the reflected wave packet $R$ has only support in the asymptotic flat region very far outside the black hole and since $|\psi\rangle$ contains no positive frequency incoming excitation the annihilation operator $a(R)$ annihilates the state $|\psi\rangle $ exactly
\begin{eqnarray}
a(R)|\psi\rangle =0.
\end{eqnarray}
If the state $|\psi\rangle $ were also annihilated by $T$ it would have been identical with the Boulware vacuum or tortoise vacuum $|0_T\rangle$ introduced in the previous lecture. But $T$ contains positive frequencies as well as negative frequencies with respect to the proper time of the freely falling observer. Thus we decompose it as follows
\begin{eqnarray}
T=T^++T^-\Rightarrow a(T)=a(T^+)+a(T^-).
\end{eqnarray}
By using the property of the Klein-Gordon inner product $(\phi_1,\phi_2)^*=-(\phi_1^*,\phi_2^*)$ we derive immediately that $a^{\dagger}(\bar{T}^{-})=-a(T^-)$. Thus
\begin{eqnarray}
a(T)=a(T^+)-a^{\dagger}(\bar{T}^-).
\end{eqnarray}
We already know that $T$ has only support near the horizon where it behaves as $T\sim \exp(-i\omega u)$. But near the horizon we have $r-r_s=\exp(-t/r_s)$ and $u\sim 2t\sim -2r_s\ln -(\tau-\tau_0)/r_s$. Thus the behavior of $T$ is of the general form ($a=\kappa=1/2r_s$)
\begin{eqnarray}
&&T\sim\exp(i\frac{\omega}{a}\ln(-\tau))~,~\tau<0\nonumber\\.
&&T=0~,~\tau>0.
\end{eqnarray}
Thus near the horizon $T$ consists of rapid oscillations which means in particular that $T^+$ and $\bar{T}^-$ are positive high frequency modes. Initially, the black  hole state $|\psi\rangle$ does not contain these high energy modes. We say that these modes are in their ground states.

As we evolve backward in time the frequencies blueshift (increase) in the same way that when evolving forward in time they will redshift (decrease). Thus, as we approach the horizon the frequency increases, with respect to the freely falling observer, until it becomes infinitely blueshifted on the horizon. In other words, these modes seem to arise deep in the UV region which is what we call the trans-Planckian reservoir. This could be a problematic issue as discussed in \cite{Jacobson:2003vx} with a proposed resolution which goes under the name of the nice-slice argument given in \cite{Polchinski:1995ta}. Both the potential problem and the proposed resolution are not very essential to us here. Indeed, we are only using the above fact regarding the very large blueshift on the horizon to conclude that the modes  $T^+$ and $\bar{T}^-$ remain high energy modes as we evolve them backward in time. Furthermore,  the earlier the infalling observer meets the mode with frequency $\omega$ the higher its proper frequency $\nu$ will be since the Schwarzschild frequency $\omega$ is redshifted with respect to the free fall frequency $\nu$ as $\nu= 2r_s\omega$.

Hence, by looking at the black hole after it had formed at times $t<<r_s$, where the Schwarzschild radius $r_s$ measures the time scale of the collapse process, the high frequency modes with $\omega>>1/r_s$ ($\nu>>2$) are not excited, which means in particular that the modes $T^+$ and $\bar{T}^-$ remain unexcited, i.e. they remain in their ground states. We conclude that the black hole state $|\psi\rangle$ does not contain positive high frequency modes throughout, viz 
\begin{eqnarray}
a(T^+)|\psi\rangle=a(\bar{T}^-)|\psi\rangle =0.
\end{eqnarray}
This is essentially the adiabatic principle. The geometry during the gravitational collapse is obviously time dependent with a time scale given by the Schwarzschild time  $r_s$. Thus, the modes with frequencies $\omega>>1/r_s$ see the change of the geometry adiabatically, i.e. very slowly, and hence they remain unexcited. 
\subsection{The Unruh Method Revisited and Grey Body Factor}
If we consider now the expectation value of the number operator $N=a^{\dagger}(P)a(P)$ in the black hole state $|\psi\rangle$ we find immediately 
\begin{eqnarray}
\langle \psi|N|\psi\rangle&=&\langle \psi|a^{\dagger}(P)a(P)|\psi\rangle\nonumber\\
&=&\langle \psi|a^{\dagger}(T)a(T)|\psi\rangle\nonumber\\
&=&\langle \psi|a(\bar{T}^{-})a^{\dagger}(\bar{T}^{-})|\psi\rangle\nonumber\\
&=&\langle \psi|[a(\bar{T}^{-}),a^{\dagger}(\bar{T}^{-})]|\psi\rangle.
\end{eqnarray}
However, we can explicitly expand the field operator in a positive frequency basis $\{f_i\}$ as
\begin{eqnarray}
\phi=\sum_i(a_if_i+a_i^{\dagger}f_i^*).
\end{eqnarray}
Also, the positive frequency wave packet $\bar{T}^{-}$ can be expanded similarly as 
\begin{eqnarray}
\bar{T}^{-}=\sum_it_i^*f_i.
\end{eqnarray}
The annihilation operator $a(\bar{T}^{-})$ is then given explicitly by 
\begin{eqnarray}
a(\bar{T}^{-})=\sum_ia_it_i.
\end{eqnarray}
We compute then 
\begin{eqnarray}
  \langle \psi|[a(\bar{T}^{-}),a^{\dagger}(\bar{T}^{-})]|\psi\rangle&=&\langle \psi\sum_i\sum_jt_it_j^*[a_i,a_j^{\dagger}]|\psi\rangle\nonumber\\
&=&\sum_i t_it_i^*\nonumber\\
&=&(\bar{T}^{-},\bar{T}^{-})\nonumber\\
&=&-(T^{-},T^{-}).
\end{eqnarray}
Thus the expectation value of the number operator becomes
\begin{eqnarray}
\langle \psi|N|\psi\rangle&=&-(T^{-},T^{-}).
\end{eqnarray}
The transmitted wave packet is given by 
\begin{eqnarray}
&&T(\tau)=\exp(i\frac{\omega}{a}\ln(-\tau))~,~\tau<0\nonumber\\.
&&T=0~,~\tau>0.\label{sl}
\end{eqnarray}
Since $\tau<0$ this is defined only outside the horizon. Thus this function contains positive and negative frequency modes with respect to the freely falling observer (recall that $T$ is a positive frequency mode with respect to the Schwarzschild observer).  This is the analogue of $g_k^{(1)}$ in the case of Rindler which was defined as a positive frequency solution only with respect to the Rindler observer in quadrant I but with respect to the Minkowski observer it contains both positive and negative frequencies. As we did in reaching equations (\ref{h1}) and (\ref{h2}) in the Rindler case, by using the method of \cite{Unruh:1976db}, we will now extend the solution (\ref{sl}) to the region inside the horizon ($\tau>0$) and obtain in the course the positive frequency and the negative frequency extensions $T^+$ and $T^-$.

First, recall that a positive frequency mode can be expanded in terms of $\exp(-i\omega \tau)$, $\omega>0$. The functions $\exp(-i\omega \tau)$ clearly vanishes in the limit $|\tau|\longrightarrow \infty$ in the lower half complex $\tau$ plane for $\omega>0$. Thus the positive frequency extension of $T$ should be obtained by analytic continuation in the lower half complex plane. This extension of $T$ from $\tau<0$ to $\tau>0$ is obtained by analytic continuation of $\ln(-\tau)$ from $\tau<0$ to $\tau>0$ in the lower half complex plane provided the branch cut of the logarithm is chosen in the upper half complex plane. This continuation of $\ln(-\tau)$ with $\tau<0$ is given by $\ln \tau+i\pi$ with $\tau>0$ 
\footnote{The function $\ln z$ is multi-valued in the complex plane. To get a single-valued function we introduce a cut line between its two branch points $z=0$ and $z=\infty$. 

For positive frequency modes we will need to extend in the lower half complex plane and choose the branch cut in the upper half complex plane. The function $\ln (-\tau)$ with $\tau<0$ is analytically continued to $\tau>0$ by writing $z=-\tau\exp(i\theta)$. Since the branch cut is in the upper half complex plane we can only go from $z=\tau$ to $z=-\tau$ counter clockwise in the lower half plane, i.e. from $\theta=\pi$ to $\theta=2\pi$. At $\theta=\pi$ we have $z=\tau<0$ and $\ln z-i\pi=\ln (-\tau)$ whereas at $\theta=2\pi$ we have $z=-\tau=\tau^{\prime}>0$ and $\ln z-i\pi=\ln z^{\prime}+i\pi$. Thus the analytic continuation of $\ln(-\tau)$, $\tau<0$, in the lower half complex plane is given by $\ln \tau+i\pi$, $\tau>0$, if the branch cut is in the upper half complex plane.}. By replacing in $T(\tau)$ with $\tau<0$ we get $T(-\tau)\exp(-\pi\omega/a)$ with $\tau>0$. The wave packet solution inside the horizon is then given by
 \begin{eqnarray}
&&\tilde{T}(\tau)=T(-\tau)=\exp(i\frac{\omega}{a}\ln(\tau))~,~\tau>0\nonumber\\.
&&\tilde{T}=0~,~\tau<0.\label{sle}
\end{eqnarray}
The total wave packet 
 \begin{eqnarray}
T^+=c_+(T+\tilde{T}\exp(-\frac{\pi\omega}{a}))
\end{eqnarray}
is clearly analytic in the lower half complex plane and bounded as $|\tau|\longrightarrow \infty$ and as such it can only contain positive frequencies. In other words, $T^+$ is the desired positive frequency extension of $T$.

The negative frequency extension of $T$ should be obtained by analytic continuation in the upper half complex plane. This extension of $T$ from $\tau<0$ to $\tau>0$ is obtained by analytic continuation of $\ln(-\tau)$ from $\tau<0$ to $\tau>0$ in the upper half complex plane provided the branch cut of the logarithm is chosen in the lower half complex plane. This continuation of $\ln(-\tau)$ with $\tau<0$ is given by $\ln \tau-i\pi$ with $\tau>0$ 
\footnote{For negative frequency modes we will need to extend in the upper half complex plane and choose the branch cut in the lower half complex plane. The function $\ln (-\tau)$ with $\tau<0$ is again analytically continued to $\tau>0$ by writing $z=-\tau\exp(i\theta)$. Since the branch cut now is in the lower half complex plane we can only go from $z=\tau$ to $z=-\tau$ counter anti-clockwise in the upper half plane, i.e. from $\theta=\pi$ to $\theta=0$. At $\theta=\pi$ we have $z=\tau<0$ and $\ln z-i\pi=\ln (-\tau)$ as before whereas at $\theta=0$ we have $z=-\tau=\tau^{\prime}>0$ and $\ln z-i\pi=\ln z^{\prime}-i\pi$. Thus the analytic continuation of $\ln(-\tau)$, $\tau<0$, in the upper half complex plane is given by $\ln \tau-i\pi$, $\tau>0$, if the branch cut is in the lower half complex plane.

}. By replacing in $T(\tau)$ with $\tau<0$ we get $T(-\tau)\exp(\pi\omega/a)$ with $\tau>0$. The total wave packet 
 \begin{eqnarray}
T^-=c_-(T+\tilde{T}\exp(\frac{\pi\omega}{a}))
\end{eqnarray}
is clearly analytic in the upper half complex plane and bounded as $|\tau|\longrightarrow \infty$ and as such it can only contain negative frequencies. In other words, $T^-$ is the desired negative frequency extension of $T$.

The boundary conditions are given by 
 \begin{eqnarray}
&&T^++T^-=T~,~\tau<0\Rightarrow c_++c_-=1\nonumber\\
&&T^++T^-=0~,~\tau>0\Rightarrow c_+\exp(-\frac{\pi\omega}{a})+c_-\exp(\frac{\pi\omega}{a})=0.
\end{eqnarray}
This gives immediately 
\begin{eqnarray}
c_+=\frac{1}{1-\exp(-\frac{2\pi\omega}{a})}~,~c_-=\frac{1}{1-\exp(\frac{2\pi\omega}{a})}.
\end{eqnarray}
By using now the negative frequency extension $T^-$ we can immediately compute the expectation value of the number operator to be given by (using also $(T,T)=-(\tilde{T},\tilde{T})$ and $(T,\tilde{T})=0$)
\begin{eqnarray}
\langle \psi|N|\psi\rangle
&=&\frac{(T,T)}{\exp(\frac{2\pi\omega}{a})-1}.
\end{eqnarray}
This is again a blackbody spectrum with the Hawking temperature $T_H=a/2\pi=1/4\pi r_s$. However, this result is actually reduced by the so-called greybody factor 
\begin{eqnarray}
\Gamma=(T,T).
\end{eqnarray}
This has the normal quantum mechanical interpretation of being the transmission probability, i.e. the probability that the wave packet $P$ when evolved backward in time  will become squeezed up against the event horizon.
\subsection{Unruh Vacuum State $|U\rangle$}
We will look now at the vacuum conditions $a(T^+)|\psi\rangle =0$, $a(\bar{T}^-)|\psi\rangle =0$ more closely. We have (using $(\phi,T)=a(T)$, $-(\phi,\bar{T})=a^{\dagger}(T)$, $(\phi,\tilde{T})=a(\tilde{T})$, $-(\phi,\bar{\tilde{T}})=a^{\dagger}(\tilde{T})$)
\begin{eqnarray}
a(T^+)&=&(\phi,T^+)=c_+a(T)+c_+e^{-\frac{\pi\omega}{a}}a(\tilde{T}).
\end{eqnarray}
\begin{eqnarray}
a(\bar{T}^-)&=&(\phi,\bar{T}^-)=-c_-a^{\dagger}(T)-c_-e^{\frac{\pi\omega}{a}}a^{\dagger}(\tilde{T}).
\end{eqnarray}
But $\tilde{T}$ is a negative norm solution. Thus, $a(\tilde{T})=-a^{\dagger}(\bar{\tilde{T}})$ and $a^{\dagger}(\tilde{T})=-a(\bar{\tilde{T}})$. The vacuum conditions 
  $a(T^+)|\psi\rangle =0$, $a(\bar{T}^-)|\psi\rangle =0$ become 
\begin{eqnarray}
\bigg(a(T)-e^{-\frac{\pi\omega}{a}}a^{\dagger}(\bar{\tilde{T}})\bigg)|\psi\rangle =0.\label{Un1}
\end{eqnarray}
\begin{eqnarray}
\bigg(-a^{\dagger}(T)+e^{\frac{\pi\omega}{a}}a(\bar{\tilde{T}})\bigg)|\psi\rangle=0.\label{Un2}
\end{eqnarray}
The operator $a(T)$ is the analogue of the operator $b_{\omega}$ in (\ref{next}) which is the exterior annihilation operator. The operator $a(\tilde{T})$ is therefore the interior annihilation operator which we will denote by $\tilde{b}_{\omega}$. The first equation in (\ref{next}) should then be corrected as

\begin{eqnarray}
a_{\nu}=\int_0^{\infty}d\omega (\alpha_{\omega\nu}b_{\omega}+\beta_{\omega\nu}^*b_{\omega}^{\dagger}+\tilde{\alpha}_{\omega\nu}\tilde{b}_{\omega}+\tilde{\beta}_{\omega\nu}^*\tilde{b}_{\omega}^{\dagger}).
\end{eqnarray}
The equations (\ref{Un1}) and (\ref{Un2}) define the so-called Unruh vacuum $|U\rangle$. As noted before, the Boulware vacuum which we will denote here by $|B\rangle$ should be annihilated by the transmission annihilation operators $a(T)$ and $a(\bar{\tilde{T}})$, viz
\begin{eqnarray}
a(T)|B\rangle=a(\bar{\tilde{T}})|B\rangle=0.
\end{eqnarray}
This state is different from the initial black hole state $|\psi\rangle$. By using the facts $[a(T),a^{\dagger}(T)]=1$ and $[a(\bar{\tilde{T}}),a^{\dagger}(\bar{\tilde{T}})]=1$ (we are assuming that the wave packets $T$ and $\bar{\tilde{T}}$ are normalized) we can represent the annihilation operators as $a(T)=\partial/\partial a^{\dagger}(T)$ and $a(\bar{\tilde{T}})=\partial/\partial a^{\dagger}(\bar{\tilde{T}})$ and as a consequence we can rewrite equations (\ref{Un1}) and (\ref{Un2}) in the form
\begin{eqnarray}
a(T)|U\rangle=e^{-\frac{\pi\omega}{a}}a^{\dagger}(\bar{\tilde{T}})|U\rangle\Rightarrow \frac{\partial}{\partial a^{\dagger}(T)}|U\rangle=e^{-\frac{\pi\omega}{a}}a^{\dagger}(\bar{\tilde{T}})|U\rangle.
\end{eqnarray}
\begin{eqnarray}
a(\bar{\tilde{T}})|U\rangle=e^{-\frac{\pi\omega}{a}} a^{\dagger}(T)|U\rangle\Rightarrow \frac{\partial}{\partial a^{\dagger}(\bar{\tilde{T}})}|U\rangle=e^{-\frac{\pi\omega}{a}} a^{\dagger}(T)|U\rangle.
\end{eqnarray}
A solution is immediately given by the so-called squeezed state
\begin{eqnarray}
|U\rangle={\cal N}\exp\bigg(e^{-\frac{\pi\omega}{a}}a^{\dagger}(T)a^{\dagger}(\bar{\tilde{T}})\bigg)|B\rangle
\end{eqnarray}
Thus the vacuum state of the black hole is the Unruh vacuum $|U\rangle$ and not the Boulware vacuum $|B\rangle$. The Unruh vacuum $|U\rangle$ should be though of as the in state in the same way that the original black hole state $|\psi\rangle$ should be thought of as the out state.

This squeezed state $|U\rangle$ is a $2$-mode entangled state. The modes correspond to $T$ (outside horizon) and ${\tilde{T}}$ (inside horizon). Since the black hole background is invariant under time translations the Hamiltonian must commute with $a^{\dagger}(T)a^{\dagger}(\bar{\tilde{T}})$. The Killing vector outside the event horizon corresponds to the usual time translation generator and thus $a^{\dagger}(T)$ must raise the Killing energy in the usual way, viz $[H,a^{\dagger}(T)]=\omega a^{\dagger}(T)$ where $\omega$ is positive. But inside the black hole the Killing vector reverses signature and it becomes like a momentum and thus its sign can be either positive or negative. We can check that 
$a^{\dagger}(\bar{\tilde{T}})$ must in fact lower the energy as $[H,a^{\dagger}(\bar{\tilde{T}})]=-\omega a^{\dagger}(\bar{\tilde{T}})$ if we want $[H, a^{\dagger}(T)a^{\dagger}(\bar{\tilde{T}})]=0$ which is required by invariance under time translations. This can also be seen from the fact that the interior mode enters through $\bar{\tilde{T}}$, which has a negative frequency, and not through $\tilde{T}$, which has a positive frequency as the exterior mode $T$. In conclusion, the total Killing energy of the entangled particle pair $T$ and $\tilde{T}$ is zero. 

The Unruh vacuum is an entangled pure state which can also be rewritten, by expanding the exponential, as follows
\begin{eqnarray}
|U\rangle&=&{\cal N}\sum_n \frac{1}{n!}e^{-\frac{n\pi\omega}{a}}(a^{\dagger}(T))^n(a^{\dagger}(\bar{\tilde{T}}))^n|B\rangle\nonumber\\
&\simeq &\sum_n e^{-\frac{n\pi\omega}{a}}|n_R\rangle|n_L\rangle.\label{pure}
\end{eqnarray}
The states $|n_R\rangle$ and $|n_L\rangle$ are the level $n$-excitations of the exterior modes $T$ and the interior modes $\bar{\tilde{T}}$ given respectively by
\begin{eqnarray}
|n_R\rangle\simeq \frac{1}{\sqrt{n!}}(a^{\dagger}(T))^n|B_R\rangle~,~|n_L\rangle\simeq \frac{1}{\sqrt{n!}}(a^{\dagger}(\bar{\tilde{T}}))^n|B_L\rangle.
\end{eqnarray}
Hence, this pure state if reduced to the outside of the event horizon we end up with a mixed state given by the density matrix 
\begin{eqnarray}
\rho_R&=&{\rm Tr}_L|U\rangle\langle U|\nonumber\\
&=&\sum_n e^{-\frac{2n\pi\omega}{a}}|n_R\rangle\langle n_R|.\label{mixed}
\end{eqnarray}
This is a thermal canonical ensemble. This the most precise statement, in my opinion, of the information loss problem: a correlated entangled pure state near the horizon gives rise to a thermal mixed state outside the horizon.

\section{The Information Problem in Black Hole Hawking Radiation}
The best presentation of the information problem remains that of Page \cite{Page:1993up}. This is a very difficult and mysterious topic and we will follow the pedagogical presentation  of \cite{Harlow:2014yka} and the elegant book \cite{Susskind:2005js}. We also refer to \cite{Polchinski:2016hrw,Mathur:2009hf}.

\subsection{Information Loss, Remnants and Unitarity}
The transition from a pure state to a mixed state observed in the Hawking radiation and black hole evaporation can be quantified as follows. We start with the Schr\"odinger equation 
\begin{eqnarray}
i\frac{\partial}{\partial t}|\psi>=H|\psi>.
\end{eqnarray}
The integrated form of this equation reads in terms of the unitary scattering matrix 
\begin{eqnarray}
|\psi^{\rm final}\rangle=S|\psi^{\rm initial}\rangle\Rightarrow \psi^{\rm final}_n=S_{nm}\psi^{\rm initial}_m.
\end{eqnarray}
The Schr\"odinger equation will evolve pure quantum states to pure quantum states. However, black hole radiation takes the pure state (\ref{pure}) to the mixed state (\ref{mixed}). Thus it takes an initial pure state of the form 
\begin{eqnarray}
\rho^{\rm initial}=|\psi^{\rm initial}\rangle\langle\psi^{\rm initial}|
\end{eqnarray}
to a final mixed state of the form 
\begin{eqnarray}
\rho^{\rm final}=\sum_ip_i|\psi^{\rm final}\rangle\langle\psi^{\rm final}|.
\end{eqnarray}
This can be expressed in terms of the so-called dollar matrix $\$ $ as follows

\begin{eqnarray}
\rho^{\rm final}_{mm^{\prime}}=\$_{mm^{\prime},nn^{\prime}}\rho^{\rm initial}_{nn^{\prime}}.
\end{eqnarray}
In the case of the Schr\"odinger equation we have 
 \begin{eqnarray}
\$_{mm^{\prime},nn^{\prime}}=S_{mn}S^*_{n^{\prime}m^{\prime}},
\end{eqnarray}
whereas in the case of the black hole radiation we have a general dollar matrix which takes pure states to mixed states. 

The opinions regrading whether or not black hole radiation corresponds to information loss divides into three possibilities:
\begin{itemize}
\item {\bf Information Loss:} This is the original stand of Hawking which is based on the conclusion that (\ref{mixed}) is correct and that the black hole will evaporate completely. In this case, the dollar matrix $\$$ is not given by Schr\"odinger equation and there is indeed information loss due to pure states (gravitational collapse and black hole formation) evolving into mixed states (Hawking radiation and black hole evaporation). Since the outgoing Hawking radiation is largely independent of the initial state, i.e. different initial states result in the same final state, black hole evaporation does not conserve information.

If information is really lost then quantum mechanics must be changed in some way. However, there are tight constraints on quantum gravity effects arising from the modification of the axioms of field theory \cite{Ellis:1983jz}, and furthermore any such modification will lead to violation of either locality or energy-momentum conservation \cite{Banks:1983by}. 

\item {\bf Unitarity:} The other possibility is therefore information conservation, i.e. there is a unitary map between the initial state of the collapse to the final state of the outgoing radiation. The black hole will also evaporate completely but (\ref{mixed}) is only correct in a coarse-grained sense. This means that the final state of the radiation becomes purified and information is carried out with the Hawking radiation in subtle quantum correlations between late and early particles. The final pure state of the radiation is presumably very complicated that any subsystem will look thermal and as a consequence equation (\ref{mixed}) is a good approximation \cite{Harlow:2014yka}.

These pure states are the microstates of the black hole and their counting is given by the exponential of the Bekenstein-Hawking formula. 

The black hole microstates may also be identified with the states of the field (or infalling matter) accumulating on the nice-slice, which is a spacelike surface interpolating between a fixed $t$ surface outside the black hole to a fixed $r$ surface inside the black hole, and which gets longer on the inside as the black hole gets older \cite{Polchinski:2016hrw}.

This solution, in which unitarity is maintained and information is conserved, if correct implies, however, a breakdown of the semi-classical description and the machinery of effective field theory.
\item {\bf Remnant:} In this case black hole evaporation stops when the decreasing black hole size becomes Planckian. The remaining Planck-sized object is what we call a remnant. This must be characterized by an extremely large entanglement entropy in order for the total state to remain pure. Thus, this is an object with a finite energy but effectively an infinite number of states and thus the connection between Bekenstein-Hawking entropy and number of states is lost.
\end{itemize}
This situation is the black hole information problem. 
\subsection{Information Conservation Principle}
In this section we will only follow the beautiful presentation of \cite{Susskind:2005js}.
\begin{itemize}

\item {\bf Von Neumann Entropy:} Information is conserved in classical mechanics (Liouville's theorem)\footnote{The volume of the initial phase space region representing the largely unknown state of the system is conserved in time under Hamilton's equations.} and in quantum mechanics (unitarity of the $S$-matrix)\footnote{In quantum mechanics the initial state of the system if unknown will be represented by a projector on some subspace. The dimension of this subspace, i.e. the rank of the projector, is conserved under Schr\"odinger equation. }. As we have already discussed, the Von Neumann entropy is the measure of information (or lack of it) which is defined by 
\begin{eqnarray}  
S=-\int dp dq \rho(p,q)\ln\rho(p,q).
\end{eqnarray}
If $\rho=1/V$, where $V$ the volume of some region in phase space, then $S=\ln V$, i.e. $V=\exp(S)$. In quantum mechanics we use instead the definition 
\begin{eqnarray}  
S=-{\rm Tr}\rho\ln\rho.
\end{eqnarray}
If $\rho=P/{\rm Tr}P$, where $P$ is a projector of rank $n$, then  $S=\ln n$. In other words, the number of states is given by the exponential of the entropy, viz 
\begin{eqnarray}  
n=\exp(S).
\end{eqnarray}
\item {\bf Pure States:} We will generally need to separate the system into two subsystems $A$ and $B$ with quantum correlations, i.e. entanglement, between them. The total system is assumed in a pure state $\psi(\alpha,\beta)$. Thus the Von Neumann entropy is zero identically, viz
\begin{eqnarray}  
S_{A+B}=0.
\end{eqnarray}
The subsystems considered separately are described by the corresponding density matrices $\rho_A(\alpha)$ and $\rho_B(\beta)$ in which the degrees of freedom of the other system are integrated out. These are generally not pure states.

The density matrix $\rho_A$ is such that: i) It is Hermitian $\rho_A^{\dagger}=\rho_A$, ii) It is positive semi-definite, viz $(\rho_A)_i\geq 0$, ii) It is normalized, viz ${\rm Tr}\rho_A=1$. 

Thus, if just one of the eigenvalues of $\rho_A$ is $1$ the rest will vanish identically. In this case the subsystem $A$ is in a pure state which means that the total system pure state factorizes as
\begin{eqnarray}  
\psi(\alpha,\beta)=\psi_A(\alpha)\psi_B(\beta).
\end{eqnarray}
The subsystem $B$ is then also in a pure state.
\item {\bf Entanglement Entropy:} A far more important identity for us here is the equality of the Von Neumann entropies of the two subsystems $A$ and $B$ if the total system is described by a pure state, viz
\begin{eqnarray}  
S_A=S_B=S_E.
\end{eqnarray}
$S_E$ is precisely the entanglement entropy.

{\bf Proof:} The density matrix $\rho_A$ is given explicitly by 
\begin{eqnarray}  
(\rho_A)_{\alpha\alpha^{\prime}}=\sum_{\beta}\psi^{\star}(\alpha,\beta)\psi(\alpha^{\prime},\beta).
\end{eqnarray}
Let $\phi$ be an eigenvector of $\rho_A$ with eigenvalue $\lambda$, viz
\begin{eqnarray}  
(\rho_A)_{\alpha\alpha^{\prime}}\phi(\alpha^{\prime})=\sum_{\beta}\psi^{\star}(\alpha,\beta)\psi(\alpha^{\prime},\beta)\phi(\alpha^{\prime})=\lambda\phi(\alpha).
\end{eqnarray}
We will assume that $\lambda\ne 0$. Similarly, we write explicitly the density matrix $\rho_B$ as
\begin{eqnarray}  
(\rho_B)_{\beta\beta^{\prime}}=\sum_{\alpha}\psi^{\star}(\alpha,\beta)\psi(\alpha,\beta^{\prime}).
\end{eqnarray}
We propose the eigenvector of $\rho_B$ to be of the form 
\begin{eqnarray}  
\chi(\beta^{\prime})=\sum_{\alpha^{\prime}}\psi^{\star}(\alpha^{\prime},\beta^{\prime})\psi^{\star}(\alpha^{\prime}).
\end{eqnarray}
Indeed, we compute 
\begin{eqnarray}  
\sum_{\beta^{\prime}}(\rho_B)_{\beta\beta^{\prime}}\chi(\beta^{\prime})&=&\sum_{\beta^{\prime}}\sum_{\alpha}\sum_{\alpha^{\prime}}\psi^{\star}(\alpha,\beta)\psi(\alpha,\beta^{\prime})\psi^{\star}(\alpha^{\prime},\beta^{\prime})\phi^{\star}(\alpha^{\prime})\nonumber\\
&=&\sum_{\alpha}\sum_{\alpha^{\prime}}(\rho_A)_{\alpha^{\prime}\alpha}\psi^{\star}(\alpha,\beta)\phi^{\star}(\alpha^{\prime})\nonumber\\
&=&\lambda\sum_{\alpha}\psi^{\star}(\alpha,\beta)\phi^{\star}(\alpha)\nonumber\\
&=&\lambda\chi({\beta}).
\end{eqnarray}
In the above we have also used the result that $(\rho_A)_{\alpha^{\prime}\alpha}\phi^{\star}(\alpha^{\prime})=\lambda\phi^{\star}(\alpha)$. Thus $\rho_A$ and $\rho_B$ have the same non-zero eigenvalues. Immediately we conclude that 
\begin{eqnarray}  
S_A=-\sum_i(\rho_A)_i\ln(\rho_A)_i=-\sum_i(\rho_B)_i\ln(\rho_B)_i=S_B.
\end{eqnarray}
Since $S_{A+B}=0$ and $S_A+S_B=2S_E$ this shows explicitly that the Von Neumann entanglement entropy is not additive. It is a fundamental microscopic fine grained entropy as opposed to the thermodynamic Boltzmann entropy.
\item {\bf Thermal Entropy:} The thermodynamic entropy is additive and it can be defined as follows. Let us assume a total system $\Sigma$ divided into many subsystems $\sigma_i$, i.e. a coarse graining. Again we will assume that the total system is in a pure state with vanishing entropy. The subsystems $\sigma_i$ are supposed to be thermal, i.e. with matrix densities $\rho_i$ given by the Blotzmann distribution   
\begin{eqnarray}  
\rho_i=\frac{e^{-\beta H_i}}{Z_i},
\end{eqnarray} 
where $H_i$ and $Z_i$ are the Hamiltonian and the partition function of the subsystem $\sigma_i$. This is the distribution which maximizes the entropy. The thermodynamic coarse grained entropy of the total system is then given by the sum of the entropies of the subsystems $\sigma_i$, viz
 \begin{eqnarray}  
S_{\rm therm}=\sum_iS_i.
\end{eqnarray} 
This coarse grained entropy $S_{\rm therm}$ as opposed to the fine grained entropy is not conserved. To see this we assume that initially the pure state of the total system factorizes completely, i.e. the subsystems are in a pure state. Then in this case $S_i=0$ and hence $S_{\rm therm}=0$. After interaction the pure state of the total system will fail to factorize, i.e. $S_i\neq 0$ and hence $S_{\rm therm}\neq 0$.  

Another important property is the fact that the thermodynamic entropy of a subsystem $\Sigma_1$ is always larger than its entanglement entropy, viz
 \begin{eqnarray}  
S_{\rm therm}(\Sigma_1)=\sum_iS_i\geq S(\Sigma_1)=-{\rm Tr}\rho\ln\rho.
\end{eqnarray} 
This is almost obvious since  from one hand $S_i\geq 0$ and thus $S_{\rm therm}\geq 0$, while from the other hand as $\Sigma_1\longrightarrow \Sigma$ the entanglement entropy approaches zero.

\item {\bf Information:} The amount of information in a subsystem $\Sigma_1$ is defined as the difference between the coarse grained entropy (thermodynamic) and the fine grained entropy (Von Neumann), viz 
 \begin{eqnarray}  
I=S_{\rm therm}(\Sigma_1)-S(\Sigma_1)=\sum_iS_i+{\rm Tr}\rho\ln\rho.
\end{eqnarray} 
As an example we take $\Sigma_1$ to be the total system $\Sigma$. In this case $S(\Sigma)=0$ and thus the information is given by the thermodynamic entropy. But for a very small subsystem $\Sigma_1=\sigma_i$ we get $I=0$ since obviously $S_{\rm therm}=S$ for such a system. In fact this is true for all subsystems which are smaller than one half the total system. A nice calculation which attempts to convince us of this result is found in \cite{Susskind:2005js}.

Let us assume that the total system is composed of two subsystems $\Sigma_1$ and $\Sigma-\Sigma_1$. Immediately, we conclude that the Von Neumann entropies are equal, viz
\begin{eqnarray}  
S(\Sigma-\Sigma_1)=S(\Sigma_1).
\end{eqnarray} 
The amounts of information contained in $\Sigma_1$ and $\Sigma-\Sigma_1$ are given by 
\begin{eqnarray}  
I(\Sigma_1)=S_{\rm therm}(\Sigma_1)-S(\Sigma_1)~,~I(\Sigma-\Sigma_1)=S_{\rm therm}(\Sigma-\Sigma_1)-S(\Sigma-\Sigma_1).
\end{eqnarray} 
If $\Sigma_1<<\Sigma/2$ then 
\begin{eqnarray}  
I(\Sigma_1)=S_{\rm therm}(\Sigma_1)-S(\Sigma_1)=0.
\end{eqnarray}
If $\Sigma_1>>\Sigma/2$ then $\Sigma-\Sigma_1<<\Sigma/2$ and as a consequence $I(\Sigma-\Sigma_1)=S_{\rm therm}(\Sigma-\Sigma_1)-S(\Sigma-\Sigma_1)=0$, i.e. there is no information in the smaller subsystem $\Sigma-\Sigma_1$. Also we will have in this case (with $f$ being the fraction of the total degrees of freedom contained in $\Sigma_1$) 
\begin{eqnarray}  
I(\Sigma_1)&=&S_{\rm therm}(\Sigma_1)-S(\Sigma_1)\nonumber\\
&=&S_{\rm therm}(\Sigma_1)-S(\Sigma-\Sigma_1)\nonumber\\
&=&S_{\rm therm}(\Sigma_1)-S_{\rm therm}(\Sigma-\Sigma_1)\nonumber\\
&=&fS_{\rm therm}(\Sigma)-(1-f)S_{\rm therm}(\Sigma)\nonumber\\
&=&(2f-1)S_{\rm therm}(\Sigma).
\end{eqnarray} 
This result can be clearly continued from $f\simeq 1$ to $f=1/2$. It vanishes identically for $\Sigma_1=\Sigma/2$, i.e. $f=1/2$. Since the amount of information vanishes also for $\Sigma_1<<\Sigma$ we conclude, again by continuity, that indeed $I=0$ for all $\Sigma_1\le \Sigma/2$.
\item {\bf Bomb in a Box:} We conclude this section by the illuminating example of \cite{Susskind:2005js}. We consider a system $\Sigma_1$ consisting of a bomb placed in a box $B$ with reflecting walls and a hole from which electromagnetic radiation can escape. The system $\Sigma-\Sigma_1$ is obviously the environment which will be denoted by $A$. The bomb will explode and we will watch the system+environment until all thermal radiation inside the box leaks out to the environment. In order to simplify tracking the evolution we will divide it into four stages:
\begin{itemize}
\item Before the explosion of the bomb, the systems $A$ and $B$ are in their ground (pure) states. The Von Neumann fine grained entropies as well as the Boltzmann thermal coarse grained entropies all vanish identically and thus the entanglement entropy and the information in the outside radiation also vanish identically, viz
 \begin{eqnarray}  
S(A)=S(B)=S_E=0~,~S_{\rm therm}(A)=S_{\rm therm}(B)=0\Rightarrow I(A)=0.
\end{eqnarray} 
\item The bomb explodes and thermal radiation fills the box (no photon has leaked out yet). The thermal entropy inside increases. All others are still zero identically, viz
 \begin{eqnarray}  
S(A)=S(B)=S_E=0~,~S_{\rm therm}(A)= 0~,~S_{\rm therm}(B)\ne 0\uparrow\Rightarrow I(A)=0.
\end{eqnarray} 
The initial information is
\begin{eqnarray}  
I(B)=S_{\rm therm}(B).
\end{eqnarray} 
\item The photons start to leak out. The Von Neumann entropies increase and thus the entanglement entropy increases, i.e. entanglement between $A$ and $B$ increases. The thermal entropy of the Box clearly decreases while that of the environment increases. But information in the outside radiation remains negligible.  
\begin{eqnarray}  
S(A)=S(B)=S_E\ne 0~,~S_{\rm therm}(A)\ne 0\uparrow~,~S_{\rm therm}(B)\ne 0\downarrow\Rightarrow I(A)= 0.
\end{eqnarray}   
At some point the thermal entropies become equal. This is called the information retention time. This is the time where the entanglement between $A$ and $B$ becomes decreasing and the information starts increasing, i.e. it is the time at which information starts coming out with the radiation. Before the information retention time only energy has come out with the radiation with no or little information. At the information retention time around one half of the radiation inside the box has come out which corresponds to one bit $(\ln 2$) of information encoded in the initial state.

\item When all photons are out the inside thermal entropy vanishes and since there is no entanglement anymore the Von Neumann entropies vanish, viz
\begin{eqnarray}  
S(A)=S(B)=S_E=0~,~S_{\rm therm}(A)\ne 0~,~S_{\rm therm}(B)=0\Rightarrow I(A)= S_{\rm therm}(A).\nonumber\\
\end{eqnarray}  
From the second law of thermodynamics the final value of the outside thermal entropy must be larger than the initial value of the interior thermal entropy, viz $S_{\rm therm}(A)>S_{\rm therm}(B)$, i.e. the information in the outgoing radiation is more than the information in the initial state of the box. 

\item Throughout the process the entanglement entropy is always smaller than the thermal entropy of $A$ or $B$. This can be seen as follows. At the beginning most information is in the thermal entropy of $B$. The information in $A$ is zero which means that the entanglement entropy is equal to the thermal entropy of $A$ which is less than the thermal entropy of $B$. At the end we have the reverse situation.  Most information is in the thermal entropy of $A$. The information in $B$ is zero which means that the entanglement entropy is equal to the thermal entropy of $B$ which is less than the thermal entropy of $A$. In summary, we have
\begin{eqnarray}  
S_E\leq S_{\rm therm}(A)~{\rm or}~S_{\rm therm}(B).
\end{eqnarray}  
\end{itemize}
Thus information is conserved which means in particular that the final state of the radiation outside the box is pure although it might look thermal at smaller scales. This very clear physical picture is summarized in the figure (\ref{information}).
\end{itemize}

\begin{figure}[H]
\begin{center}
\includegraphics[width=8.0cm,angle=0]{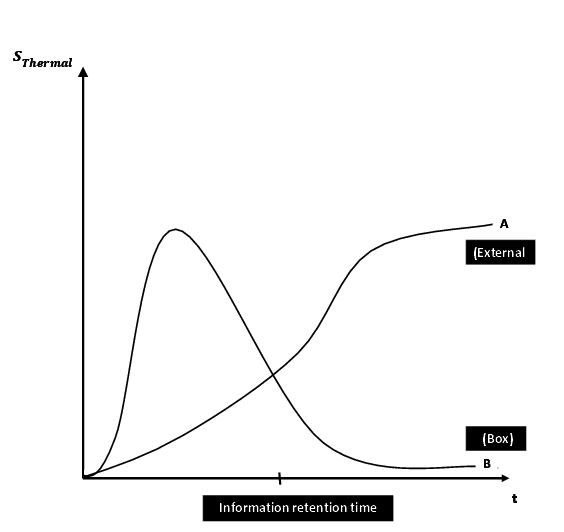}
\includegraphics[width=8.0cm,angle=0]{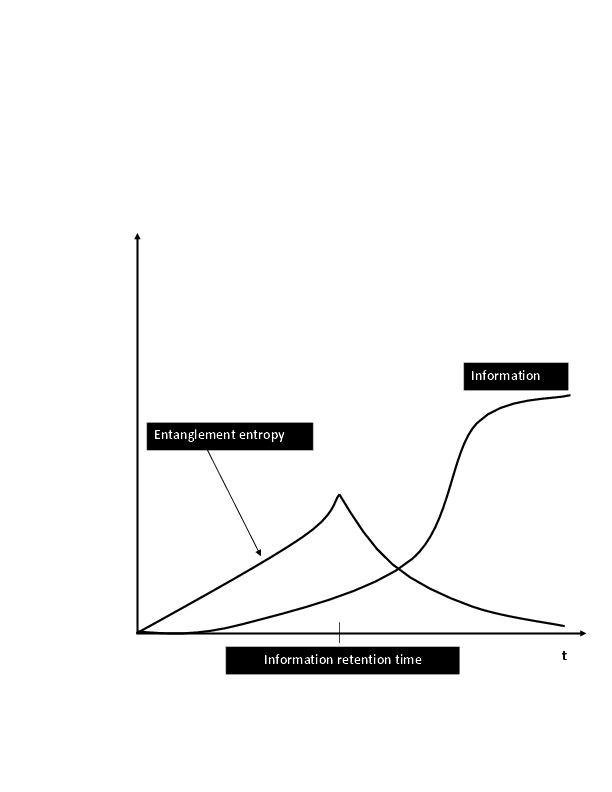}
\end{center}
\caption{The information retention time, the entanglement entropy and the information in the "bomb in a box" problem.}\label{information}
\end{figure}

\subsection{Page Curve and Page Theorem}

We have a quantum system consisting of a black hole and its corresponding Hawking radiation. We split the outgoing Hawking radiation into early and late with corresponding Hilbert spaces ${\cal H}_R$ and ${\cal H}_{\rm BH}$, viz
\begin{eqnarray}  
{\cal H}_{\rm out}={\cal H}_{\rm R}\otimes {\cal H}_{\rm BH}.
\end{eqnarray}
The notation ${\cal H}_{\rm BH}$ indicates explicitly that the late Hawking radiation is nothing else but the remaining black hole. The plot of the entanglement entropy $S_E$ of the early radiation as a function of time is called the Page curve \cite{Page:1993wv,Page:2013dx}. Obviously, $S_E=S(R)=S(BH)$.

The initial state of the black hole is pure.  Initially the thermal entropy of the black hole is non-zero, i.e.  $S_{{\rm therm}}(BH)=0$, the entanglement entropy is zero, viz $S_E=0$, and the information in the Hawking radiation $I(R)$ is zero, i.e. $I(R)=0$. Hawking radiation starts coming out. The entanglement entropy $S_R$ between the Hawking radiation and the black hole starts increasing, the thermal entropy of the black hole $S_{\rm therm}(BH)$ decreases while the thermal entropy of the radiation $S_{\rm therm}(R)$ increases. 

At the retention time, also called Page time, the two thermal entropies become identical, viz
\begin{eqnarray}  
S_{\rm therm}(BH)=S_{\rm therm}(R).
\end{eqnarray}
At this time the entanglement reaches its maximum and starts decreasing, and the information $I(R)$ at the Page time starts increasing, i.e. it starts coming out in the Hawking radiation. The final state is a pure state of the radiation with vanishing entanglement entropy and information at its maximum value. 

The expected picture is shown in the second figure of (\ref{information}). However, this is only a sketch of the actual physics by assuming unitarity while the actual calculation of the Page curve remains a major challenge.

Indeed, as reported concisely by Harlow in his lectures \cite{Harlow:2014yka} he says that "Andy Strominger has argued that being able to compute the Page curve in some particular theory is what it means to have solved the black hole information problem; even in AdS/CFT or the BFSS model we are far (Harlow stating) from being able to really do this".

The above picture can however be fleshed out a little more by using the elegant Page theorem \cite{Page:1993df}. This says that for a given bipartite system ${\cal H}_{AB}={\cal H}_A\otimes{\cal H}_B$ with $|A|={\rm dim}A<|B|={\rm dim}B$ a randomly chosen pure state $\rho_{AB}$ in ${\cal H}_{AB}$ is likely to be very close to a maximally entangled state if $|A|<<|B|$. In other words, if  $|A|<<|B|$, the pure state $\rho_{AB}$ will correspond to a totally mixed state $\rho_A={\rm Tr}_B\rho_{AB}$, i.e. $\rho_A\propto {\bf 1}_A$. 

More precisely,  we write this theorem as the inequality 
\begin{eqnarray}  
\int dU ||\rho_A(U)-\frac{{\bf 1}_A}{|A|}||_1\leq \sqrt{\frac{|A|^2-1}{|A||B|+1}}.
\end{eqnarray}
The norm $||..||_1$ is the $L_1$ operator trace norm defined by $||M||_1={\rm Tr}\sqrt{M^{\dagger}M}$. The integration over the Haar measure represents a randomly chosen pure state $|\psi(U)\rangle=U|\psi\rangle$, i.e. $\rho_{AB}(U)=|\psi(U)\rangle\langle\psi(U)|$ and $\rho_A(U)={\rm Tr}_B|\psi(U)\rangle\langle\psi(U)|$. It is clear from the above equation that if $|A|<<|B|$ then $\rho_A$ is very close to a totally mixed state and as a consequence $|\psi\rangle$ or equivalently $\rho_{AB}$ is a maximally entangled pure state.

Let us compute the behavior of the entanglement entropy. We have (with $\Delta\rho_A=\rho_A-{\bf 1}_A/|A|$ and ${\rm Tr}\Delta\rho_A=0$)
\begin{eqnarray}  
\int dU S_A&=&-\int dU {\rm Tr}\rho_A\ln\rho_A\nonumber\\
&=&\ln|A|-\frac{1}{2}|A|\int dU{\rm Tr}\Delta\rho_A^2+O(\Delta\rho^3).
\end{eqnarray}
The remaining integral over $U$ can be done exactly using unitary matrix technology (see equation $(5.13)$ of \cite{Harlow:2014yka}) to find for $|A|<<|B|$ the result 
\begin{eqnarray}  
\int dU S_A
&=&\ln|A|-\frac{1}{2}\frac{|A|}{|B|}+....
\end{eqnarray}
We now apply this theorem to the entanglement entropy of the black hole. 

We know that Hawking radiation consists mostly of $s$-wave quanta, i.e. modes with $l=0$. These can be described by a $1+1$ dimensional free scalar field at a Hawking temperature $T_H=1/4\pi r_s$. These modes can escape the black hole because the Schwarzschild potential is not fully confining as in the Rindler case. Indeed, the barrier height for $s$-wave particles is of the same order of magnitude as the Hawking temperature. Thus, each particle which escapes is carrying energy given by Hawking temperature $\nu\sim T_H=1/(8\pi GM)$. 

Further, we will assume that one single quanta will escape (since $l=0$) per one unit of Rindler time $\omega=t/2r_s$. Thus, $1/2r_s$ quanta per unit Schwarzschild time will escape the barrier. The total energy carried out of the black hole per unit Schwarzschild time is then given by $1/(8\pi GM)\times 1/2r_s\sim 1/G^2M^2$. We write this as
\begin{eqnarray}  
\frac{dE_R}{dt}=\frac{C}{G^2M^2}.\label{lj}
\end{eqnarray}
By energy conservation the energy per unit Schwarzschild time lost by the black hole is immediately given by  
\begin{eqnarray}  
\frac{dM}{dt}=-\frac{C}{G^2M^2}\Rightarrow C dt=-G^2M^2dM.
\end{eqnarray}
In the above two equations $C$ is some constant of proportionality.

In order to apply Page's theorem we will first need to assume that the pure state of the Hawking (early) radiation $R$ and the black hole (late radiation) $BH$ is random. Obviously, at early times $|R|<<|BH|$. The entanglement entropy is then given immediately by the theorem to be given by 
\begin{eqnarray}  
S_E= S_R\sim \ln |R|.
\end{eqnarray}
On the other hand, the energy carried by the radiation during a small time interval $t$ is obtained by integrating equation (\ref{lj}) assuming that the mass $M$ remains constant. This gives
 \begin{eqnarray}  
E_R=\frac{C}{G^2M^2}t.
\end{eqnarray}
However, from equations (\ref{below1}) and (\ref{below2}) below, the entropy and energy of the radiation are related by 
 \begin{eqnarray}  
\frac{E_R}{S_R}\sim \frac{1}{r_s}.
\end{eqnarray}
By taking the ratio of the above two results we obtain 
\begin{eqnarray}  
S_R\sim tT.
\end{eqnarray}
This should be valid only for times such that $S_R<<S_{\rm BH}\sim M^2$, i.e. $t<<M^3$. During these times it is also expected that $S_{\rm BH}\sim \ln |BH|$. At early times we have then the linear behavior of the entanglement entropy as a function of time, viz
\begin{eqnarray}  
S_E\sim tT~,~t<<M^3.
\end{eqnarray}
After the Page time $t_{\rm Page}$ defined by 
\begin{eqnarray}  
\ln |R|\sim \ln |BH|,
\end{eqnarray}
we should apply Page's theorem in the opposite direction since we can assume now that $|BH|<<|R|$. Thus in this case 
\begin{eqnarray}  
S_E=S_{BH}\sim \ln |BH|.
\end{eqnarray}
However, by integrating equation (\ref{lj}) between $t$ and $t_{\rm evap}$ we obtain 
\begin{eqnarray}  
C\int_t^{t_{\rm evap}}dt^{\prime}=-G^2\int_M^0M^{\prime 2}dM^{\prime}\Rightarrow (t_{\rm evap}-t)^{2/3}\sim M^2.
\end{eqnarray}
However, for the black hole the entropy is proportional to the area which is proportional to its mass squared, thus we obtain immediately 
\begin{eqnarray}  
S_{BH}\sim (t_{\rm evap}-t)^{2/3}.
\end{eqnarray}
At late times we have then the  behavior of the entanglement entropy as a function of time given by 
\begin{eqnarray}  
S_E\sim (t_{\rm evap}-t)^{2/3}~,~t_{\rm Page}\leq t\leq t_{\rm evap}.
\end{eqnarray}

\section{Black Hole Thermodynamics}

Again we will follow \cite{Harlow:2014yka} and the book \cite{Susskind:2005js}. 

\subsection{Penrose Diagrams}
\begin{figure}[H]
\begin{center}
\includegraphics[width=10.0cm,angle=0]{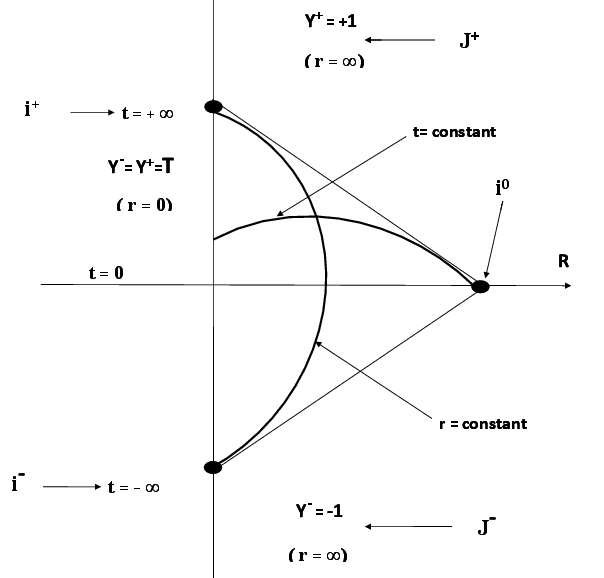}
\end{center}
\caption{Penrose diagram of Minkowski spacetime. The infinities $i^{\mp}$ are at $t=\mp\infty$, the infinity $i^0$ is at $r=\infty$ while the infinities ${\cal J}^{\mp}$ are at $r=\infty$.}\label{PD1}
\end{figure}

The idea of Penrose diagrams relies on the theorem that any two conformally equivalent metrics will have the same null geodesics and thus the same causal structure. Thus, Penrose diagrams represent essentially the causal structure of spacetimes and they involve the so-called conformal compactification. Let us take the example of flat Minkowski spacetime given by the metric 
\begin{eqnarray}
ds^2=-dt^2+dr^2+r^2d\Omega^2.
\end{eqnarray}
The light cone is defined by $dt=\pm dr$. The form of the light cone is therefore preserved if we transform $t$ and $r$ to $T$ and $R$ such that 
 \begin{eqnarray}
Y^+=T+R=f(t+r)~,~Y^-=T-R=f(t-r).
\end{eqnarray}
We can map the Minkowski plane $0\leq r\le \infty$ and $-\infty\leq t\leq +\infty$ to a finite region of the plane with boundaries at finite distance by choosing $F$ to be the function $\tanh$, viz
\begin{eqnarray}
Y^+=\tanh (t+r)~,~Y^-=\tanh(t-r).
\end{eqnarray}
We have the limiting behaviors 
\begin{eqnarray}
Y^+=+1~,~Y^-=-1~,~r\longrightarrow\infty~,~\forall t.
\end{eqnarray}
\begin{eqnarray}
Y^+=Y^-=\frac{e^t-e^{-t}}{e^t+e^{-t}}~,~r\longrightarrow 0.
\end{eqnarray}
\begin{eqnarray}
Y^+=Y^-=1~,~t\longrightarrow +\infty.
\end{eqnarray}
\begin{eqnarray}
Y^+=Y^-=-1~,~t\longrightarrow -\infty.
\end{eqnarray}
The new coordinates have now the range $|T\pm R|< 1$ and $R\geq 0$. The boundary $|T\pm R|=1$ can be included by dropping the diverging prefactor in the metric when expressed in terms of $T$ and $R$ (by using the above theorem). In other words, spacetime is compactified. The range becomes  $|T\pm R|\leq 1$ and $R\geq 0$. This is a triangle in the $TR$ plane defined by 
\begin{eqnarray}
Y^+=Y^-~,~Y^+=+1~,~Y^-=-1.
\end{eqnarray}
We can discern the following infinities (see figure (\ref{PD1})):
\begin{itemize}
\item The usual future and past time like infinities at $t=\pm\infty$ denoted by $i^-$ and $i^+$ respectively. All time like trajectories begin at $
i^-$ and ends at $i^+$. 
\item The usual space like infinity at $r=\infty$ denoted by $i^0$. All space like trajectories end there.  In general relativity, conserved charges such as the energy are written as boundary integrals at this spatial infinity $i^0$.
\item Also we observe two light like infinities at $Y^-=-1$ and $Y^+=+1$ denoted by $J^-$ and $J^+$ respectively. All light like trajectories begin at $J^-$ (incoming null rays) and end at $J^+$ (outgoing null rays). Thus the $S$-matrix will map incoming states defined on $i^-\cup J^-$ to outgoing states defined on $i^+\cup J^+$.
\end{itemize}

\begin{figure}[H]
\begin{center}
\includegraphics[width=10.0cm,angle=0]{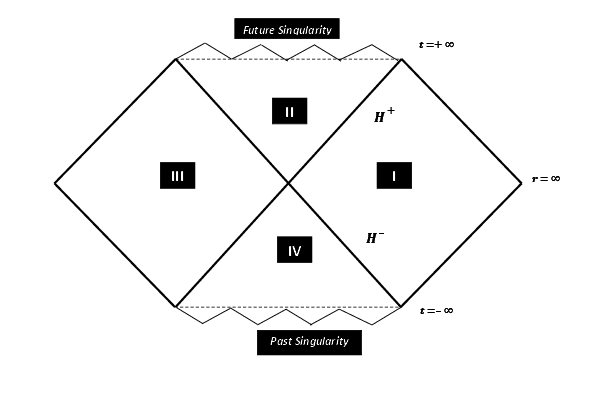}
\includegraphics[width=6.0cm,angle=0]{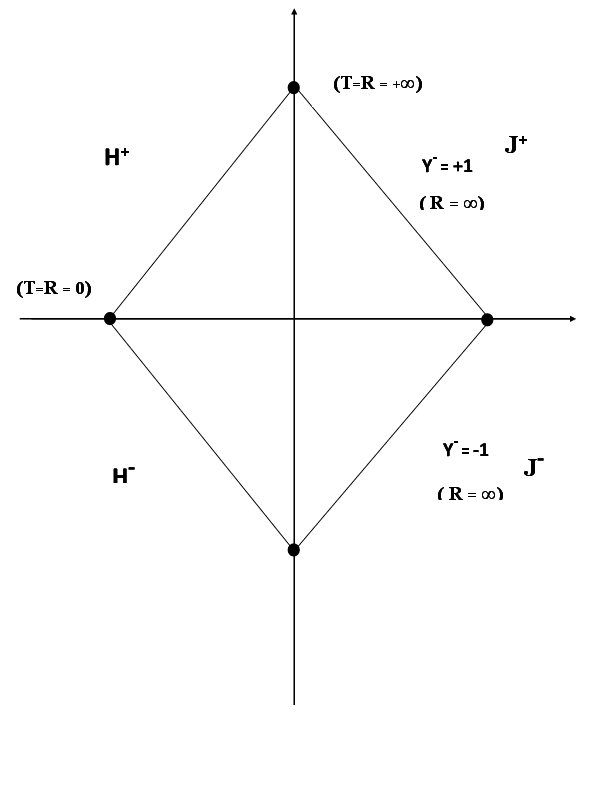}
\end{center}
\caption{Penrose diagram of Schwarzschild metric. The infinities $i^{\mp}$ are at $t=\mp r=\mp \infty$, the infinity $i^0$ is at $r=\infty$ while the infinities ${\cal J}^{\mp}$ are at $r=\infty$, $Y^{\mp}=\mp 1$. The center of the diagram is at $T=R=Y^{\pm}=0$ while the horizons $H^{\pm}$ are at $T=\mp R$.}\label{PD2}
\end{figure}

Let us consider the more interesting example of Schwarzschild geometry given by the Kruskal-Szekeres metric 
\begin{eqnarray}
ds^2=\frac{32G^3M^3}{r}\exp(-\frac{r}{2GM})(-dT^2+dR²)+r^2d\Omega^2.
\end{eqnarray}
The only difference between the Schwarzschild coordinates $(T,R,\Omega)$ and the Minkowski coordinates $(t,r,\Omega)$ is their range. For the Minkowski coordinates we have $0\leq r\leq \infty$ and $-\infty\leq t\leq +\infty$. For Schwarzschild we have instead (in region I)
\begin{eqnarray}
0\leq R\leq \infty~,~ -R\leq T\leq +R.
\end{eqnarray}
The horizon is at $T=\pm R$. Thus, as before, we consider the deformation 
\begin{eqnarray}
Y^+=T^{\prime}+R^{\prime}=\tanh (T+R)~,~Y^-=T^{\prime}-R^{\prime}=\tanh(T-R).
\end{eqnarray}
We still obtain in the limit $R\longrightarrow +\infty$ the two light like infinities $J^+$ ($Y^+=1$) and $J^-$ ($Y^-=-1$) and the space like infinity $i^0$. We do not now have the boundary $Y^+=Y^-$ since $T$ does not take the unrestricted values between $-\infty$ and $+\infty$. Since $T$ takes the values between $-R$ and $+R$ we have in the limit $T\longrightarrow R$ the surface 
\begin{eqnarray}
Y^+=\frac{e^{2R}-e^{-2R}}{e^{2R}+e^{-2R}}~,~Y^-=0.
\end{eqnarray}
This is the future horizon $H^+$ which is parallel to $J^-$ and it varies from $Y^+=1$ at $R\longrightarrow\infty$ to $Y^+=0$ at $R\longrightarrow 0$. The time like infinity $i^+$ is at $T=R=+\infty$. Similarly, in the limit $T\longrightarrow -R$ we get the past horizon $H^-$ which is parallel to $J^+$. The time like infinity $i^-$  is at $T=R=-\infty$.   The Penrose diagram of the full Schwarzschild geometry is shown on figure (\ref{PD2}). 

Let us now consider a real black hole as it forms from the gravitational collapse of a thin spherical shell of massless matter. We start with the Penrose diagram of Minkowski spacetime. The infalling shell is represented by an incoming light like line which divides the diagram into region A (interior) and region B (exterior). The incoming light like line starts at the null infinity $J^-$ ($r=\infty$) and ends at $Y^+=Y^-$ ($r=0$). Region $A$ is physical but region $B$ needs to be modified in order to take into account the effect of the gravitational field created by the shell on the spacetime geometry. 

\begin{figure}[H]
\begin{center}
\includegraphics[width=5.0cm,angle=0]{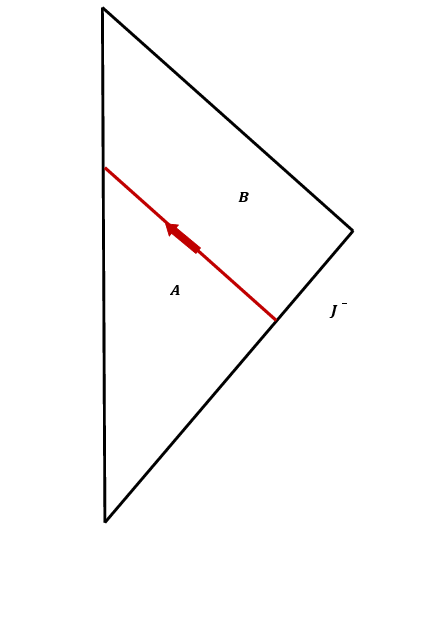}
\includegraphics[width=9.0cm,angle=0]{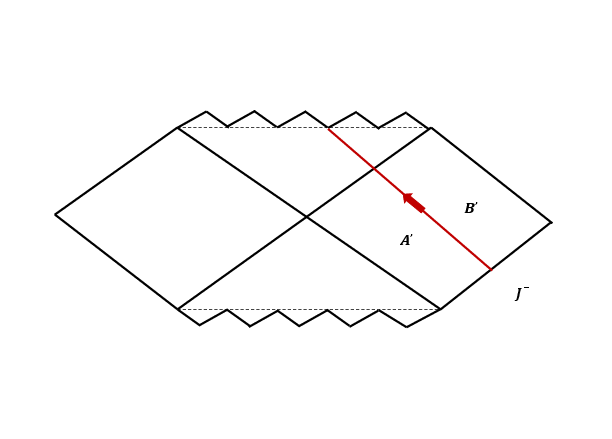}
\includegraphics[width=7.0cm,angle=0]{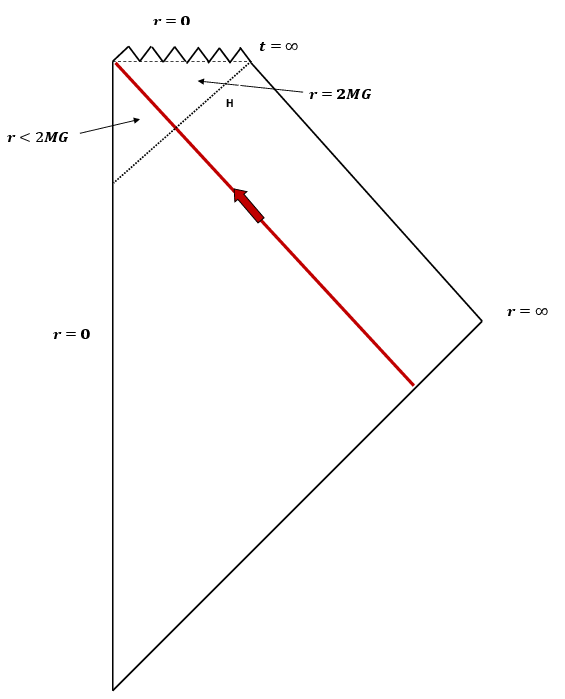}
\end{center}
\caption{Penrose diagram of the formation of a black hole from gravitational collapse. In the last graph the region $A$ is below the red line (infalling shell) and the region $B^{\prime}$ is above the line. Outside the shell (above the red line) the horizon is at $2GM$ while inside the shell (below the red line) the horizon will start at $r=0$ and reaches the value $r=2GM$ at the end of the collapse. Thus, the horizon is a global concept and not a local one, since it forms before the shell reaches the center.}\label{PD3}
\end{figure}

By Birkoff's theorem the geometry outside the spherical shell is nothing else but Schwarzschild geometry. We consider therefore Penrose diagram of Schwarzschild geometry divided by the incoming light like into regions $A^{\prime}$ (interior) and $B^{\prime}$ (exterior). Now, it is the region $A^{\prime}$ which is unphysical and needs to be replaced in a continuous way by the region $A$ above. This is explained nicely in \cite{Susskind:2005js} and the end result is the Penrose diagram in figure (\ref{PD3}). 

The horizon $H$ in region $B^{'}$ coincides with the horizon $H^+$ of Schwarzschild geometry and thus it is located at $r=2GM$. The horizon in region $A$ is however at a value  $r< 2GM$ and it will only reach the value $r=2GM$ at the end of the collapse.    

\subsection{Bekenstein-Hawking Entropy Formula}
To a distant observer the Schwarzschild black hole appears as a thermal body with energy given by its mass $M$ and a temperature $T$ given by Hawking temperature 
\begin{eqnarray}
T=\frac{1}{8\pi GM}.
\end{eqnarray}
The thermodynamical entropy $S$ is related to the energy and the temperature by the formula $dU=T dS$. Thus we obtain for the black hole the entropy 
\begin{eqnarray}
dS=\frac{dM}{T}=8\pi GM dM\Rightarrow S=4\pi GM^2.
\end{eqnarray}
However, the radius of the event horizon of the Schwarzschild black hole is $r_s=2MG$, and thus the area of the event horizon (which is a sphere) is 
\begin{eqnarray}
A=4\pi (2MG)^2.
\end{eqnarray}
By dividing the above two equations we get 
\begin{eqnarray}
S=\frac{A}{4G}.
\end{eqnarray}
The entropy of the black hole is proportional to its area. This is the famous Bekenstein-Hawking entropy formula.

The Bekenstein-Hawking entropy is a thermodynamical macroscopic coarse-grained entropy which counts the microstates of the black hole. It should satisfy the so-called generalized second law of thermodynamics: "When common entropy goes down a black hole, the common entropy in the black-hole exterior plus the black-hole entropy never decreases" \cite{Bekenstein:1973ur,Bekenstein:1974ax}. But, since the  Bekenstein-Hawking entropy is $S_{}=A/4G$ where $A$ is the area of the event horizon, we can see that the area of the event horizon can not decrease (if there was no radiation) \cite{Hawking:1971tu}.

Now, the black hole according to general relativity is only characterized by its temperature and its mass. Thus immediately we conclude that by creating a black hole we loose most of the information about its past, since clearly the initial state can not be recovered by running the dynamic backward in time starting from the black hole state which is, as we said, is characterized only by the mass and the temperature. Thus, the black hole must also be characterized by its microstates which are counted exactly by the exponential of the Bekenstein-Hawking entropy formula.

However, black hole also evaporates, and thus the above is not sufficient to maintain the principle of information conservation (the first law of nature in the words of \cite{Susskind:2005js}).

\subsection{Brick Wall and Stretched Horizon}
In this final section we will follow the presentation of \cite{Harlow:2014yka,Susskind:2005js}.

We have found that the vacuum of the scalar field in the Schwarzschild geometry is not given by the vacuum state $|B\rangle$, which is annihilated by $a(T)$ and $a(\bar{\tilde{T}})$, but it is given by a thermal density matrix of the form (with $\beta=2\pi/a$)

\begin{eqnarray}
\rho_R&=&\bigotimes_{\omega,l,m}\rho_R(\omega,l,m)\nonumber\\
&=&\bigotimes_{\omega,l,m}\bigg[(1-e^{-\beta\omega})\sum_n e^{-n\beta\omega}|n_R\rangle\langle n_R|_{\omega,l,m}\bigg].
\end{eqnarray}
This is diagonal where $n$ is the occupation number and $1-\exp(-\beta\omega)$ is a normalization constant inserted so that ${\rm Tr}\rho_R(\omega,l,m)=1$ for each mode. The Hamiltonian is given immediately by
\begin{eqnarray}
H_R&=&\bigotimes_{\omega,l,m}H_R(\omega,l,m)\nonumber\\
&=&\bigotimes_{\omega,l,m}\bigg[(1-e^{-\beta\omega})\sum_n n\omega e^{-n\beta\omega}|n_R\rangle\langle n_R|_{\omega,l,m}\bigg].
\end{eqnarray}
Thus the energy is given by 
\begin{eqnarray}
E=<H_R>&=&{\rm Tr}\rho_RH_R\nonumber\\
&=&\sum_{\omega,l,m}\bigg[(1-e^{-\beta\omega})\sum_n n\omega e^{-n\beta\omega}\bigg]\nonumber\\
&=&\sum_{\omega,l,m}\frac{\omega}{e^{\beta\omega}-1}.
\end{eqnarray}
The above state corresponds to the canonical ensemble, i.e. $\rho_R$ can also be rewritten as (see how this was done in Rindler case)
\begin{eqnarray}
\rho_R=\bigotimes_{\omega,l,m}\frac{e^{-\beta H(\omega,l,m)}}{Z(\omega,l,m)}.
\end{eqnarray}
\begin{eqnarray}
{\rm Tr}\rho_R(\omega,l,m)=1~,~Z(\omega,l,m)={\rm Tr}e^{-\beta H(\omega,l,m)}=\sum_{n=0}e^{-\beta\omega n}=\frac{1}{1-e^{-\beta\omega}.}
\end{eqnarray}
The entanglement entropy (this is an entanglement entropy because it was obtained by integrating out the interior modes) is then given immediately by 
\begin{eqnarray}
S=-{\rm Tr}\rho_R\ln\rho_R=-\sum_{\omega,l,m}\rho_{\omega,l,m}\ln\rho_{\omega,l,m}.
\end{eqnarray}
We use the identities 
\begin{eqnarray}
\frac{\partial}{\partial N}(\rho_i)^N{\bigg |}_{N=1}=\rho_i\ln\rho_i.
\end{eqnarray}
\begin{eqnarray}
\frac{\partial}{\partial N}e^{-N\beta H_i}{\bigg |}_{N=1}=-\beta H_i e^{-\beta H_i}.
\end{eqnarray}
\begin{eqnarray}
\frac{\partial}{\partial N}Z_i^{-N}{\bigg |}_{N=1}=-\frac{\ln Z_i}{Z_i}.
\end{eqnarray}
The entropy then takes the form 
\begin{eqnarray}
S_i=\beta E_i+\ln Z_i=\beta E_i-\beta F_i.
\end{eqnarray}
The total entropy, total energy and total free energy are then given simply by
\begin{eqnarray}
S=\sum_{\omega,l,m}S_{\omega,l,m}~,~E=\sum_{\omega,l,m}E_{\omega,l,m}~,~F=\sum_{\omega,l,m}F_{\omega,l,m}.
\end{eqnarray}
We already have compute $E$. The entropy is given on the other hand by
\begin{eqnarray}
S=\sum_{\omega,l,m}\frac{\omega}{e^{\beta\omega}-1}-\sum_{\omega,l,m}\ln(1-e^{-\beta\omega}).
\end{eqnarray}
This entropy is clearly an entanglement entropy since it arised  from a reduced density matrix. 

The expressions for the energy and the entropy are IR divergent due to the infinite volume of space as well as UV divergent due to the presence of the horizon. The $r\longrightarrow\infty$ IR divergent is regulated in the usual way by putting the system in a box while the $r\longrightarrow r_s$ near horizon UV divergence should be regulated by some new unknown physics at the Planck scale near the horizon. Following t'Hooft \cite{'tHooft:1984re} we will regulate this UV behavior by imposing Dirichelet boundary condition on the scalar field near the horizon, viz
\begin{eqnarray}
\phi=0~{\rm at}~r=r_{\rm min}.
\end{eqnarray}
In terms of the proper distance $\rho$ this minimum distance from the horizon reads
\begin{eqnarray}
\rho=\sqrt{r_{\rm min}(r_{\rm min}-r_s)}+r_s\sinh\sqrt{\frac{r_{\rm min}}{r_s}-1}\simeq 2\sqrt{r_s(r_{\rm min}-r_s)}\Rightarrow r_{\rm min}=r_s+\frac{\rho^2}{4r_s}.
\end{eqnarray}
This is the so-called brick wall introduced by t'Hooft. In terms of the tortoise coordinate $r_{*}$ it is situated at
\begin{eqnarray}
r_{*\rm min}=r_{\rm min}-r_s+r_s\ln(\frac{r_{\rm min}}{r_s}-1)\simeq 2r_s\ln \frac{\rho}{2r_s}.
\end{eqnarray}
Recall now that every mode $\psi_{lm}$ in the expansion $\psi=\sum_{lm}Y_{lm}\psi_{lm}$ is subjected to the Schrodinger equation 
\begin{eqnarray}
(\partial_t^2-\partial_{r_*}^2+V(r_*))\psi_{lm}=0,
\end{eqnarray} 
with a potential function in the tortoise coordinates $r_*$ of the form 
\begin{eqnarray}
V(r_*)=\frac{r-r_s}{r}(\frac{r_s}{r}+\frac{l(l+1)}{r^2}).
\end{eqnarray} 
We have the behavior 
\begin{eqnarray}
V(r_*)=\frac{l(l+1)}{r^2_*}~,~r_*\longrightarrow\infty~,~r\longrightarrow\infty.
\end{eqnarray} 
\begin{eqnarray}
V(r_*)=\frac{l(l+1)+r_s^2}{r_s^2}\exp(\frac{r_*-r_s}{r_s})~,~r_*\longrightarrow -\infty~,~r\longrightarrow r_s.
\end{eqnarray} 
The mode $\psi_{lm}$ comes from the brick wall at $r_{*\rm min}$ until it hits the potential at the turning point $r_{*\rm tur}$ defined by the condition 
 \begin{eqnarray}
\frac{l(l+1)+r_s^2}{r_s^2}\exp(\frac{r_*-r_s}{r_s})=\omega^2.
\end{eqnarray} 
Since we are near the horizon, i.e. $r_*\longrightarrow -\infty$, we have $\omega\longrightarrow 0$ unless $l>>1$. The modes with small $l$ are also suppressed from entropy consideration, i.e. low degeneracy.  Thus, for modes with $l>>1$ we obtain the turning point 
 \begin{eqnarray}
r_{*\rm tur}=2r_s\ln\frac{r_s\omega}{l}.
\end{eqnarray} 
Each mode then moves between the brick wall $r_{*\rm min}$ and its own turning point. These are the zone modes (zero modes with support in the near-horizon region only) which dominates the canonical statistical ensemble. The IR box corresponds to a length 
\begin{eqnarray}
L=\Delta r_*=r_{*\rm tur}-r_{*\rm min}=2r_s\ln \frac{2r_s^2\omega}{\rho l}.
\end{eqnarray} 
The quantization of a particle in a box of size $L$ leads immediately to the quantization condition 
\begin{eqnarray}
k_n=\frac{n\pi}{L}\Rightarrow \omega_n\simeq \frac{n\pi}{2r_s\ln \frac{2r_s^2\omega_n}{\rho l}}
\end{eqnarray} 
Obviously, the size of the box shrinks as we increase $l$ until it vanishes when $l=2r_s^2\omega/\rho$. Since $L$ now depends on the modes, we should  make the usual replacement $\sum_{\omega l m}/L\longrightarrow \int  d\omega/2\pi$ as follows
\begin{eqnarray}
\sum_{\omega l m}f(\omega)\simeq 2\int_0^{\infty} \frac{d\omega}{2\pi}f(\omega)\int_0^{2r_s^2\omega/\rho}dl (2l+1)2r_s\ln \frac{2r_s^2\omega}{\rho l}.
\end{eqnarray} 
The factor of $2$ in front is due to the fact that we only integrate over positive frequencies. We get immediately 
\begin{eqnarray}
\sum_{\omega l m}f(\omega)&\simeq &2\int_0^{\infty} \frac{d\omega}{2\pi}f(\omega)\bigg(-\frac{16 r_s^5\omega^2}{\rho^2}\int_0^1 dx x\ln x\bigg)\nonumber\\
&\simeq &\frac{8r_s^5}{\rho^2}\int_0^{\infty} \frac{\omega^2d\omega}{2\pi}f(\omega).
\end{eqnarray} 
As we can see most contribution comes from large angular momenta $l\sim 2r_s^2\omega/\rho$. The energy and the entropy are then given by the estimation (with $\beta=2\pi/a=4\pi r_s$)
\begin{eqnarray}
E&\simeq&\frac{8r_s^5}{\rho^2}\int_0^{\infty}\frac{\omega^3 d\omega}{2\pi}\exp(-\beta\omega)\nonumber\\
&\simeq &\frac{24r_s^5}{\pi\rho^2\beta^4}\nonumber\\
&\simeq &\frac{24 r_s}{\pi\rho^2(4\pi)^4}.\label{below1}
\end{eqnarray} 
\begin{eqnarray}
S&\simeq&E+\frac{8r_s^5}{\rho^2}\int_0^{\infty}\frac{\omega^2 d\omega}{2\pi}\exp(-\beta\omega)\nonumber\\
&\simeq &E+\frac{8r_s^5}{\pi\rho^2\beta^3}\nonumber\\
&\simeq &\frac{8 r_s^2}{\pi\rho^2(4\pi)^3}.\label{below2}
\end{eqnarray} 
In the above equations we are assuming that Hawking temperature $T_H$ is very small and thus $\beta\longrightarrow\infty$. The energy is proportional to $\beta$ while the entropy is proportional to $\beta^2$. We obtain divergent (as expected) expression in the horizon limit $\rho\longrightarrow 0$.  However, if we assume the existence of a stretched horizon away from the mathematical horizon by a distance of the order of the Planck length, then 
\begin{eqnarray}
\rho^2\leq l_P^2=8\pi G.
\end{eqnarray}
We can fix $\rho$ by demanding that the entropy of the field is equal to the full Bekenstein-Hawking entropy, viz 
\begin{eqnarray}
S\equiv\frac{A}{4G}=\frac{8 r_s^2}{\pi\rho^2(4\pi)^3}\Rightarrow \rho^2=\frac{G}{8\pi^5}.
\end{eqnarray} 
The energy becomes with this choice
\begin{eqnarray}
E=\frac{3r_s}{4G}=\frac{3M}{2}.
\end{eqnarray} 
Thus, indeed, one should take $\rho\leq l_P$ in order for the field to carry no more energy and entropy than the black hole itself.

In summary, we have from one hand a divergent entropy in the near-horizon limit $\rho\longrightarrow 0$, while from the other hand the entropy must be, without any doubt, finite equal to the Bekenstein-Hawking value $S=A/4G$. In other words, quantum free field theory gives an overestimation of the entropy. As it turns out, adding interaction will not help but in fact it will make things worse. Indeed, in a $3+1$ dimensional interacting scalar field theory the entropy density is always given by a formula of the form 
\begin{eqnarray}
S(T)=\gamma(T)T^3,
\end{eqnarray} 
where $\gamma(T)$ is the effective number of degrees of freedom at the temperature $T$ and it is a monotonically increasing function of $T$. Hence, since the proper temperature $T(\rho)=1/2\pi \rho$ diverges near the horizon we see that QFT gives always a divergent entropy. Furthermore, since the local temperature diverges in the limit $\rho\longrightarrow 0$ the entropy is indeed mostly localized on the horizon.

In the correct quantum theory of gravity it is therefore expected that the number of degrees of freedom decreases drastically as we approach the horizon. In other words, QFT theory should only describe the degrees of freedom at distances much greater than a Planck distance  away from the horizon, while at distances less than a Planck distance away from the horizon the degrees of freedom may become sparse or they may even disappear altogether. This separation between QFT degrees of freedom and Quantum Gravity degrees of freedom can be achieved by a stretched horizon, i.e. a physical dynamical membrane, at a distance of one Planck length $l_P=\sqrt{G\hbar}$ from the actual horizon, where the temperature gets very large and most of the black hole entropy accumulates. Thus the stretched horizon is a time like surface where real dynamics can take place, and where most of the black hole energy and entropy are localized. It is in thermal equilibrium with the thermal atmosphere, and thus it absorbs and then re-emits infalling matter continuously, while evaporation is seen in this case only as a tunneling process. 
\subsection{Conclusion}


We consider a black hole formed by gravitational collapse as given by the Penrose diagram (\ref{PD3}). The Hilbert space ${\bf H}_{\rm in}$ of initial states $|\psi_{\rm in}\rangle$ is associated with null rays incoming from ${\cal J}^{-}$ at $r=\infty$, i.e. ${\bf H}_{\rm in}={\bf H}_{-}$. The Hilbert space ${\bf H}_{\rm out}$ of final states $|\psi_{\rm out}\rangle$ is clearly a tensor product of the Hilbert space ${\bf H}_{+}$ of the scattered outgoing radiation which escapes to the infinity ${\cal J}^+$ and the Hilbert space ${\bf H}_S$ of the transmitted radiation which falls behind the horizon into the singularity. This is the assumption of locality. Indeed, the outgoing Hawking particle and the lost quantum behind the horizon are maximally entangled, and thus they are space like separated, and as a consequence localized operators on ${\cal J}^+$ and $S$ must commute. We have then 
\begin{eqnarray}
H_{\rm in}=H_-~,~H_{\rm out}=H_+\otimes H_S.
\end{eqnarray}
From the perspective of observables at ${\cal J}^+$ (us), the outgoing Hawking particles can only be described by a reduced density matrix, even though the final state $|\psi_{\rm out}\rangle$ is obtained from the initial state $|\psi_{\rm in}\rangle$ by the action of a unitary $S$-matrix. This is the assumption of unitarity. This reduced density matrix is completely mixed despite the fact that the final state is a maximally entangled pure state. Eventually, the black hole will evaporate completely and it seems that we will end up only with the mixed state of the radiation. This the information paradox. There are six possibilities here:
\begin{enumerate}
\item Information is really lost which is Hawking original stand.
\item Evaporation stops at a Planck-mass remnant which contains all the information with extremely large entropy.
\item Information is recovered only at the end of the evaporation when the singularity at $r=0$ becomes a naked singularity. This contradicts the principle of information conservation with respect to the observe at ${\cal J}^+$ which states that by the time (Page or retention time) the black hole evaporates around one half of its mass the information must start coming out with the hawking radiation.

\item Information is not lost during the entire process of formation and evaporation. This is the assumption of unitarity. But how?
\item Horizon is like a brick wall which can not be penetrated. This contradicts the equivalence principle in an obvious way.
\item Horizon duplicates the information by sending one copy outside the horizon (as required by the principle of information conservation) while sending the other copy inside the horizon (as required by the equivalence principle). This is however forbidden by the linearity of quantum mechanics or the so-called quantum xerox principle \cite{Susskind:2005js}. 
\end{enumerate}


\paragraph{Acknowledgments:}

This research was supported by CNEPRU: "The National (Algerian) Commission for the Evaluation of University Research Projects"  under the contract number ${\rm DO} 11 20 13 00 09$. 

I woul like also to acknowledge the generous funding from the International Center for Theoretical Physics ICTP (Trieste) within the associate scheme $2015-2020$. 


All illustrations found in this review were created by Dr. Khaled Ramda.

\end{document}